\documentclass[aastex]{emulateapj}
\usepackage{graphicx,longtable}
\usepackage{array}
\usepackage{amsmath}
\usepackage{amssymb}
\usepackage{subfigure}
\shorttitle{Supernova Relic Neutrinos and the Supernova Rate}
\shortauthors{Mathews et al.}

\begin{document}
%-----------------------------------------------------------------%
\title{Supernova Relic Neutrinos and the Supernova Rate Problem: Analysis of Uncertainties and Detectability of  ONeMg and Failed Supernovae}
%-----------------------------------------------------------------%

\author{Grant J. Mathews,\altaffilmark{1,2} Jun Hidaka\altaffilmark{2}, Toshitaka Kajino,\altaffilmark{2,3} and Jyutaro Suzuki\altaffilmark{2}  }

\altaffiltext{1}{Center for Astrophysics, Department of Physics, University of Notre Dame, Notre Dame, IN 46556, USA\\}

\altaffiltext{2}{National Astronomical Observatory of Japan, 2-21-1
Osawa, Mitaka, Tokyo, 181-8588, Japan\\}

\altaffiltext{3}{Department of Astronomy, Graduate School of Science, The
University of Tokyo, 7-3-1 Hongo, Bunkyo-ku, Tokyo, 113-0033, Japan\\}

%-----------------------------------------------------------------%
%\date{\today}
%-----------------------------------------------------------------%
\begin{abstract}
Direct measurements of the core-collapse supernova rate ($R_{SN}$) in the redshift range $0 \le z \le1$ 
appear to be  about a factor of two smaller than the rate inferred from the measured cosmic massive-star formation rate (SFR).
This discrepancy would imply that about one half of the massive stars that have been  born
in the local observed comoving volume did not explode as luminous supernovae.
In this work we explore the possibility that one could
clarify the source of  this "supernova rate problem" by detecting 
the energy spectrum of supernova relic neutrinos with a next generation $10^6$ ton
water \v{C}erenkov detector like Hyper-Kamiokande.
First, we re-examine the supernova rate problem.  We make a conservative alternative compilation of the measured SFR 
data over the redshift range 0 $\le z \le$ 7.  We show that, by only including published SFR data for which the dust obscuration has been directly determined,
 the ratio of the observed massive SFR to the
 observed supernova rate $R_{SN}$ has large uncertainties $\sim1.8^{+1.6}_{-0.6}$, and is statistically consistent with no supernova rate problem.  
 If we further consider that a significant fraction of massive stars will end their liives as faint ONeMg SNe or as failed SNe leading to a black hole remnant, then the ratio reduces to  $\sim1.1^{+1.0}_{-0.4}$ and the rate problem is essentially solved.  We next examine the prospects for detecting this solution to the supernova rate problem.
We  first study the sources of uncertainty involved in the theoretical 
estimates  of the neutrino detection rate and analyze whether the
spectrum of relic neutrinos can be used to independently identify  the existence of a supernova rate problem and its source. 
We consider an ensemble of  published and unpublished core collapse supernova  simulation models 
to estimate the uncertainties in the anticipated  neutrino luminosities and temperatures.   We illustrate how the spectrum of detector events might be used to establish the  average neutrino temperature and constrain SN models.
We also consider  supernova $\nu$-process nucleosynthesis to deduce  constraints on  the temperature  of the various neutrino flavors.  
We study the effects of neutrino oscillations on the detected neutrino energy spectrum 
and also show that one might distinguish the equation of state (EoS) as well as  the cause  of the possible missing luminous supernovae
from the detection of supernova relic neutrinos.
We  also analyze a possible  enhanced  contribution  from   failed supernovae  leading to a black hole remnant as a solution to the supernova rate problem.
We conclude that indeed  it might  be  possible (though difficult) to measure the neutrino temperature, neutrino oscillations, the EOS,  and confirm this source of missing luminous supernovae by the detection of the spectrum of relic neutrinos. 
\end{abstract}
%-----------------------------------------------------------------%
\medskip

\keywords{cosmology: theory - diffuse radiation - neutrinos - stars: formation - stars: massive - supernovae: general}

\maketitle
%------------------------------------------------------------------%

\vskip 1.3cm

%-----------------------------------------------------------------%

%%%%%%%%%% THE BODY %%%%%%%%%%

%%%%%%%%%%%%%%%%%%%%%%%%%%%%%%%%%%
\section{Introduction}
%%%%%%%%%%%%%%%%%%%%%%%%%%%%%%%%%%

Massive stars ($M \ge 8$ M$_{\odot}$) culminate their evolution as core collapse supernovae (CC-SNe).  
 Such supernovae  are of particular interest because they are unique sources 
for all three flavors of energetic neutrinos.  
An intensive flux of neutrinos with total energy of order $\sim 10^{53}$ ergs is  emitted over an interval of $\sim 10$ s
during core-collapse supernovae.
This energy corresponds  to almost 99$\%$ of the released gravitational binding energy during the formation of  a 1.4 $M_\odot$
proto-neutron star, and this neutrino luminosity is nearly equally partitioned among  the three  neutrino flavors,  i.e. 
$\nu_{{\rm e}, \mu, \tau}$ and their antiparticles.  
In the case that a  black hole is formed instead of a neutron star in the core-collapse of more massive stars,  a similar energy $\sim 10^{53}$ ergs is expected to be carried away
by more energetic neutrinos during a shorter time interval  $\le 1$ s.   If however, a massive star ends its life as an ONeMg supernova, the neutrino flux and energies are diminished.
In this paper we investigate the possibility of detecting the various contributions to the supernova relic neutrino background in a $10^6$ ton next-generation water \v{C}erenkov detector such as the proposed Hyper-Kamiokande.

Neutrinos are weakly interacting elementary particles, and therefore emerge 
from deep within the interior of core collapse supernovae.  As such, supernova neutrinos 
have the potential to provide  information regarding  the physical processes that take place inside the star.
For the same reason, the emitted neutrinos can almost freely stream  from the time of the early galaxy formation epoch
until the present time without  absorption in the intervening  intergalactic material.  
Since massive stars have lifetimes  ($\sim 10^6$ to $10^7$ y) that are much shorter than 
the cosmic age ($\sim 10^{10}$y),  supernova neutrinos can be considered to have been emitted almost continuously since  the very earliest phase of galaxy formation. 
This  diffuse galactic background of  accumulated neutrinos due to SNe is referred to (e.g.~\cite{tot96} and refs.~therein) as the  supernova relic neutrinos (SRNs).

The detection of these supernova relic neutrinos and their energy spectrum could be used  to study the supernova history from the beginning of 
 galaxy formation \citep{hop06}.  Such data could also provide information on
neutrino properties such as flavor oscillations \citep{cha11,lun12} and/or the neutrino temperatures  produced in supernova explosions. One may even be able to discern \citep{Nak13} the shock revival time from the detected spectrum of relic neutrinos.
Indeed, a next generation Hyper-Kamiokande detector with a mega-ton of pure water is currently  being planned with a goal of measuring the relic neutrino background spectrum.  Moreover, results from the Super-Kamiokande collaboration \citep{SK12} already place interesting limits on the total neutrino energies and temperatures from the background spectrum of supernova relic neutrinos \citep{SK12,Sek13}.

However, theoretical predictions of the detection rate of supernova relic neutrinos are still subject to 
a number of  uncertainties. 
These uncertainties include   the star formation rate and/or  supernova rate and 
the spectral energy distributions of the three flavors of supernova neutrinos. 
These  in turn  depend upon the explosion mechanism, whether the final remnant is a neutron star or black hole,
 the equation of state (EoS) for proto-neutron stars, and  the possibility of neutrino oscillations and self interactions, etc.
 One goal of the present work is to better clarify these uncertainties.

In this context, it is of particular interest to the present work that an apparent   discrepancy  has been noted  \citep{hor11}    between 
the supernova rate deduced from observations of the cosmic SFR [e.g. \cite{hop06}] at multiple wavelengths, and the detected core-collapse supernova rate ($R_{SN}$)
\citep{lea11,li11a, li11b,maoz11}.   
The inferred $R_{SN}$ in the redshift range 0 $\le z \le$ 1 
is about a factor of two smaller than that deduced from the measured SFR.  This has been dubbed \citep{hor11} as  the "supernova rate problem."
This discrepancy suggests that either about  half of all massive stars are obscured when they explode, or that
they do not explode as luminous supernovae.

One explanation for this latter possibility  that we explore in this paper is that some of the expected luminous supernovae actually end their lives as faint 
ONeMg supernovae.  Such supernovae should result from progenitor masses  in the range 8$M_{\odot} \le M \le 10M_{\odot}$ \citep{Isern91}.
%If there is a wider mass range for these SNe and/or an enhanced initial mass function (IMF) for this mass range in the redshift range of $0 \le z \le 1$,   
%this might explain the discrepancy.
%Since the luminosity of  ONeMg supernovae is known to be an order of magnitude smaller than
%that of ordinary SNe, they may not be observed efficiently.  The contribution of ONeMg supernovae has been considered in \cite{lun12}.  Here we expand on that work by analyzing the uncertainties and possible role and detectability for an enhanced fraction of ONeMg SNe as a solution to the supernova rate problem.
%However, this scenario may be unlikely because  the required  IMF has not been detected  in  external galaxies.

Another aspect   
is that a significant fraction of relatively massive stars with progenitor masses
in the range $M \ge  25$ M$_{\odot} $
evolved to failed supernovae leaving black hole remnants rather than to become SNIb,c.  In this case,  the ejecta  falls back 
onto the central black hole.  These failed supernovae
have small  explosion kinetic energies  and are much less luminous.  In a series of papers \citep{lun09,Yang11,kee12} the contribution of failed SNe to
the relic neutrino signal has been analyzed.  Here, we expand upon that study 
by considering the  role and associated uncertainties and detectability  of failed SNe and also the possibility of an enhanced fraction of such events 
as a means to resolve the supernova rate problem.

Theoretical studies \citep{sum05,sum08,Nak08} of  such failed supernovae, however,   show slightly higher 
neutrino luminosity with much higher neutrino temperatures than those of ordinary supernovae.
Depending upon  the EoS employed, the difference in neutrino temperatures among the different flavors is
even larger than the difference in ordinary supernovae.
%Although it takes only about 500 ms to 1 s before the apparent horizon 
%of the black hole appears,
%the proto-neutron star still ejects an enormous   flux of neutrinos and their total energy 
%is even larger than that of ordinary core-collapse supernovae.
This  opens up the possibility that the spectrum of energetic 
relic supernova neutrinos could be  dominated by a contribution  from failed supernovae.
A detection of the spectrum of these relic neutrinos could thus provide insight into the supernova rate problem as well as neutrino oscillation effects and the effects of 
the EoS on  supernova collapse \citep{lun12}.

One  purpose of the present work, therefore,  is to explore the possibility  
that one  can solve the "supernova rate problem" by detecting 
the energy spectrum of supernova relic neutrinos.
For this purpose we first analyze possible uncertainties in  both the measured core-collapse supernova rate ($R_{SN}$) 
and star formation rate (SFR) 
over the redshift range of 0 $\le z \le$ 7. 
We then study the sources of uncertainty in  theoretical 
calculations of the relic neutrino detection rate.  

If we consider  only observations for which  the dust correction was taken  into account,
we find that  the supernova rate problem has a large uncertainty, i.e.  $\sim1.8^{+1.6}_{-0.6}$ for the ratio of the supernova rate deduced from the observed massive star formation rate (SFR) to  the directly observed core-collapse supernova rate $R_{SN}({\rm Obs}$).  This is statistically consistent with no supernova rate problem.  
 Moreover, if we consider that a significant fraction of massive stars will end their lives as faint ONeMg SNe or as failed SNe leading to a black hole remnant, then the ratio reduces to  $\sim1.1^{+1.1}_{-0.4}$ and the rate problem is essentially solved. 

We next analyze the uncertainties in detecting this solution to the supernova rate problem.  
We adopt an ensemble of published and unpublished  supernova simulation models 
in order to estimate the uncertainty in the source neutrino luminosities and temperatures.   
These are taken from core-collapse supernova models leaving either a neutron star or black hole (including  possible collapsar gamma-ray burst models) as a remnant, 
or leading to ONeMg supernovae.
We also incorporate several recent studies of  supernova nucleosynthesis
in order to constrain the neutrino temperatures.

After clarifying and minimizing the temperature uncertainties from the theoretical calculations, 
we study the effects of different SN models, equations of state, and the effects of the neutrino oscillations on the detected spectrum.  We explore  
whether one can  measure the neutrino temperature, distinguish the EoS, 
and even constrain the unknown neutrino
mass hierarchy,
through the sensitivity of the resultant detected neutrino spectrum.
Finally, we  explore whether  an enhanced  contribution of  "failed supernovae" 
might be detectable and thereby  confirm whether these events  are the source of the  "supernova rate problem"   \citep{hor11}.

 \section{star formation rate}
 \label{SFR}
 As noted in the introduction, it is an intriguing conundrum in observational cosmology 
that the measured core-collapse supernova rate [$R_{SN}({\rm Obs})$]  in the redshift range 0$\le z \le$1 
appears to be about a factor of two smaller than the core-collapse supernova rate deduced from the measured cosmic massive-star formation rate (SFR).

For the present study we consider  three different star formation rates as a means to estimate the overall uncertainty in the SFR dependence.  For one,  
we deduce a revised  SFR
 based upon a piecewise linear  fit to the observed star formation rate vs. redshift as shown on Figure \ref{fig:1}.  
 For illustration we also consider the theoretical  SFR model (not shown) of \cite{kob00} with and without a correction for  dust extinction.
 % as  shown in
% Fig.\ref{fig13}. 
 \subsection{SFR Data}
In view of the importance of clarifying the possible uncertainty in  the supernova rate problem we have made an alternative fit to the observed SFR data.  As a starting point we use the star formation rate data set employed in \cite{hor11}.  This data set was based upon the compilation of \cite{hop06}.  A critical part of the SFR data, however, is the correction for extinction by dust.  The data of \cite{hop06} is heavily weighted in the interval of $z = 0-1$ by the UV data of \cite{bal05}, \cite{wol03} and \cite{arn05}.  However, these data were not corrected for obscuration.  To implement effective obscurations they utilized the FIR SFR densities \citep{flo05} up to $z = 1$.   Although this seems reasonable, the associated dust correction is quite large (up to 80\% at $z = 1$) and may lead to large systematic errors \citep{kob13}.  Hence, as an alternative here we consider a more conservative data set that includes only published SFR data for which the dust correction was made by direct  observation.  This limits our data  to a set of  123 data points in X-ray, $\gamma$-ray, UV, Optical, NIR-H$\alpha$, and FIR  SFR observations that include dust corrections.  In addition to a subset of points from \cite{hop06} these data include some newer  data  \citep{red08,jam09}.

As can be seen on Figure \ref{fig:1}, the limitation of the data set  to only those that included dust corrections implies a larger uncertainty and lower normalization in the critical redshift range of $z = 0-1$ than the compilation of  \citet{hop06}. This has also been  noted in  other works \citep{ouc09,kob13} that discuss the large uncertainties in the dust corrections.  Our claim is not that this is a better data set, but that it is a more conservative choice of the data compilation that gives an alternative assessment of the uncertainty in the supernova rate problem.
  
\subsection{fit to the data}
  For our $\chi^2$ fitting, we use the  continuous linear [in $\log{ (1 + z)}$] piecewise star formation rate 
$ \psi_*(z)$  as a function of redshift as given in  \citet{yuk08},  i.e. 
\begin{equation}
 \psi_*(z) = \dot{\rho_0} \left[ (1 + z)^{\alpha\eta} + 
\left(\frac{1 + z}{B}\right)^{\beta\eta} + 
\left(\frac{1 + z}{C}\right)^{\gamma\eta}\right]^{\frac{1}{\eta}} , 
\label{4-3}
\end{equation}
where our  deduced parameters for the best fit and $\pm 1 \sigma$ upper and
lower limits are given  in Table \ref{tbl.1} .  When deducing the $\pm 1\sigma$ errors in the fit, we use the best-fit value of $\alpha = 4.22$.  This keeps the curves parallel at low redshift ($z = 0-1$) and is consistent with the dispersion in the data.  By keeping the curves parallel, the determination of the standard deviation in the fits is just given by a change of normalization to give $\Delta \chi^2 = 1$.  The CC-SN rate evolution is assumed to follow the same $(1+z)^\alpha$ behavior  where $\alpha$ is fixed to 4.22 obtained from SFR fits.  This makes a comparison of the overall normalization possible.  

In Eq.~(\ref{4-3}), we have chosen a value of the smoothing parameter $\eta=-10$ consistent with that used in previous works [cf. \cite{hop06,yuk08,hor11}]. The best fit corresponds to $\chi^2 = 272$ for 118 degrees of freedom implying a reduced $\chi_r^2 = 2.3$.  Because our adopted data set has a larger dispersion, our $\chi^2$ is larger than the fits of \cite{hop06}, \cite{yuk08}, and \cite{hor11}.  In addition, the  data at low redshift cause the best fit slope to be somewhat steeper than in previous works, i.e. $\alpha = 4.2$ vs. $\alpha = 3.4$ in \cite{hop06}, \cite{yuk08} and\cite{hor11}.  This larger value for $\alpha$ also implies that the absolute fit to the SFR at $z=0$ is lower, i.e. we deduce $\dot \rho_0 = 0.010^{+0.01}_{-0.004} $ M$_\odot$ y$^{-1}$ (see Table 1) vs. $\dot \rho_0 = 0.017 $ M$_\odot$ y$^{-1}$ in \cite{hor11}.
This slope, however, also reduces the observed supernova rate so that the supernova rate problem remains at about the same factor as noted below.

We also note that we have not included a correction for the flux sensitivity at high redshift as pointed out in \cite{Kistler09}.  Hence, one should not take our fit seriously in the redshift range $z > 4$.  As noted below, the SFR at high redshift contributes negligibly to the observed SRN spectrum.  Thus, for this paper, we are almost entirely concerned with the SFR at low redshift where this correction is not necessary.

\begin{figure}[h]
%\epsscale{.30}
\begin{center}
\includegraphics[width=4.5in]{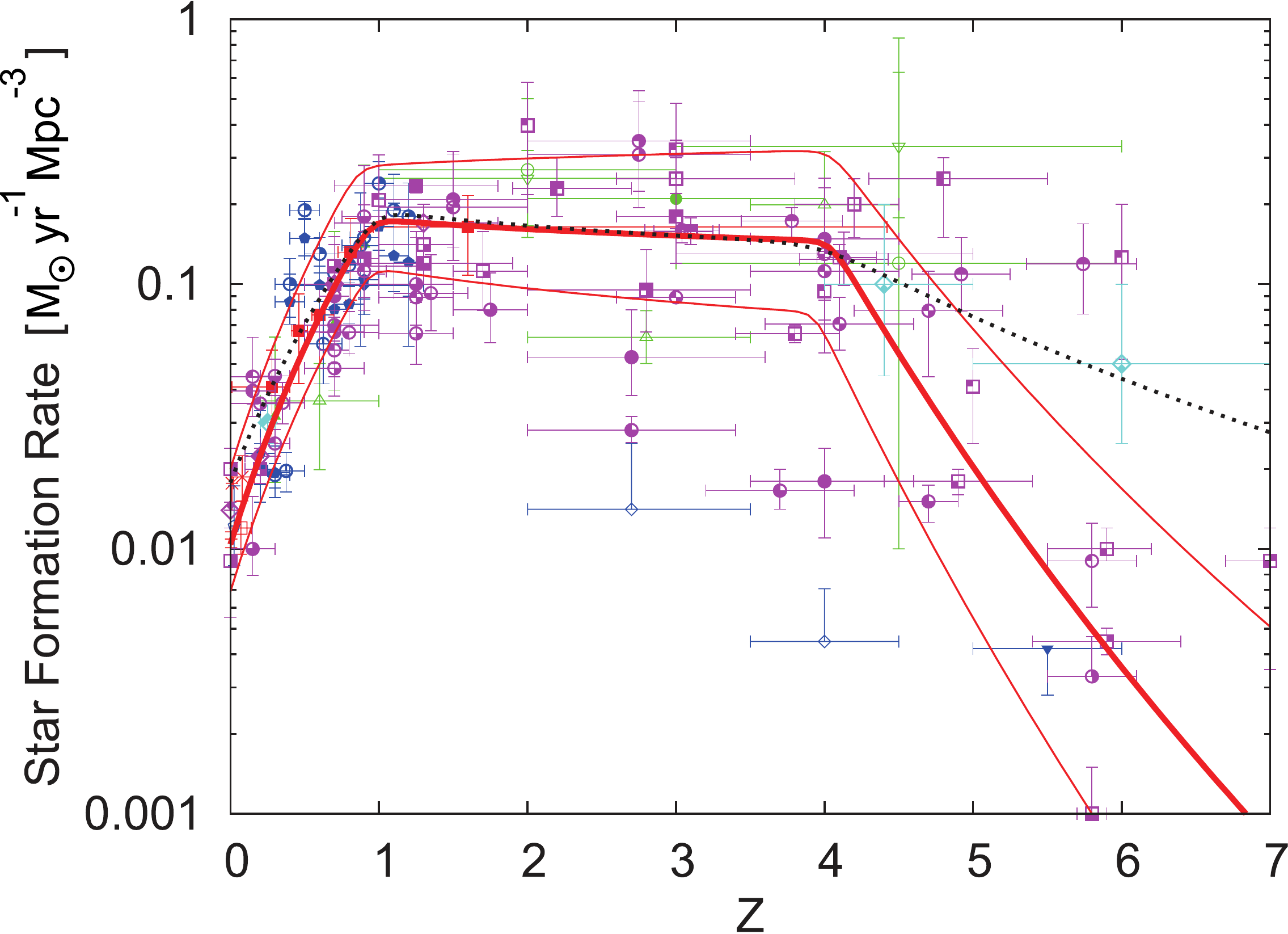}
\caption{[Color online] Piecewise linear  star formation rate function from our fit to the subset of  observed dust-corrected data.
Red, blue, magenta, light blue, and green points show the observed data in IR,
optical, UV,  X-ray / $\gamma$-ray, and  radio bands, respectively. 
 Red lines show the SFR as  function of redshift $z$ deduced from $\chi^2$
fitting, along with the $ \pm 1\sigma$ upper and lower limits to the  SFR.  The reduced $\chi^2_r$ for the fit is 2.3. 
The black dotted line shows the SFR based upon the data set used in \citet{yuk08}.
\label{fig:1}}
\end{center}
\end{figure}

%\subsection{Observed supernova rate}
Figure \ref{fig:1} shows the data-points from rest-frame X-ray, $\gamma$-ray, UV,
Optical, NIR-H$\alpha$, FIR to sub-mm observations compared with our deduced   SFR $\psi_*$(z) and that given in \cite{yuk08}.
 The   thick line shows the best fit SFR.  The    thin lines show  the $\pm 1 \sigma$ upper and lower limits to the  SFR.  As noted above  a key difference between  our star formation rate data  and that of \cite{yuk08} is the limitation to data with published extinction corrections.  
 
 As an illustration of the systematic error due the dust extinction corrections,   in our analysis below we also consider the theoretical star formation rate deduced in \cite{kob00} from a fit to a limited data set both with and without dust corrections.  As shown in that paper, the magnitude of the dust correction  is comparable to the error bars in the piecewise linear star formation rate of Figure \ref{fig:1}.  
 
 On the other hand, as shown below,  there is little difference between the relic neutrino detection rate with either form for the SFR.   For our purposes we will mainly utilize the piecewise linear SFR as the best direct representation of the observed massive star formation rate.  However, for illustration, we  also make some comparisons with the rate of \cite{kob00} with and without dust correction.

\subsection{Core-Collapse Supernova Rate}
To deduce the core-collapse supernova rate from the total SFR one must determine the fraction of total stars that produce observable supernovae.  In the estimated visible supernova  rate of \cite{hor11} it was assumed that all massive progenitors  from 8 to 40 M$_\odot$ lead to visible SNe.  The higher end of this mass range corresponds to SNIb,c, while the lower end corresponds to normal SNII.   

We similarly include   contributions from both normal core-collapse SNII and  SNIb,c,  the latter of which accounts for about $\sim 25 \pm 10$\% of observed core-collapse supernova rate at redshift near $z=0$ \citep{Smartt09}. However, we treat the possibility of failed SNe forming a black hole and dim NeOMg supernovae in the low mass end.
Hence, for the total observed core-collapse  SN rate we write:
\begin{equation}
 R_{SN}(z) = R_{SNII}(z) + R_{SNIb,c}(z)
\label{RSNtot}
\end{equation}
The mass range for Type-II supernova progenitor stars is taken to be from 8 to 25 M$_\odot$.  The minimum mass for observed CC-SN progenitors appears to be $\approx 8 \pm 1$ M$_\odot$ \citep{Smartt09}.  Nevertheless, there is large uncertainty in the degree to which progenitors have experienced mass loss and the true initial progenitor mass could be greater.  Moreover, it is believed  \citep{Heger03} that initial progenitor stars in the mass range of 8-10 M$_\odot$ collapse as an electron degenerate ONeMg core and do not produce bright supernovae.  Hence, we also consider a lower limit of 10 M$_\odot$ as discussed  below when allowing for the possibility of ONeMg SNe.  

Regarding SNIb,c, although it is known \citep{Smartt09} that they are associated with massive star forming regions, there is some ambiguity as to their progenitors.  There are two theoretical possibilities.  \cite{Heger03} and \cite{Fryer99} have studied the possibility that massive stars ($M \sim 25-100$ M$_\odot$) can  shed their outer envelope by radiative driven winds leading to bright SNIb,c supernovae.  This source for SNIb,c was adopted in \cite{hor11} by assuming that all stars with  ($M \sim (25-40) $ M$_\odot$) explode as SNIb,c.  We also adopt this as one possibility. We note that  \cite{hop06} adopted a range of (25-50) M$_\odot$ for SNIb,c.  Changing the upper mass limit from 40 50 M$_\odot$ to 50 M$_\odot$ makes only a slight difference (~1\%) in the inferred total core-collapse supernova rate. 

Theoretically,  however, this single massive-star paradigm does not occur as visible SNIb,c supernovae until well above  solar metallicity \citep{Heger03}.  At lower metallicity  most massive stars with $M > 25$ M$_\odot$ collapse as failed supernovae with black hole remnants.  Hence, this mechanism is not likely to contribute to the observed SNIb,c rate at the higher redshifts of interest here.  On the other hand, it has been estimated \citep{Fryer99} that 15 to 30\% of massive stars up to $\sim 40$ M$_\odot$ are members of interacting binaries that can shed their outer envelopes by roche-lobe overflow \citep{Podsiadlowski92, Nomoto95} leading to bright SNIb,c supernovae.  We also consider  that hypothesis here.  To do so we  take  a  fraction $f_b$ of all massive stars in the range of (8 to 40) M$_\odot$ to result in bright SNIb,c via binary interaction.   We adopt $f_b =0.25$ consistent with the observed SNIb,c rate \citep{Smartt09} and the theoretical estimate of \cite{Fryer99}. 

Hence, for $R_{SNIb,c}(z)$ we write two possibilities.  One is:
\begin{equation}
\begin{array}{c}
 R^s_{SNIb,c}(z) =  \psi_*(z) \times \frac{\int_{25M_\sun}^{40M_\sun}dM \phi(M)}{\int_{M_{min}}^{M_{max}}dM
   M\phi(M)} ~~,
\end{array} \label{SNIbc-1}
\end{equation}
where the superscript $s$ denotes the single-star paradigm.  The other possibility is
\begin{equation}
\begin{array}{c}
 R^b_{SNIb,c}(z) = f_b \psi_*(z) \times \frac{\int_{8M_\sun}^{40M_\sun}dM \phi(M)}{\int_{M_{min}}^{M_{max}}dM
   M\phi(M)} ~~,
\end{array} \label{SNIbc-2}
\end{equation}
where superscript $b$ denotes the interacting binary paradigm.

Similarly, then the rate of normal core-collapse SNII, $R_{SNII}(z)$, is given by
\begin{equation}
\begin{array}{c}
 R_{SNII}(z) =  (1 - f_b) \psi_*(z) \times \frac{\int_{8M_\sun}^{25M_\sun}dM \phi(M)}{\int_{M_{min}}^{M_{max}}dM
   M\phi(M)} ~~,
\end{array} \label{4-5-1}
\end{equation}
where the case of the interacting binary SNIb,c scenario, $R_{SNII}(z)$ is reduced   to allow a fraction $f_b$ of stars to be in  interacting binaries.

As in \cite{hor11} we  adopt the modified  broken power-law Salpeter A IMF \citep{Baldry03},
\begin{equation}
\phi(M) = M^{-\zeta} ~~,
\end{equation}
with $\zeta = 2.35$ for stars with $M \ge 0.5$ M$_\odot$ and $\zeta = 1.5$ for $0.1$ M$_\odot  < M < 0.5$ M$_\odot$.  The star formation rate $\psi_*(z)$ is taken to be the piecewise linear rate shown in Figure \ref{fig:1}.  For the integration normalization we adopt M$_{min}= 0.1$ M$_\odot$ and M$_{max} = 125$ M$_\odot$.  We note that \cite{hop06} and \cite{hor11} adopted M$_{max} = 100$ M$_\odot$, while a number of authors [e.g.~\cite{Ando04,lun09,lun12}] use 125 M$_\odot$ based upon the models of \cite{Heger03}.  Nevertheless, as noted in \cite{hop06},  there is almost no contribution from stars above  100 M$_\odot$ and  there is no significant difference ($\sim 1$\%) between the use of 125 or 100 M$_\odot$ as the upper limit.  Hence, there is no need to correct the data for whether the observers employed 100 or 125 M$_\odot$ in the upper limit of the IMF used to deduce the  SFR.

\subsection{Failed Supernovae}
For failed supernovae  (fSNe)  we consider  all single massive stars with progenitor masses
with $M > 25$ M$_{\odot} $, \citep{Heger03},  although this cutoff is uncertain.  For stars above this cutoff mass   the whole star collapses leading to a black hole and no detectable supernovae. 
Allowing for 25\% of stars with $M  = (25-40)$ M$_\odot$ to form SNIb,c by binary interaction as noted above, then 
the rate for failed supernovae is given by,
\begin{eqnarray}
 R({\rm fSN})  &= & (1-f_b) \psi_*(z) \times \frac{\int_{25 M_\sun}^{40M_\sun}dM \phi(M)}{\int_{M_{min}}^{M_{max}}dM
   M\phi(M)} ~~ \nonumber \\
   &&+  \psi_*(z) \times \frac{\int_{40M_\sun}^{125 M_\sun}dM \phi(M)}{\int_{M_{min}}^{M_{max}}dM
   M\phi(M)} ~~.
 \label{4-5-3}
\end{eqnarray}

\subsection{ONeMg Supernovae}
We also consider here the possibility of  ONeMg (or electron-capture) supernovae.  Such stars involve  explosion energies and luminosities that are an order of magnitude less than normal core collapse supernovae.  Hence, they  may go undetected. ONeMg supernovae are believed to arise from progenitor masses  in the range
of 8 $M_{\odot} \le M \le 10 M_{\odot}$.  Such supernovae occur  \citep{Isern91} when an electron-degenerate ONeMg  core reaches a critical density 
(at $M_{core} \sim 1.4$ M$_\odot$).  In this case the electron Fermi energy exceeds the threshold for electron  capture on $^{24}$Mg.  These electron captures  cause a loss of hydrostatic support from the electrons and also a heating of the material.   Ultimately, this leads to a dynamical collapse and the ignition of an O-Ne-Mg burning front that consumes the star.  The rate for ONeMg SNe is then taken to be,
\begin{equation}
\begin{array}{c}
 R({\rm ONeMg~SN}) = (1-f_b) \psi_*(z) \times \frac{\int_{8M_\sun}^{10M_\sun}dM \phi(M)}{\int_{M_{min}}^{M_{max}}dM
   M\phi(M)} ~~,
\end{array} \label{4-5-2}
\end{equation}
where  the integration limits in the denominator are the same as for $R_SN$  as given below Eq.~(\ref{4-5-1}), and the factor of $(1-f_b) = 0.75$ arises in the binary SNIb,c paradigm.   In this case, the lower limit for the progenitors of SNII and SNIb,c becomes 10 M$_\odot$ instead of 8 M$_\odot$.
 
With our adopted integration limits and IMF parameters, the reduction factor from the SFR to the deduced supernova rate is $R_{SN}({\rm calc})/SFR = 0.010$ for the single-star SNIb,c  paradigm.  The factor reduces to 0.0093 if failed supernovae are removed for $M>25$ M$_\odot$ in the interacting binary scenario, and if faint ONeMg supernovae are undetected  for stars with 
$8$ M$_\odot$ $< M < 10$ M$_\odot$, the factor reduces to 0.0063.

\subsection{Supernova Rate Problem}
To analyze the supernova rate problem, one must compare the above scenarios for visible core-collapse supernovae with the rate of those observationally detected.
As in \cite{hor11} we fit the seven measured core collapse supernova rates $R_{SN}({\rm Obs})$ in the redshift range of $z= 0-1$ deduced in  \cite{li11b}, \cite{cap99}, \cite{bot08},
\cite{cap05}, \cite{baz09} and  \cite{dah04} as given in Table 1 of \cite{hor11}.   Also, as in \cite{hor11} we  use the functional form
\begin{equation}
R_{SN}({\rm Obs}) = a (1+z)^b~~.
\end{equation}
However,  we find a much better fit for $b = 4.22$ ($\chi^2 = 1.8$) from our fit to the SFR rather than $b=3.4$ ($\chi^2 = 2.5$) as in \cite{hor11}.   With $b = 4.22$, the best fit normalization is for $a = 0.60 \pm 0.07$.

With this observed core collapse SN rate we obtain the following characterizations of the supernova rate problem.
\begin{eqnarray}
\frac{R_{SN}({\rm calc})}{R_{SN}({\rm obs})} &=& 1.8^{+1.6}_{-0.6}  ~~~{\rm single~star~SNIb,c} ~, \\
=&& 1.6^{+1.5}_{-0.6}  ~~~{\rm binary~SNIb,c + fSNe} ~, \\
=&& 1.1^{+1.0}_{-0.4}  ~~~{\rm binary~SNIb,c+fSNe} \nonumber \\
&&{\rm +~ONeMg~SNe} ~.
\end{eqnarray}

 If we adopt the likely possibility that failed black-hole producing supernovae and faint ONeMg supernovae account for the unobserved supernovae, then
 there is essentially no supernova rate problem.   
 %Figure \ref{fig:2} then shows our version of the supernova rate problem.  This figure compares the measured total core-collapse supernova rates  $R_{SN}$ in the redshift range of $z= 0-1$ deduced in 
%\citet{li11b}, \citet{cap99}, \citet{bot08},
%\citet{cap05}, \citet{baz09}, \citet{dah04}  with the inferred cosmic $R_{SN}$ based upon the Salpeter IMF and our piecewise-linear SFR as summarized in Figure \ref{fig:1}.
%One can see that our inferred supernova  rate based upon a fit to the dust corrected observations only  is closer to the observed supernova rate.
Indeed, with our inferred uncertainties, all three possibilities are consistent with no supernova rate problem at the $1\sigma$ level  as also noted in \cite{kob13}.  

On the other hand, as much as a factor of 2 discrepancy as deduced in  \cite{hor11} is not ruled out.  The reason for the difference between the ratio deduced here and that in \cite{hor11} is that we have only included data with published corrections for dust obscuration.  One can see from the dotted line in Figure \ref{fig:1} that the SFR of \cite{hop06}, \cite{yuk08} and\cite{hor11} is systematically higher in the redshift intervals $z = 0-1$ and $4-7$.  The discrepancy at high redshift is not particularly relevant to the present work. The star formation rate at high redshift requires  a correction \citep{Kistler09} for unseen galaxies and is probably best deduced from GRBs at high redshift.   The main focus of this study, however, is the SFR in the interval $z = 0-1$.   

In what follows we will examine whether this solution to the supernova rate problem can be determined observationally.  
A key point is that, although black-hole forming and ONeMg  SNe are optically dim, they are quite luminous in neutrinos.  
Hence, as a means to test this hypothesis we  analyze whether the relic neutrino detection rate and spectrum can be used to identify one or the other  possible solutions to the putative supernova rate problem.  To do that we next analyze the relic neutrino background, the detector response, and the associated uncertainties. 

\section{SRN background}
In this section we  analyze the detection rate and spectrum along with various theoretical uncertainties in the predicted  SRN background. 
As a fiducial detector, we consider a Hyper-Kamiokande next-generation \v{C}erenkov detector. 
This  detector is proposed \citep{abe11} to consist of a mega-ton of pure water laden with 0.2$\%$ GdCl$_3$ to reduce the background \citep{bea04}.
Our goal is to clarify  the uncertainties involved in estimates of the detection rate.

Most  of the uncertainties arise from the energy flux of the arriving SRNs.
 The presently observed SRN flux spectrum  ${dN_\nu}/{dE_\nu} $ can be derived \citep{str05,yuk07} from an integral over the  cosmic
 redshift $z$, of the neutrinos emitted per supernova and the cosmic supernova rate per unit redshift $R_{SN}(z)$:
% Eq.9:
\begin{eqnarray}
 \frac{dN_\nu}{dE_\nu} &=&
 \frac{c}{H_0}\int_0^{z_{max}}R_{SN}(z)\frac{dN_\nu(E^\prime_\nu)}{dE^\prime_\nu} \nonumber \\
 &\times&
 \frac{dz}{\sqrt{\Omega_m(1 + z)^3 + \Omega_\Lambda}} \label{4-15}~~,
\label{nuflux}
\end{eqnarray}
where $z_{max}$ is the redshift at which star formation begins. For our purposes  we take $z_{max}= 5$.  $R_{SN}$ is the number of
supernovae per unit redshift  per comoving volume as discussed in the previous section.  The quantity $dN_\nu(E^\prime_\nu)/dE^\prime_\nu$ is the emitted neutrino spectrum at the source, where the energy  $E'_\nu= (1+z)E_\nu$ is the energy at emission, while $E_\nu$ is the redshifted energy observed in the detector. In the present work we improve upon previous estimates based only upon the neutrino emission from a typical type II core collapse supernovae.  As described below, we also average over  the neutrino emission from ONeMg electron-capture  supernovae and the truncated neutrino emission from the collapse of massive stars to form black holes in addition to averaging over the  range of progenitor models for normal Type II supernovae. 

Here and throughout we assume a standard $\Lambda$CDM cosmology.
The quantities $\Omega_m$ and $\Omega_\Lambda$ are the contributions to the closure density of total baryonic plus dark matter and the cosmological dark energy, respectively.  
In this work, we adopt  a standard $H_0= 70$ km s$^{-1}$ Mpc$^{-1}$, $\Omega_m = 0.3,~\Omega_\Lambda = 0.7$ cosmology [as in \cite{hor11}].  This is consistent with the best-fit WMAP9 parameters \citep{hin13}, i.e. $H_0 = 68.92^{+0.94}_{-0.95},~\Omega_m = 0.2855^{+0.0096}_{-0.0097}$ and $ \Omega_\Lambda = 0.717 \pm 0.011$.
 
 In this study, we calculate SRN detection rate for
a water \v{C}erenkov detector with a fiducial 
volume of 1.0 $\times$ 10$^6$ ton  (laden with 0.2$\%$ GdCl$_3$). As shown below, the threshold  energy
for  SRN detection is $\sim10$ MeV due  to the existence of background $\bar{\nu_e}$ emitted
from terrestrial nuclear reactors, while  the upper detection  limit is $\sim33$ to 37 MeV due  to the 
existence of the atmospheric $\bar{\nu_e}$ background. The SRN energy spectrum at
detection can  then be written  as:
\begin{equation}
\frac{dN_{event}}{dE_{e^+}} = N_{target} \cdot \varepsilon(E_\nu) \cdot
\frac{1}{c} \cdot \frac{dF_\nu}{dE_\nu}
\cdot \sigma(E_\nu) \cdot \frac{dE_\nu}{dE_{e^+}} \label{3-1-1}
\end{equation}
where $N_{target}$ is the number of target particles in the water
\v{C}erenkov detector, $\varepsilon(E_\nu)$ is the efficiency for neutrino
detection, ${dF_\nu}/{dE_\nu}$ is the incident flux of SRNs,
and $\sigma(E_\nu)$ is the cross section for  neutrino absorption:
$\bar{\nu_e} + p \to e^+ + n$, and  $E_\nu = E_{e^+} + 1.3~{\rm MeV}$. For simplicity we set
$\varepsilon(E_\nu)$ = 1.0, and we use the  cross
sections given  in \citet{str03b} when calculating the reaction rate of SRN with
target particles in the detector.

 We only consider the detection of $\bar{\nu_e}$, because the cross section  in a
water \v{C}erenkov detector for
the $\bar{\nu_e} + p \to e^+ + n$ reaction is about 10$^2$ times larger
than that for $\nu_e$ detection via  $\nu_e + n \to e^- + p$. 

\section{Neutrino spectra from Core Collapse  SNe}
%{SRN sources (SN, failed SN, GRB)}
Next, we analyze the uncertainties in the emitted neutrino spectrum, i.e.  ${dN_\nu(E^\prime_\nu)}/{dE^\prime_\nu} $ in Eq.~(\ref{4-15}).
\subsection{Neutrino spectrum for CC-SNe}
 Although  the neutrino temperature T$_\nu$ in core collapse SNe may also  depend upon progenitor mass \citep{lun12},
 we note that the dependence on progenitor mass is rather small compared to the uncertainty in the neutrino temperatures themselves. 
 The reason for this is  that the mass  of most observed neutron stars is rather  tightly
constrained to be $\sim 1.4$ M$_\sun$ (with a maximum mass of $\sim 2 - 3$ M$_\sun$).
Because the mass is also  proportional to the degeneracy pressure, the narrow  range of observed remnant neutron star masses
suggests that the associated neutrino temperatures  should  also
be tightly constrained. Hence,  for our purposes we  ignore the dependence on progenitor mass.
In this study, therefore, we adopt a fiducial  SN 1987A model [i.e. progenitor mass $\simeq$ 16.2
M$_{\sun}$, remnant mass $\simeq$ 1.4M$_\sun$, liberated binding energy
$\simeq$ 3.0 $\times$ 10$^{53}$ erg  \citep{arn89}].  We assume that this model  is representative of  every core-collapse SN with progenitor masses
in the range of 8 to  25 M$_\sun$ and also every SNIb,c over the adopted mass range for 8-40 M$_\odot$ in any of the possible paradigms.  
Furthermore, we assume  that the
proto-neutron star formed during  all core-collapse SN explosions is in thermodynamic equilibrium.
Hence, the liberated binding energy is equally divided among the  6 neutrino species (3 flavors and their
anti-particles).  Even in the context of this single progenitor model, however, there is a range of predicted neutrino temperatures and chemical potentials in various supernova core-collapse 
simulations in the literature.  We now survey them as a means to estimate the uncertainty in the neutrino detection rate 
due to  these parameters.
%\subsubsection{Neutrino temperature in supernova explosion}
%\begin{description}
%\subsubsection{Constraint of T$_\nu$ - T$_{\nu_x}$ and T$_{\bar{\nu_e}}$}
%\subsubsubsection{Constraint of T$_\nu$ :(1) T$_{\nu_x}$}

\subsection{Constraint on T$_{\nu_e}$ T$_{\bar{\nu_e}}$, T$_{\nu_\mu}$, T$_{\nu_\tau}$}
\label{temps}
As discussed below, the uncertainty in the neutrino temperatures constitutes the largest present uncertainty in the expected relic neutrino detection rate.  Indeed, recent results from Super-Kamiokande \citep{SK12,Sek13} already place some constraints on the relic neutrino temperatures.  
In this section we summarize the  independent constraints and their uncertainties on theoretical models for the temperatures of emitted neutrinos.
One result we make use of is from   \citet{yos08} where it was concluded   that the temperature (T$_{\nu_x}$) of $\nu_\mu$, $\nu_\tau$ and their
anti-neutrinos  should   be in the range 4.3  to 6.5 MeV for SN
1987A models. This constraint 
%of T$_{\nu_x}$ in \citet{yos08}
is based \citep{yos97}  upon %a galactic chemical evolution (GCE) model 
 the observed solar-system meteoritic abundance ratio of boron isotopes.  Similarly,  
 it has  been demonstrated \citep{hay10}  that both T$_{\nu_e}$ and T$_{\bar{\nu_e}}$ 
should be $\sim 4.0$ MeV  to produce the correct abundance ratio of $^{180}$Ta/$^{138}$La.  

The range of uncertainty in all neutrino types from core collapse models is indicated in 
 Figures \ref{fig:3}a and \ref{fig:3}b along with Tables \ref{tbl.2-1}-\ref{tbl.2-3}.  The tables  summarize values derived in previous models for T$_{\nu_e}$, T$_{\bar{\nu_e}}$, and 
T$_{\nu_x}$, the dimensionless chemical potentials $\eta_{\nu_i}$, and whether the temperatures are derived from the average  $<e>$, or the RMS energy $<e^2>^{1/2}$.   Results are shown from both  original
published numerical simulations (filled circles in the figures) as summarized in Tables \ref{tbl.2-1}-\ref{tbl.2-3}, and
unpublished private communications (empty circles).  When the chemical potentials at the neutrino sphere were not apparent in the simulations we assumed a small chemical potential and deduced temperatures from the approximation  $T_i \approx \langle e \rangle/3.15$.
From these data we have adopted  the best  central
temperature values and the  $\pm$ 1$\sigma$ uncertainty based upon  the  $\chi^2$ distribution of these data.  This is denoted by the orange ellipses in Figures \ref{fig:3}a and \ref{fig:3}b.
  
%Figure \ref{fig14} shows the three-dimensional correlation among T$_{\nu_e}$, T$_{\bar{\nu_e}}$, and 
%T$_{\nu_x}$ for various supernova simulations. 
 Figure \ref{fig:3}a shows the projected correlation between T$_{\bar{\nu_e}}$ and T$_{\nu_e}$ for various supernova simulations, while Figure \ref{fig:3}b shows the correlation between T$_{\nu_x}$ and T$_{\bar{\nu_e}}$. 
Red crosses represent T$_{\nu_x}$ and T$_{\bar{\nu_e}}$ summarized in the
Tables \ref{tbl.2-1}-\ref{tbl.2-3} with assumed errors of $\pm 10\%$ as deduced below. %Thick crosses correspond to the same data described by the  red filled circles in Figure \ref{fig14}, and the thin crosses correspond to the data with red open circles in Figure \ref{fig14}. 
%The light orange ellipses in  denote  the average value
%$\pm$ 1 $\sigma$ for  these data. 
 
 The neutrino temperature hierarchy  requires that    T$_{\nu_x}$ $\ge$
T$_{\bar{\nu_e}}$, while the constraint of \cite{yos08} requires  T$_{\nu_x} \sim  4.3$ to   6.5 MeV. 
We therefore adopt  T$_{\bar{\nu_e}} \sim $3.8 to  6.0 MeV as shown by the horizontal  lines in Figure   \ref{fig:3}a  and 
the vertical lines and  light blue trapezoid in Figure  \ref{fig:3}b.
 We then deduce our adopted  constraints on values for the T$_{\bar{\nu_e}}$-T$_{\nu_x}$ correlation as
the  overlapping area of the orange ellipses and the blue trapezoids as shown in
Figure \ref{fig:3}b. 
 
 We selected seven  pairs of T$_{\bar{\nu_e}}$-T$_{\nu_x}$  to use  for  analyzing  the dependence of the SRN energy spectra on 
 T$_{\bar{\nu_e}}$-T$_{\nu_x}$. These are indicated by the green, red, magenta,
blue, yellow and two black circles without error bars shown in Figure  \ref{fig:3}b.  These correspond to  temperatures (T$_{\bar{\nu_e}}$,
T$_{\nu_x}$) = (6.0 MeV, 6.5MeV), (5.0 MeV, 6.0 MeV),
(4.1MeV, 6.5MeV), (3.9MeV, 5.6MeV), (4.5MeV, 4.7MeV), (6.7MeV, 7.6MeV) and
(2.5MeV, 2.5MeV), respectively.  
Note that our adopted temperature range is  consistent with the 90\% C.L.  temperatures recently deduced by the Super-Kamiokande collaboration \citep{SK12,Sek13}.

The error bars for the points shown on Figures \ref{fig:3}a and \ref{fig:3}b correspond to the uncertainty due to the variation of the neutrino temperatures in time during the explosion.
An analysis was made in \citet{tot98} of the time dependence of the supernova neutrino  temperature  based upon the SN explosion model
of \citet{may87} and assuming a Fremi-Dirac neutrino spectrum.
In that work it was  concluded that  after about 1 sec the neutrino temperatures  only vary by about 10\% over the next few seconds when most of the neutrino luminosity is emitted. On the other hand, some of the most recent neutrino transport calculations based upon modern relativistic solutions to the Boltzmann Eq.~[e.g. \cite{Fischer12,Roberts12}] imply smaller average neutrino temperatures and a more rapid decline of temperature with time.  Nevertheless, for our purposes we can adopt a 10\% as a reasonable uncertainty due to the decline of the neutrino temperatures with time in
Figures \ref{fig:3}a and \ref{fig:3}b  keeping in mind that the true neutrino flux should be derived from a numerical integration of the emitted neutrino flux from the numerical models and may not be perfectly represented by a Fermi-Dirac temperature.   

  We also note that the time dependence of
the correlation between T$_{\nu_x}$ and T$_{\bar{\nu_e}}$ was analyzed in  \citet{may87}.
That work confirms  that our adopted upper and lower limits  of T$_{\nu_e}$, T$_{\bar{\nu_e}}$ and T$_{\nu_x}$ are satisfied
throughout most of the explosion.

Tables \ref{tbl.2-1}
to  \ref{tbl.2-3} show the
calculated neutrino temperatures and dimensionless chemical potentials $\eta_\nu \equiv \mu_\nu/T_\nu$ from various  numerical simulations of core-collapse SN explosions. These
data show that there is no universal agreement as to the chemical
potential for the emitted supernova neutrinos.

 Nevertheless, in spite of this uncertainty,  it has been demonstrated [e.g. \citet{yos05}]   that the uncertainty in 
 the reaction rate  for  any neutrino processes due to  the neutrino chemical potential  is less
than 10\% as long as the dimensionless chemical potential $\eta_\nu \equiv \mu_\nu/T \le 3.0$.  Since this condition is
satisfied in Tables \ref{tbl.2-1}
to  \ref{tbl.2-3}, we are justified in ignoring the neutrino chemical potential in the present study, while keeping in mind that this can add to the overall uncertainty in the detection rate.
%\twocolumn
% The chemical potential dependence of SRN spectra (for flux and event rate)
%are discussed later.  
%\end{description}

\subsection{Dependence of the SRN energy spectra on Neutrino Oscillations}
It is by now well established that the flavor eigenstates for the neutrino
 are not identical to the  mass eigenstates.  
  The flavor eigenstates $\nu_e$,
$\nu_\mu$, $\nu_\tau$ are  related to the mass eigenstates by an unitary matrix
$U$:

\begin{equation}
 \left(
\begin{array}{c}
 \nu_e \\
 \nu_\mu\\
 \nu_\tau
\end{array}
\right) 
= U
 \left(
\begin{array}{c}
 \nu_1 \\
 \nu_2 \\
 \nu_3
\end{array}
\right) ~~,
 \label{2-16}
\end{equation}
where the matrix U can be decomposed  as:
\begin{equation}
U =
\left(
\begin{array}{ccc}
  U_{e1} & U_{e2} & U_{e3}\\
  U_{\mu1} & U_{\mu2} & U_{\mu3}\\
  U_{\tau1} & U_{\tau2} & U_{\tau3}   
\end{array}
\right) \label{2-17}
\end{equation}
\begin{eqnarray}
&\equiv&
\left(
\begin{array}{ccc}
  1 & 0 & 0 \\
  0 & c_{23} & s_{23} \\
  0 & -s_{23} & c_{23}
\end{array}
\right)
\left(
\begin{array}{ccc}
 c_{13} & 0 & ~s_{13}e^{-i\delta} \\
 0 & 1 & 0 \\
 -s_{13}e^{i\delta} & 0 & c_{13} 
\end{array}
\right) \nonumber \\
&&\times
\left(
\begin{array}{ccc}
 c_{12} & s_{12} & 0\\
 -s_{12} & c_{12} & 0\\
 0 & 0 & 1 
\end{array}
\right) ~~.
 \label{2-18}
\end{eqnarray}
%\begin{equation}
%=
%\left(
%\begin{array}{ccc}
% c_{12}c_{13} & s_{12}c_{13} & s_{13}e^{i\delta} \\
% -s_{12}c_{23} -c_{12}s_{23}s_{13}e^{i\delta} & c_{12}c_{23}
% -s_{12}s_{23}s_{12}e^{i\delta} & s_{23}c_{13}\\
% s_{12}s_{23} -c_{12}c_{23}s_{13}e^{i\delta} & c_{12}s_{23} 
% -s_{12}c_{23}s_{13}e^{i\delta} & c_{23}c_{13}
%\end{array}
%\right) \label{2-19}
%\end{equation}
Here, s$_{ij}$ $\equiv$ $\sin \theta_{ij}$, c$_{ij}$ $\equiv$ $\cos \theta_{ij}$, and $\theta_{ij}$ is
the mixing angle between neutrinos with mass eigenstates i and j, and $\delta$ represents
the CP  phase. 
%The  mass matrix is then converted to non-diagonal matrix by a unitary transformation:
%\begin{equation}
% M \rightarrow UMU^{\dagger} \label{2-20}
%\end{equation}

 For the mixing of mass eigenstates one must consider both the normal mass hierarchy, $m_1< m_2< m_3$, and the inverted mass hierarchy, $m_3< m_1< m_2$.  
Both models
have two resonance  mass regions. The resonance at higher density is  called
the H-resonance, and the one at lower density  is called the
L-resonance. In the normal mass
hierarchy, both resonance points are in the neutrino sector. In the inverted mass hierarchy, however,
one resonance is in the
neutrino sector and the other is in the anti-neutrino sector.

 The prospects for detecting effects of neutrino flavor oscillations in the spectrum of detected SRNs has been discussed in \cite{cha11} using a parameterized form for the emitted neutrino spectrum from \cite{kei03b}, and also in \cite{lun12} using the supernova simulations of the Basel group \citep{Fisch10}. In this study, we similarly  consider the possible influence and detectability of neutrino oscillations.  Rather than to adopt  a particular mixing scenario  we explore mixing in the limit of 3 neutrino oscillation paradigms
according to \citet{dig00}.  These are:  Case {\it I} - normal mass hierarchy case or an inverted
mass hierarchy with complete non-adiabatic mixing; Case {\it II} - an inverted mass hierarchy case with complete adiabatic
mixing; and Case {\it III} - no mixing.  We also explore the possible effects of a neutrino self interaction  below in \S \ref{selfint}.

Let us consider  the inverted hierarchy first.  If we assume an efficient conversion probability $P_H$ of $\bar\nu_3 \leftrightarrow \bar \nu_1$ at the H-resonance,
then the $\bar \nu_{e}$ flux emitted from the supernova becomes:
\begin{eqnarray}
\phi_{\bar \nu_e} &=&  |U_{e1}|^2 \phi_{\bar \nu_1} +  |U_{e2}|^2\phi_{\bar \nu_2} + |U_{e3}|^2 \phi_{\bar \nu_3} \nonumber \\
&=& |U_{e1}|^2 \{(1 - P_H) \phi_{\bar \nu_1}(0) + P_H \phi_{\bar \nu_3} (0)\}  \\
&+& |U_{e2}|^2 \phi_{\bar \nu_2}(0) + |U_{e3}|^2 \{P_H \phi_{\bar \nu_1}(0) + (1 - P_H) \phi_{\bar \nu_3}(0)\} \nonumber~.
\end{eqnarray}
From this, one can deduce the survival probability $\bar P$ for $\bar {\nu_e}$:
\begin{equation}
\bar P =  |U_{e1}|^2 P_H + |U_{e3}|^2 (1 - P_H) \approx 0.7 P_H ~~.
\end{equation}
The same result follows for the normal mass hierarchy.  Hence,  we define case {\it I} as: 
\begin {equation}
\bar \nu_e (0) \rightarrow 0.7 \times \bar \nu_e^0 + 0.3 \times \nu_x^0 ~~~{\rm (case}~I).
\end{equation}

On the other hand, it is now known \citep{DayaBay} that the $\theta_{1 3}$ mixing is relatively large [$\sin^2{(\theta_{1 3})}= 0.092 \pm 0.0016 (stat) \pm 0.005(syst)$].  Hence, if there is no  supernova shock wave, then 
the survival probability can be small ($P_H \rightarrow 0$) so that the conversion of $\bar \nu_e$ into $\nu_x$ is efficient.  We define this as case {\it II}:
\begin {equation}
\bar \nu_e (0) \rightarrow  \nu_x^0 ~~~{\rm (case}~II),
\end{equation}
while the case with no oscillations we will call case {\it III}.
In what follows we will consider these three possible  oscillation cases in our analysis of the uncertainties in the SRN fluxes and detection rates.

 \subsection{Dependence of the SRN energy spectra and detection rate on the star formation rate and cosmological redshift}
 The neutrino spectra arriving at a terrestrial detector will depend upon a combination of the star formation rate and the cosmological redshift
 [cf.~Eq.~(\ref{nuflux})].  Hence, we next examine the dependence of the star formation rate described in \S \ref{SFR} on the cosmological redshift.
 
%\subsubsection{Redshift dependence of the SRN flux and event rate}
 Table \ref{tbl.5}  shows the percentage  SRN flux and detection rate as a function of redshift bin from $z = 0$ to $2.5$. 
 For this table we assume (T$_{\bar{\nu_e}}$, T$_{\nu_x}$) = (5.0 MeV,
6.0 MeV) as described above.  We also adopt the linear piecewise fit for the  SFR, and neutrino  oscillation case {\it III} (i.e.~no oscillations).
 Table~\ref{tbl.5}  shows  that the redshift interval with the largest contribution to the
SRN flux corresponds to  $z \sim  0.0$ to $1.0$. On the other hand, the contribution from $z \sim  1.5$
to $ 2.5$ is only about $15\%$ of that from $z = 0$ to $1.0$.
 
 Table \ref{tbl.6}  shows the differential redshift dependence for the predicted
SRN detection rate in the no oscillation case ({\it III})  with
(T$_{\bar{\nu_e}}$, T$_{\nu_x}$) = (5.0 MeV,
6.0 MeV), based upon  the fit to the observed SFR data of \cite{kob00} with and without the  correction of the observed SFR for  dust extinction.
 
Figure \ref{fig:4}  shows  the spectrum of arriving neutrinos in various redshift bins.   As discussed in many previous works [e.g.~\cite{Ando04}] the main contribution to the present total SRN energy spectrum is
from supernovae in the  low redshift region.  
This can be seen in  Figure \ref{fig:4} along with Tables \ref{tbl.5}  and \ref{tbl.6}.   

These tables and figures show that  the arriving neutrino spectrum is dominated by events in the redshift range from $z = 0$ to  $2$. The reason for
this is that at larger distances the  neutrino energy is redshifted away.
In addition to the redshift effect, the SFR itself flattens and/or decreases for $z > 2$.  This also diminishes the importance of supernovae from higher redshifts.
At high redshift, the 
contribution to the  SRN spectrum decreases as  $\sim 1 / z^{-2}$. 
 
\subsection{Dependence of SRN detection rate  on the SFR}
Figures \ref{fig:5} and \ref{fig:7} show the dependence of the SRN detection rate on the SFRs considered here.  Figure \ref{fig:5} shows the supernova detection rate for the piecewise linear SFR for the 3 cases of neutrino oscillations and fiducial neutrino temperatures of (T$_{\bar{\nu_e}}$, T$_{\nu_x}$) = (5.0 MeV, 6.0 MeV).  Figure  \ref{fig:7} shows the neutrino detection rate for the two fit star formation rates of \cite{kob00}. Figure \ref{fig:7} shows that the difference between the two SRN fits with and
without dust correction of the data is about a factor of 2 throughout the detectable
energy range for positrons.  This highlights the importance of  the dust extinction correction when estimating the SRN
flux and detection rate.

 Regarding Figure  \ref{fig:7} , in the case with no  dust extinction correction,
the predicted SRN detection rate is $\sim 1/3$ of that obtained when employing 
the piecewise linear  SFR model used in this study (i.e. Figure \ref{fig:1}). This difference  is mainly because
 the maximum of the SFR  with the extinction correction is less than 1/2
of the SFR without extinction. % (see Figure \ref{fig13}). 
 
%  The z dependence of the SRN flux and event rates with dust correction
%case is significantly different from those without dust correction. This
%difference comes from the difference in the  SFR  for z $\gtrsim$ 1.0. 
 The thick  lines in Figure \ref{fig:5} show the event rate based upon the piecewise linear  SFR and our best guess neutrino temperatures. The  thin lines above and below the thick 
  line show the $\pm 1\sigma$ upper and lower limits to the total SRN detection rate. Hence,  the
width of the  enclosed band indicates the uncertainty in
the SRN event rate  observations due to the SFR uncertainty.
%\subsubsection{Discussion}

 Figure \ref{fig:5} shows that  the  positron energy spectra are
not particularly affected by the  different oscillation paradigms in our fiducial model. 
Also, the width of uncertainty band is consistent with  the uncertainty of  the SFR between z = 0
- 2. This means that the SFR  directly fixes magnitude of the detected SRN
flux.  This suggests  that 
one may be able to  constrain the SFR from the SRN detection rate if
uncertainties  in other parameters can be minimized.  
%The SFR in this redshift region has been already well
%defined by observed results, therefore detection of SRN energy spectrum is a
%promised tool for presuming SN neutrino temperature.

 However, one must still consider  the  possible 
influence of missing non-luminous ONeMg and failed SNe as well as the paradigm for SNIb,c as discussed below.

\subsection{Dependence of SRN flux and event rate on SNIb,c model}
In section 2 we considered two different paradigms for SNIb,c, i.e. single star as in \cite{hor11}, or the binary  interaction + fSNe + ONeMg SNe adopted here.
Figure \ref{fig:6} shows a comparison of the neutrino flux (upper panel) and detector event rate (lower panel) for the two cases in our fiducial model and for no oscillations (case {\it III}).
One can see that there is very little dependence ($< 10$\%) of the flux or event rate on the assumed SNIb,c model. This is because the spectrum is dominated by the neutrinos from
normal CC-SNe in the mass range $10$ M$_\odot < M < 25$ M$_\odot$.  Hence, even though the neutrino spectra are different from ONeMg and fSNe, their contribution to the total detected neutrino spectrum is small.  This means that the predicted detector spectrum is independent of specific assumptions about the contributions of various SNe.  One can, however detect  a difference if one  attempts to solve the supernova rate problem by significantly enhancing the fraction of fSNe as described below.
 
\subsection{Dependence of SRN energy spectra and event rate on neutrino temperatures}
 Among the parameters considered in this work, the SN neutrino temperature has  the strongest  influence on  the resultant SRN energy spectra. 
 Figure \ref{fig:8} shows the dependence of the total detection rate of SRN  on the neutrino temperature as a function of the   e$^+$ energy in the detector for the three different neutrino oscillation cases as labeled.

 Table \ref{tbl.7} shows the numerical data for  the dependence of the detected  SRN on  the neutrino temperatures, for various redshift bins, and neutrino oscillation possibilities.  Table \ref{tbl.8} illustrates the dependence  on neutrino temperatures and   oscillation parameters of the total number of detected 
 SNR events over a 10 year run time.
 Each table contains
the predicted detector event rate  for the pairs of T$_{\nu}$ (T$_{\bar{\nu_e}}$, T$_{\nu_x}$)
that  were shown in Figures \ref{fig:3}a and \ref{fig:3}b and summarized in \S \ref{temps}.
 %Table \ref{tbl.7} and \ref{tbl.9-1} give  the numerical results for  the oscillation case {\it I}
%($\bar{\nu_e}$ = 0.7 $\times$ $\bar{\nu_e}^0$ + 0.3 $\times$ ${\nu_x}^0$). Tables \ref{tbl.8} and \ref{tbl.9-2} are  for  oscillation case {\it II}
% ($\bar{\nu_e}$ = ${\nu_x}^0$). 
 
 Figure \ref{fig:8} shows that the T$_\nu$
dependence of the  SRN detection rate is strongly related to SRN energy range.
 This dependence is particularly strong  in the  high energy range. 
Furthermore, the T$_\nu$ dependence of the SRN detection rate is 
stronger  in the oscillation case {\it II} ($\bar{\nu_e}$ = ${\nu_x}^0$) than in  oscillation
case {\it I} ($\bar{\nu_e}$ = 0.7 $\times$ $\bar{\nu_e}^0$ + 0.3 $\times$ ${\nu_x}^0$). 
This result is due to the fact that the number of emitted
 $\nu_x$ is influenced more strongly 
by T$_{\nu}$ than by $\bar{\nu_e}$, especially in the high
energy range.  Also,  the $\bar{\nu_e}$ detected on  Earth  originated from
more than 99 $\%$ $\nu_x$  in oscillation case {\it II} ($\bar{\nu_e}$ = ${\nu_x}^0$).   On the other hand, only  
30$\%$ mixing occurs in  oscillation case {\it I} ($\bar{\nu_e}$ = 0.7 $\times$ $\bar{\nu_e}^0$ + 0.3 $\times$ ${\nu_x}^0$). 

In other words, T$_\nu$ in core-collapse SN explosions  influences both  the event rate  and the spectra.  In particular, it influences the value of
the positron energy that gives the maximum
event rate per unit positron energy width (hereafter E$_{peak}$). Thus, if one could more
tightly constraint  T$_\nu$ in normal core-collapse SN (CC-SN) explosions, then a better estimate  for the  E$_{peak}$ could be obtained, or conversely,
a measurement of the number of events at E$_{peak}$ compared to higher energy (say 25 MeV) could be used to measure the neutrino temperature.. 
 %This is also related to the background events by reactor $\bar{\nu_e}$,  because whether E$_{peak}$ is exceeds 
%  the lower threshold for  detection is important  for  constraining T$_\nu$
%in CC-SNe.  

Figure \ref{fig:9} illustrates how one might determine the average electron anti-neutrino temperature from the detected positron event rate.
This figure shows the ratio of the event rate at the observed spectrum peak to the rate at 25 MeV. (This energy below the energy at which atmospheric neutrino background dominates).
There is a very strong correlation between this ratio and neutrino temperature that characterizes the different representative  supernova models identified in FIgure \ref{fig:3}.  The best correlation is in the no-oscillation case (solid line) but  even in the different oscillation
cases the relationship between characteristic neutrino temperature from the supernova models and this ratio is quite robust in the low temperature region.  

Hence, such measurements could provide a valuable constraint on both the average neutrino temperature, but also the supernova models themselves.   The more recent models tend to predict a higher neutrino opacity and hence, a lower neutrino temperature at the neutrino sphere.  Hence a measurement of a low  neutrino temperature in this way could confirm whether the current supernova models are correct.

\subsection{Dependence of SRN detection on the neutrino self interaction
  effect}
  \label{selfint}
  In the case of  an inverted mass hierarchy, a ``self interaction effect'' among  neutrinos \citep{fog07}  might cause an additional difference between the energy spectrum of supernova neutrinos at detection and production \citep{lun12}.  
To illustrate the possible importance of this effect, we have calculated the  energy spectrum of SRNs in the simplest  case with a neutrino self-interaction, i.e.
a single angle interaction. 

It has been demonstrated \citet{fog07}  that a single angle  interaction  causes the
 energy spectra of $\bar{\nu_e}$ and $\nu_x$ ($\bar{\nu_\mu}$
and $\bar{\nu_\tau}$) to exchange with each other at 4.0 MeV [see Figure 5 in \citet{fog07}].
This means that the energy spectra of $\bar{\nu_e}$ and $\nu_x$ are perfectly
exchanged in the detectable energy range of water \v{C}erenkov detectors.
Hence, a neutrino self interaction could change
the SRN energy spectrum in oscillation case {\it II} into that of  case
{\it III}.  

In Figure \ref{fig:10} we show the change in the neutrino detection rate relative to the detection rate based upon the fiducial  neutrino temperatures and star formation rate for the no oscillation case  ({\it III}) and a fiducial temperature of  (T$_{\bar{\nu_e}}$, T$_{\nu_x}$) = (5.0 MeV, 6.0 MeV).  This figure shows that for the most part the correction for  neutrino self interaction produces a relatively small change  in the event rate for SRNs, and hence, corresponds to a relatively small contribution to the overall uncertainty in the predicted spectrum. A similar conclusion was reached in \cite{lun12} who also considered both neutrino self-interaction and oscillation effects on the associated neutrino spectrum.

\subsection{Energy resolution and statistical variance of the  SRN detection rate}
%It is important to also understand  the 
%error bars in the detected SRN based upon the typical energy resolution  and sensitivity of a water \v{C}erenkov detector.  
%  Figures \ref{run time} (a) through  (c) illustrate  event rate per MeV per Mt of active water \v{C}erenkov detector per time interval of: a) 1; b) 10; or c) 100 yrs. These plots are equivalent to  total events for a detector of this active mass for the specified interval  assuming  our standard star formation rate, adopted neutrino temperatures.   
%  The Red, green, and blue lines show the  SRN detection rate for oscillation case {\it I},
%case {\it II}, and case {\it III}, respectively. 
%Horizontal error bars show the
%anticipated energy resolution based upon that of  Super-Kamiokande.  The vertical error bars show  the statistical standard deviation based
%upon the indicated number of events.

The error due to energy resolution of the detector  can be estimated from the published analytic function for the energy resolution   of the
Super-Kamiokande (SK-I)  detector \citep{sup08}.   For the energy range
of interest here ($\sim 10-30$ MeV) the resolution is nearly constant (11-12\%).  The uncertainty in the overall spectrum due to the energy resolution, however, is less than the  uncertainty due to the statistics of detected events and can be neglected here.   Note also, that the   energy resolution  could be better 
in  a future  water \v{C}erenkov detector.

% The horizontal error bars in the oscillation cases {\it I} and {\it II} in
%Figures \ref{run time} (a) and (b) overlap each other over the whole detectable e$^+$  energy
%range. 
The statistical error can be deduced from  the standard deviation
 of the anticipated number of detected events. 
 Even in the ideal case that all other uncertainties are negligible, however,
the predicted energy spectra and error
bars between the cases of different oscillation parameters cannot be distinguished  in
a run time of 1 year to 10 years. Hence, one needs  many decades of run time to
distinguish these two cases even for  a 10$^6$ ton  class water \v{C}erenkov
detector. 

Moreover, we have only considered  two extreme cases for adiabatic
neutrino oscillation parameters in this study.  However, the oscillation effect on
SRN may not be completely adiabatic. Hence, the influence of neutrino  oscillations
 on the SRN spectra may be ambiguous.

\section{Contribution from dark (ONeMg) and failed  (BH-formation) supernovae to the SRN energy spectra and detection rate}
 Studies of the relic neutrino spectrum have mainly concentrated on the contribution from normal core-collapse supernovae, however, neutrinos are also emitted in other environments.  In particular, we now examine  the influence of dark supernovae, [i.e.~ONeMg
electron-degenerate supernovae  \citep{Isern91,hud10}], and failed supernovae [i.e.~black hole formation \citep{sum08}], on the SRN energy spectra and event rate.
  A summary of the model classification schemes and adopted model parameters is given  
  in Table \ref{tbl.9}.

\subsection{Neutrinos from ONeMg Supernovae}
For our purposes we will consider the ONeMg supernovae (or electron-capture supernovae) to arise from progenitor masses  in the range
of 8 $M_{\odot} \le M \le 10 M_{\odot}$.  
%Such supernovae occur  \citep{Isern91} when an electron-degenerate ONeMg  core reaches a critical density 
%(at $M_{core} \sim 1.4$ M$_\odot$).    
In the present application we have assumed neutrino fluxes and temperatures from the 
  ONeMg SN numerical simulation given in \citet{hud10} as summarized in Table \ref{tbl.9}.  The values for T$_{\bar{\nu_e}}$ and T$_{\nu_x}$ are lower than those in a
typical core-collapse SN model.  In the ONeMg SN model
of  \citet{hud10}, the neutrino temperatures are T$_{\bar{\nu_e}} = 3.0$ MeV, T$_{\bar{\nu_e}} = $T$_{\nu_x} =  3.6$ MeV.

\subsection{Neutrinos from Failed Supernovae}
For failed supernovae  (fSNe)  we consider  massive stars with progenitor masses
with $M > 25$ M$_{\odot} $, though this cutoff is uncertain.  For stars above this cutoff mass   the whole star collapses leading to a black hole. 
In general fSNe  can lead to higher 
neutrino luminosities and  temperatures than those emitted from ordinary core-collapse SNe.
Also, the difference of the neutrino temperatures among different flavors is
even larger than the difference in ordinary supernovae, depending upon  the EoS used in the collapse simulations.
Although it takes only about 500 ms to 1 s before the  horizon 
of the black hole appears,
the proto-neutron star still ejects a large  flux of neutrinos and their total energy 
can exceed that of ordinary core-collapse supernovae.

For purposes of estimating the neutrino flux we  adopt the fSN  models of \citet{sum08}.   As our fiducial  fSN model we utilize the relativistic mean field EoS  of
\citet{she98} for the collapse dynamics. On the other hand, in the work of \citet{sum08} it was noted that 
fSN  models based upon the soft ($K = 180$) EoS of \citet{lat91} have a different
 time dependence, along with higher average neutrino temperatures,  and
neutrino luminosities in each flavor.  A soft EoS results in higher neutrino luminosities and temperatures
because the neutronized hadronic matter inside the proto-neutron star is
compressed more strongly than in the case of a stiff EoS during the collapse to a black hole.
Therefore, in what follows  we will also  compare the SRN detection rates  based upon
these two different EoSs for the  fSN models.

\subsubsection{Collapsar model for GRBs}
The collapse of a star with $M > 25$ M$_{\odot} $ does not necessarily produce a visible supernova although, in the case of a rotating core and sufficient magnetic field strength  \citep{har09}, the collapse can lead to a heated accretion disk and an observable gamma-ray burst (GRB).  Although GRBs do not contribute much to the relic neutrino 
spectrum, for illustration we also consider this subcategory of fSNe.  
For progenitor stars in the mass range $M \sim (25-40)$ M$_\odot$, the combination of magnetohydrodynamic acceleration and neutrino pair heating in a funnel region above the black hole can lead to the launch \citep{mac99,woo06,har09,Nakamura13} of a relativistic jet.  This jet can be a source of GRBs and hypernovae.   To estimate the fraction of
fSNe that lead to GRBs, we write:
\begin{equation}
\frac{R(GRB)}{R(fSN)} = \frac{\epsilon(J,B,Z)  \int_{25 M_\sun}^{40 M_\sun}dM \phi(M)}{\int_{25 M_\sun}^{125 M_\sun}dM
   \phi(M)} ~~,
\end{equation}
where $\epsilon$ is a poorly known efficiency for collapsar GRB production as a function  of progenitor angular momentum $J$, magnetic field strength $B$, and metallicity $Z$.  For example among the many models considered in \cite{har09} the optimum conditions to generate a relativistic MHD driven jet were found to involve  a combination of: 1)  a rotating model with initial specific angular momentum
$J = 1.5J_{iso}$ (where $J_{iso}$ is the specific angular momentum of material in the inner-most stable circular orbit of the nascent black hole); and 2)  the highest initial magnetic fields corresponding to $B >  10^{10}$ G for
the core of the progenitor.

For purposes of illustration we can estimate this quantity phenomenologically.  To begin with we write:
\begin{equation}
\frac{R(GRB)}{R(fSNe)} = \frac{R(GRB)}{R(SN)}  \times \frac{R(SN)}{R(fSN)} ~~,
\end{equation}
The observed core collapse supernova rate from the Stockholm VIMOS Supernova Survey \citep{Melinder12} in the interval $z = 0.1 -1$ is $ 3.3^{+3.1+2.0}_{-1.8-1.4} \times 10^{-4}$ Mpc$^{-3}$ y$^{-1}$.  On the other hand, the inferred overall local GRB rate is $3.3 \pm 1.1 \times 10^{-8} h^3_{65}$ Mpc$^{-3}$ y$^{-1}$ \citep{Guetta05}.
 This gives,
\begin{equation}
\frac{R(GRB)}{R(SN)}   \approx 1.0 \times 10^{-4} ~~,
\end{equation}
 The ratio $R(SN)/R(fSNe)$ on the other hand only depends upon the IMF
 \begin{equation}
\frac{R(SN)}{R(fSN)}   \approx 4 ~~,
\end{equation}
so we estimate that GRBs only contribute about 0.04\% to the total fSNe rate.  This fraction is consistent with other estimates [e.g.~\cite{Fryer99}] that the collapsar rate is a small fraction of the core-collapse supernova rate. 

On the other hand there is also a known preference for long duration GRBs (collapsars) to occur in low metallicity galaxies \citep{Wolf07}. As pointed out in that paper,  this could be an consequence  of the fact that rapidly rotating cores are needed to form the collapsar accretion disk \citep{woo06,har09}. Angular momentum, however, implies larger mass loss rates. Since the mass loss rate scales with metallicity \citep{Vink05}, it may be that lower metallicity progenitors are more likely to have sufficiently rotating cores at the time of collapse. 

Clearly, all of the above discussion  highlights the uncertainties in  fSNe models, and collapsar GRB models in particular, due to the unknown efficiency for forming rotating cores with high magnetic field strength. For this reason we consider below the possibility that the fSNe rate can be scaled in such a way as to resolve the possible supernova rate problem. Indeed, we argue that the detection of the unique relic neutrino signal from fSNe may be a way to gain new insight into the history of black hole formation over the lifetime of the Galaxy.

 \subsection{SRN flux}
Figure \ref{fig:11} shows the relative contributions of each source (core collapse SNe, ONeMg SNe,  fSNe, and collapsar GRBs)  to the total SRN flux for our fiducial temperatures and the three neutrino oscillation cases considered here.  The thick solid  line in each figure shows
the total flux of SRN. The dashed  line shows the contribution of SRN emitted from
normal core-collapse SNe.  The dotted line is from failed SNe, the dot-dashed  line is from ONeMg SNe, and the  lower grey line shows the contribution from collapsars (GRBs).
From this figure we see that  ONeMg SNe contribute $< 20\%$ to the arriving neutrino flux, while collapsar GRBs contribute a negligibly small fraction $\sim 0.1\%$.

Table \ref{tbl.9}  shows that  E$_{\nu_e}^{total}$ and E$_{\bar{\nu_e}}^{total}$ in failed  SNe
are almost equal. The physical reason for this is that a 
proto-neutron star which leads  to a fSN has an accretion
disk and a large  amount of mass is supplied 
from this accretion disk.  This is enough to keep the number of neutrons and protons equal, (i.e. electron fraction, Y$_e$
$\approx$ 0.5 ) during explosion. Hence,  the cooling rates for both  electron capture and
positron capture are  almost the same, so that the emitted energy  in $\nu_e$ and $\bar \nu_e$ are similar.  
 
Table \ref{tbl.9}  also shows that, unlike core collapse SNe,   E$_{\nu_x}^{total}$ is an order of magnitude  less than
E$_{\nu_e}^{total}$ and E$_{\bar{\nu_e}}^{total}$ in GRBs \citep{har09}. The
physical reason for this 
is  that in the collapsar hot accretion disk the cooling
rate $\dot q_2$ for $e^+$ or  $e^-$ capture by free nucleons exceeds   the rate $\dot q_4$ for $e^+ e^-$ pair
annihilation. The cooling rates are given by
\begin{equation}
 \dot{q_2} \approx 2.27N_AT^6_{MeV} \displaystyle MeV s^{-1}
 g^{-1} \label{5-1} 
\end{equation}
\begin{equation}
 \dot{q_4} \approx 0.144N_A\frac{T^9_{MeV}}{\rho^8} \displaystyle MeV s^{-1} 
   g^{-1} ~~.
   \label{5-2} 
\end{equation}
In the collapsar hot accretion disk  the temperature and density near the neutrino-sphere are more extreme than in the proto-neutron star of 
a core collapse supernova.  The temperature 
is $\sim 30$ MeV and the density is $\sim10^{14}$g cm$^{-3}$. 
For these conditions the cooling
rate from capture by free nucleons [Eq.~(\ref{5-1})]  exceeds   the pair
annihilation rate [Eq.~(\ref{5-2})] since the annihilation  rate  depends upon rather strongly on the temperature and density,  
 while $e^+ e^-$ capture is independent of the number density.  Hence, the pair annihilation rate is suppressed.  
 This means  that the production of the  $\nu_x$ by pair annihilation is suppressed. 
 Therefore, the
neutrino luminosity  hierarchy can be altered  in the case of neutrino
emission from the optically thin accretion disk .

The ratio of the cooling
rate $\dot{q_2}/\dot{q_4} \sim 10$ according to  Eqs.~(\ref{5-1}) and
(\ref{5-2}). This
ratio also directly affects the difference in the contribution from GRBs to the  whole
energy spectrum between  the non-adiabatic case and  the adiabatic case
of neutrino oscillations (see Figs.~\ref{fig:5}, \ref{fig:12}, and Table \ref{tbl.10}).  
 
 The T$_{\nu_x}$ in  the GRB model of \citet{har09}
is 4.4 MeV, which is lower than T$_{\bar{\nu_e}}$ in the same model, in
contrast to the usual neutrino temperature hierarchy. One possible   reason for this is
that neutrinos are mainly
produced in an optically thin region of the accretion
disk.

\subsection{Detection Rates}
 Figures \ref{fig:12} and \ref{fig:13}, along with Table \ref{tbl.10}  show SRN detection rates for the different contributions shown in Figure \ref{fig:11} and for the three oscillation scenarios considered in this work. The solid  line in Figure \ref{fig:12} shows
the total detection rate  of SRN. The dashed line shows the dominant contribution from
normal core collapse supernovae (CC-SNe).  The dotted  line shows the contribution from  failed SNe, while the dot-cashed line shows the contribution from ONeMg SNe which is only significant at low energies.  The lowest  thin dotted line is from collapsars/GRBs.   The contribution from GRB events is negligibly small ($<1 \%$) 
because the ratio of the  GRB rate to the  number of stellar explosion is  $< 0.1\%$ even though
the gravitational binding energy released in SRN during  GRBs  is a few times larger  than that for normal core collapse SNe.

Figure \ref{fig:13} illustrates the uncertainty in the total number of detected events  in our fiducial model after 10 years running.  The shaded regions indicate the combined uncertainties from detector statistics and the uncertainty in the SFR.  The sensitivity to the contribution from fSNe is illustrated by the difference between the events detected for the  stiff  EoS of \cite{she98}  (red region) and the soft ($K=180$) EoS of \cite{lat91} (green region). 

 Figure \ref{fig:12} and Table \ref{tbl.10}  show that the contribution from failed
SNe to the SRN detection rate is  less than 10$\%$ after integration over the whole energy range. This is
roughly the ratio of  failed SNe compared to the total  number of stellar explosions. 
Even so, there is a  difference in the SRN detected events from fSNe between  the soft EoS of \cite{lat91} and the stiff EoS of \cite{she98}.
This is because of the differences between these two EoSs affect strongly the neutrino luminosities and temperatures as shown in Table \ref{tbl.9}.
Independently of the uncertainty due to neutrino oscillations, the star formation rate or the detector statistics, the ratio between events at the peak to events at 30 MeV
is about 6 for the stiff Shen EoS and about 3 for the soft LS EoS.  Hence, if uncertainties due to the neutrino temperature can be minimized, it might be possible to infer
both the presence of failed SNe and their associated EoS from the spectrum of relic neutrinos.

\section{Summary of Uncertainties in SRN detection rates}
 Ignoring the uncertainties in the  SRN energy spectrum due to  neutrino oscillations and the possible  neutrino
self-interaction, the total uncertainty in  the SRN energy spectrum
is dominated  by  three contributions:  1) the uncertainty in  the SFR; 2) the uncertainty in 
T$_\nu$ in CC-SN explosions; and 3) the statistical variance in SRN detection. 
The relative contributions  of these three constituents
affect  the energy spectrum of detected SRNs.  

Among these, the SFR has
the largest contribution to the uncertainty, 45$\%$ to $ 99\%$, except
in the case of an inverted mass hierarchy in the high energy region. So the most effective way
to reduce the total uncertainty in the SRN detection rate, especially in the low energy
region, is to derive a more precise SFR from 
the observational data. Even so, the SFR only affects the overall normalization.  Hence, ratios between peak events and energetic events as discussed here
are independent of this uncertainty as long as the events are above the background.

 The second largest contributor to the uncertainty is  the range of possible  T$_\nu$ from normal CC-SN explosions.  
 This  contributes $\sim 10\%$ to $ 60\%$ and also affects  the SRN energy spectrum. To reduce
this uncertainty, it will be  necessary to develop definitive supernova explosion
models, or utilize supernova neutrino nucleosynthesis to better constrain the neutrino temperatures. 
For example, the numerical data plots used to
derive the ellipsoid in Figure \ref{fig:3} includes some old models and the
dispersion in the T$_{\bar{\nu_e}}$ - T$_{\nu_x}$ plane among the models has a
tendency to shrink into a narrower region  for the more recent models. This suggests that the variance
of data plots in the T$_{\bar{\nu_e}}$ - T$_{\nu_x}$ plane may become
smaller in the future as neutrino transport in SN explosions is better understood.

Of particular use might be the long anticipated Galactic supernova event, especially if one were close enough that an accurate measurement could be made of the
neutrino spectrum and light curve for the different neutrino types.  Such a detection would go a long way toward clarifying the neutrino temperatures and flavor 
mixing, particularly if the progenitor star can be well characterized. 

Another uncertainty is the detailed relationship
between the progenitor mass and T$_\nu$ (or E$_\nu$) in the mass range
of (10 - 25) M$_\sun$.  If this relation can be  better understood, the
contribution to the total SRN energy spectrum from CC-SNe can be obtained with
better precision,
especially in the high energy range.

 The uncertainty due to the  statistical variance of SRN detection in water
\v{C}erenkov detectors contributes only $\sim 1\%$ to $ 40\%$ and depends upon the neutrino energy. 
Although it would be useful to reduce this
uncertainty  especially in the low energy region, this
contribution  is smaller than the other two
components, especially in the high energy range. So the effectiveness of
reducing this uncertainty is limited.
In addition, improving this component will require  a technological innovation
in SRN detection. For example, a water
\v{C}erenkov detector with larger effective volume than those of Hyper-K or
a  detection run time of more than 50 years would be needed to significantly reduce the statistical
uncertainty. 

Among other possible improvements, the  neutrino  mass hierarchy remains unknown, but should be determined 
within in the next five years.  In addition, there is ongoing  observational research into the 
star formation rate. Once  these parameters are better  determined along with the supernova parameters T$_\nu$
and the progenitor-mass boundaries between ONeMg SNe, CC-SNe, SNIb,c,   and
fSNe, it will be possible to better predict the spectrum of SRN.

\section{Solution to the star formation rate problem and SRN Detection rates} 

Although we have suggested that there may be no supernova rate problem, this was based upon a very liberal  estimate of the errors in the observed star formation rate.  
The analysis of  \cite{hor11}, however, is still quite valid, and indeed, probably a better analysis of the supernova rate.  If we adopt the SFR of \cite{hop06} and \cite{hor11}, then even after allowing for a significant fraction of stars becoming ONeMg SNe or fSNe, a supernova problem remains.  Hence, it is worthwhile to  continue to entertain  the possibility 
that there is a supernova rate problem at the level of a factor of 2   between the supernova rate expected based upon the observed SFR and
the direct SN remnant detections.  

In this section we postulate that enhancing the fraction  of fSNe relative to CC-SNe in the overall SFR may
explain the missing SNe problem, and moreover, may be detectable in the spectrum or relic supernova neutrinos.  
To do this we consider that there is enough uncertainty in the mass range  that forms failed SNe 
and/or enough uncertainty in the IMF itself to simply rescale the relative fraction contributed by fSNe  to  achieve the required ratio.  

One must keep in mind, however,
that the deduction of the SFR from observations is dependent upon an assumed IMF for the stars producing the observed UV flux.  
The UV flux is dominated by OB associations containing stars in the mass range of (15-40) M$_\odot$.  Since the observed SFR is dependent upon the assumed IMF in this region, one should be cautious about applying a modified IMF.  In particular, the overall normalization of total stars in that mass range must be preserved. Also we note, that the progenitor mass range of observed OB associations does not overlap the mass range of ONeMg SNe.  Hence, changing the fraction of stars that from ONeMg SNe cannot solve the supernova rate problem, i.e. the number of visible supernovae from progenitors with $M > 10$ M$_\odot$ is still fixed by the observed UV flux.

On the other hand, one is free to modify the fraction of stars that form failed SNe in the mass range $15 < m < 40$ M$_\odot$.  This can be achieved either by changing the lower threshold for fSNe, decreasing the fraction of luminous SNIb,c,  or by modifying the IMF in such a way as to keep the normalization the same.  Here,  we consider the consequences that such modifications of the relative fraction of luminous vs.  fSNe would have on the 
 detection rate of the relic neutrino background.  In particular, we wish to determine whether
 an excess fSNe could be detected.  For a best case scenario we will assume that  the
enhanced fraction of failed supernovae produce  a factor of two reduction in the expected  rate of observed supernova remnants.  We also presume 10
years of detector run time in the fiducial  SRN detector.  

Figures \ref{fig:14}ab  and \ref{fig:15}
%through \ref{fig:15} 
show the calculated detection event rates with an enhanced contribution from   fSNe.  
These results  are similar to  those of other studies 
\citep{lun09,Yang11,kee12, lun12}.  These figures, however also delineate  
the effect of some of the  uncertainties. As a best case Figures \ref{fig:14} and \ref{fig:15} show the
combination of the smaller uncertainty in the SFR of \cite{hor11} and the detector statistics with 10 years run time. 
%As a worse case the error bands in
%Figures \ref{fig:14} and \ref{fig:15} show the
%combination of uncertainties  from  the presently deduced SFR, detector statistics, and also including the uncertainties due to T$_\nu$ in normal CC-SNe.
%[Note, however, that the temperature uncertainty may be reduced as modern simulations progress, and/or a Galactic supernova is detected.]

 The two panels on Figure \ref{fig:14} show the dependence of the positron detection rate on the neutrino oscillation scenario for the two choices of EoS for fSNe.
 The three bands on each panel are for the three different neutrino oscillation scenarios considered here.
  Panel  (a) shows the case that all of the missing SNe are fSN modeled with the soft EoS of
\citet{she98}.  Panel (b) shows  the case that all of the missing SNe are fSN modeled with the EoS of
\citet{lat91}.  Plotted detection
rates are based on the assumption that one can increase the fraction of fSNe until the expected number of luminous SNe is diminished by a factor of two.

The star formation rate for missing fSNe and
its uncertainty  are  based upon the calculations of \citet{hor11}  as noted above.
 Among the possibilities  shown in Figure \ref{fig:14}, case (b)  shows that for SNe based upon
 the soft LS EoS, the error bars for the oscillation case {\it I}
($\bar{\nu_e}$ = 0.7 $\times$ $\bar{\nu_e}^0$ + 0.3 $\times$ ${\nu_x}^0$) and those for the
oscillation case {\it II} ($\bar{\nu_e}$ = ${\nu_x}^0$) in
the range of E$_{e^+}$ = 10 - 25 MeV do not overlap. Hence, one might 
 be able to distinguish oscillation case {\it I} from the
oscillation case {\it II} if all of the fSNe follow the LS EoS and if one can detect the SRN energy spectrum
in a 10$^6$ ton class detector. 

 The three panels in Figure \ref{fig:15} show detection rates for each of the oscillation scenarios in separate panels.
 The two sets of  colored bands on each figure are for different EoSs as labeled.  
  Note that in each oscillation paradigm in Figure \ref{fig:15}, one can easily distinguish the
EoS from the value of E$_{e^+ peak}$. This
means that one might be able to distinguish the EoS leading to  fSNe if we
can detect SRN energy spectra in a 10$^6$ ton class detector and with any
conditions of neutrino oscillation parameters.   

 In the spectra shown in Figures \ref{fig:14} and \ref{fig:15}, we have assumed that  the
influence of the uncertainty in T$_\nu$ from CC-SNe can be ignored.
Under these assumptions,
these figures suggest that one might  be able to constrain neutrino
oscillation parameters and this candidate of missing SNe.
On the other hand, 
 %Figures \ref{fig:14} and \ref{fig:15} show the more realistic expectation based upon the present work.  For these figures 
 if the uncertainty in
 T$_\nu$ is included 
 %and we presume that there is only a factor of 1.4 enhancement of the supernova rate deduced from the SFR compared to 
% the observed supernova rate. 
%   It is clear from these figures that the neutrino temperature  uncertainty particularly affects the  high energy positron detection. 
it becomes more difficult to distinguish whether failed SNe are the source of the supernova rate problem.

% The three panels in Figure \ref{fig:14} show the detection rate for each of the possible supernova sources.
% The three colored bands on each panel are for the three neutrino oscillation cases.   The rate for missing SNe and
%its uncertainty are  based on our calculations shown in 
%Figure \ref{fig:2}. In every panel  in Figure \ref{fig:14}
%the error bands overlap each other over a wide range and the
% $e^+$ energy spectrum for each oscillation scenario
%is indistinguishable. In these cases, it will be  difficult to distinguish
%neutrino oscillation cases  or candidates for missing SNe from SRN detection.
 
% The three panels in Figure \ref{fig:15} show detection rates for each of the oscillation scenarios considered here.
% The three colored bands on each figure are for each possible source for the missing neutrinos.  
% From these figures we deduce that one might  be able to distinguish the
%candidate for missing SNe by the detection of  the SRN energy spectra
%within a 10$^6$ ton class detector.

%\clearpage
% Page 
\section{Conclusion}

It is an intriguing conundrum \citep{hor11}  in observational cosmology 
that the measured core-collapse supernova rate ($R_{SN}$)  in the redshift range 0$\le z \le$1 
may be about a factor of two smaller than that inferred from the supernova rate deduced from the measured cosmic massive-star formation rate (SFR).
We have studied the supernova relic neutrino background  theoretically in order to clarify this issue.
In particular, we analyze   the energy spectrum and the detection rate 
for a  next-generation \v{C}erenkov detector like Hyper-Kamiokande.  We have shown that some insight into the supernova rate problem
can be gained by measuring the ratio of events at the peak of the spectrum to  the number of energetic (25-30 MeV) events.
Our second goal has been to clarify and minimize the uncertainties involved in
 estimates of the  spectrum of arriving  SRNs.
%This flux is derived from an integral over  time and/or redshift 
%of the source SRN energy spectrum (redshifted from the epoch of the  supernova explosion to the present) 
%multiplied by the supernova rate in a comoving volume at that epoch.
%We have assumed a standard cosmological $\Lambda$CDM model.

\subsection{SFR analysis}
In the first part of this study we made and alternative fit to  the measured massive star formation rate over the redshift range $0 \le z \le 7$
and derived a new  supernova rate over the redshift range $0 \le z \le 1$, respectively.
 Our alternate analysis of the supernova rate problem  only considered the star formation data with directly detected extinction corrections.  
Our estimated uncertainty  in  the massive SFR 
is then +100$\% (+1\sigma)$ and -35$\% (-1\sigma)$ from the $\chi^2$ of the fitting function.
This  uncertainty contributes to the uncertainty in the normalization of the total 
SRN energy flux and the detection rate when analyzing the  fraction of missing non-luminous supernovae.

 Our new supernova rate also allows  for the fact  low-mass SN progenitors end their lives as ONeMg SNe, while the highest-mass progenitors can lead to either SNIb,c events
 or failed SNe.   Our  estimated uncertainty in the core-collapse supernova rate ($R_{SN}$) 
 is  then +7$\% (+1\sigma)$ and -30$\% (-1\sigma)$.
In our adopted  paradigm, the predicted luminous supernova  rate
is  only a factor of  $1.1^{+1.0}_{-0.4}$ larger than  the observed core-collapse supernova rate.  This is consistent with no supernova rate problem and
also consistent with the results in \cite{kob13}.
%We adopted our core-collapse supernova rate ($R_{SN}$) 
%as a "default case" when we discuss the effects of a possible missing part
%of $39\%-100\%$ enhancement as an extra component.

We also studied the sensitivity of the SRN detection rate
to a range of neutrino temperatures from a number of published supernova models.
For each supernova model we  deduced  the most plausible
neutrino temperatures T$_{\nu_e}$, T$_{\bar \nu_e}$ and T$_{\nu_x}$
($\nu_x$ = $\nu_{\mu, \tau}$ or $\bar{\nu_{\mu, \tau}}$) and their uncertainties 
because they affect strongly the SRN detection rate. 
%For standard CC-SNe, we considered  many previous  supernova simulations to extract 
%an  ensemble of average of the neutrino temperatures.  
We adopted a natural temperature hierarchy 
T$_{\nu_e} \le$T$_{\bar \nu_e} \le$T$_{\nu_x}$
from the consideration of both charged and neutral current interactions
of neutrinos with neutron-rich material in the core.
We also incorporated  studies \citep{yos08,hay10} of $\nu-$process  nucleosynthesis 
in order to constrain the neutrino temperatures.

We showed that the ratio of the number of peak events to energetic events (25-30 MeV) in the spectrum can be used as a robust
 measure of the temperature of arriving election anti-neutrinos.  Such a study would thereby provide a valuable  constraint on supernova models.

\subsection{Neutrino Oscillations}
We also  considered  the effects of the neutrino flavor oscillations.
In our fiducial model, there is not much dependence of the detected spectrum on the oscillation paradigms considered here.
However, in the case that there is an enhanced contribution for fSNE the dependence  SRN energy flux and detection
rate on the neutrino oscillation parameters can be clearly seen. 
In this case there  difference between number of peak events in the completely adiabatic  case  (I) and the no oscillation case (III).  This because in the complete adiabatic 
mixing  case
all SRN $\bar{\nu_e}$ that reach a terrestrial detector  originate  as more energetic $\nu_x$.
Nevertheless, this may be difficult to detect because  the  uncertainties in  the SFR and T$_{\bar{\nu_e}}$ are comparable to  the effects of oscillations.

\subsection{EoS Sensitivity}
We also studied the effects of the EoS on the detection of SRN.  
Variations in the EoS  
particularly  affect the  neutrino luminosities and temperatures
in  failed SNe.  We used two models:  One is the rather  soft EoS 
taken from that of the  ($K=180$) model of \cite{lat91},
and the other is a hard relativistic-mean-field EoS taken from \cite{she98}.
A soft EoS results in higher neutrino luminosities and temperatures
because the neutralized hadronic matter inside the proto-neutron star is
compressed more strongly than in the case of a hard EoS during  the  collapse to a black hole.
Because  the contribution from the failed SNe is $\le 10\%$   for our fiducial model, the EoS difference is small, but possibly detectable. 

\subsection{Contributions from failed and ONeMg  SNe}
In another aspect of this work  we considered contributions from
  additional  supernova models. In addition to those leading to a neutron star (CC-SN), 
  we considered  the contribution to the observed relic neutrino spectrum from massive stars leading to  a black hole (failed SN) or
gamma-ray burst (GRB), and the contribution from low-mass SN progenitors  
leading to an electron-degenerate supernova (ONeMg SN).  
For this work  we  adopted the modified  Salpeter A initial mass
function model  \citep{Baldry03} in order to evaluate the relative  fractions
for which these processes contribute to the  total massive-star formation rate and/or core-collapse supernova rate.

We have considered  a number of  models for  failed SNe \citep{sum08} and GRBs \citep{har09}, as well as dark OMgNe SNe \citep{hud10}  
assuming  that the energy spectrum of each 
neutrino species follows a Fermi-Dirac distribution with zero chemical potential.
We estimate that  the  effect of ignoring the chemical potential is  less than
$5-10\%$ in temperature based upon  previous studies \citep{yos05}. 
We find as expected that the CC-SNe make the dominant contribution
to the SRN energy spectra and the detection rate,
and that the contribution from  ONeMg SNe or failed SNe and GRBs is $\le 10\%$
if we do not impose  a larger fraction of failed SNe contributions to solve the  "supernova rate problem".
Even so, the contribution  from and EoS of failed SNe in the detector spectrum is possibly detectable through the ratio of peak to energetic events.
%One can, however, identify the contribution from the failed SNe
%in the  high energy region  of the SRN flux and detection rate.  This is  
%because  failed SNe have much higher neutrino temperatures
%for all species.  

\subsection{A testable solution to the possible supernova rate problem}
Finally, we have studied a possible solution to the "supernova rate problem," i.e. 
if the observed cosmic supernova rate is as much as  a factor of  2 smaller than 
the supernova rate expected based upon the measured cosmic massive-star formation rate  and a modified Salpeter A IMF.
One possible interpretation of the difference between the inferred and observed  $R_{SN}$ is to assume an enhanced 
fraction of   failed SNe leading to  black hole formation.
In this work we have demonstrated  that the neutrino luminosities and temperatures among
all neutrino species from  such failed SNe are very different from that of other SNe.
In particular, for the  Lattimer-Swesty soft EoS, the resultant neutrino luminosities and temperatures
are much higher.
Hence, one may be able to both solve the supernova rate problem and distinguish the EoS 
from the detection of the supernova relic neutrinos within ten years of run time on a detector like Hyper-Kamiokande.

We have  also found that in this case one may be able to  constrain the neutrino
oscillation type, i.e. whether the high density resonance is adiabatic
or non-adiabatic.   It may even  be possible to place a constraint on the neutrino  mass hierarchy.
We note, however, that such  conclusions depend upon the uncertainties in the anticipated neutrino temperatures, 
 the EoS, and the neutrino oscillation parameters.

\vskip .1 in 
We finally conclude that the detected SNR \v{C}erenkov positron energy spectrum in the region 10$\le E_{e^+} \le$35 MeV in a next generation Hyper-Kamiokande-like detector can  can provide a robust measure of  the neutrino  temperatures emanating from supernovae and thereby constrain the supernova models and oscillation parameters.  At the same time such a detection can provide valuable insight into the supernova rate problem and the 
contribution of failed supernovae in particular.

\acknowledgments

  Work at the University of Notre Dame (GJM) supported by   
the U.S. Department of Energy under Nuclear Theory Grant DE-FG02-95-ER40934.  
This work was supported in part by Grants-in-Aid for Scientific Research of JSPS (26105517, 24340060).

\clearpage

\clearpage

\begin{figure}[h]
\begin{center}
\includegraphics[angle=0,width=3.0in]{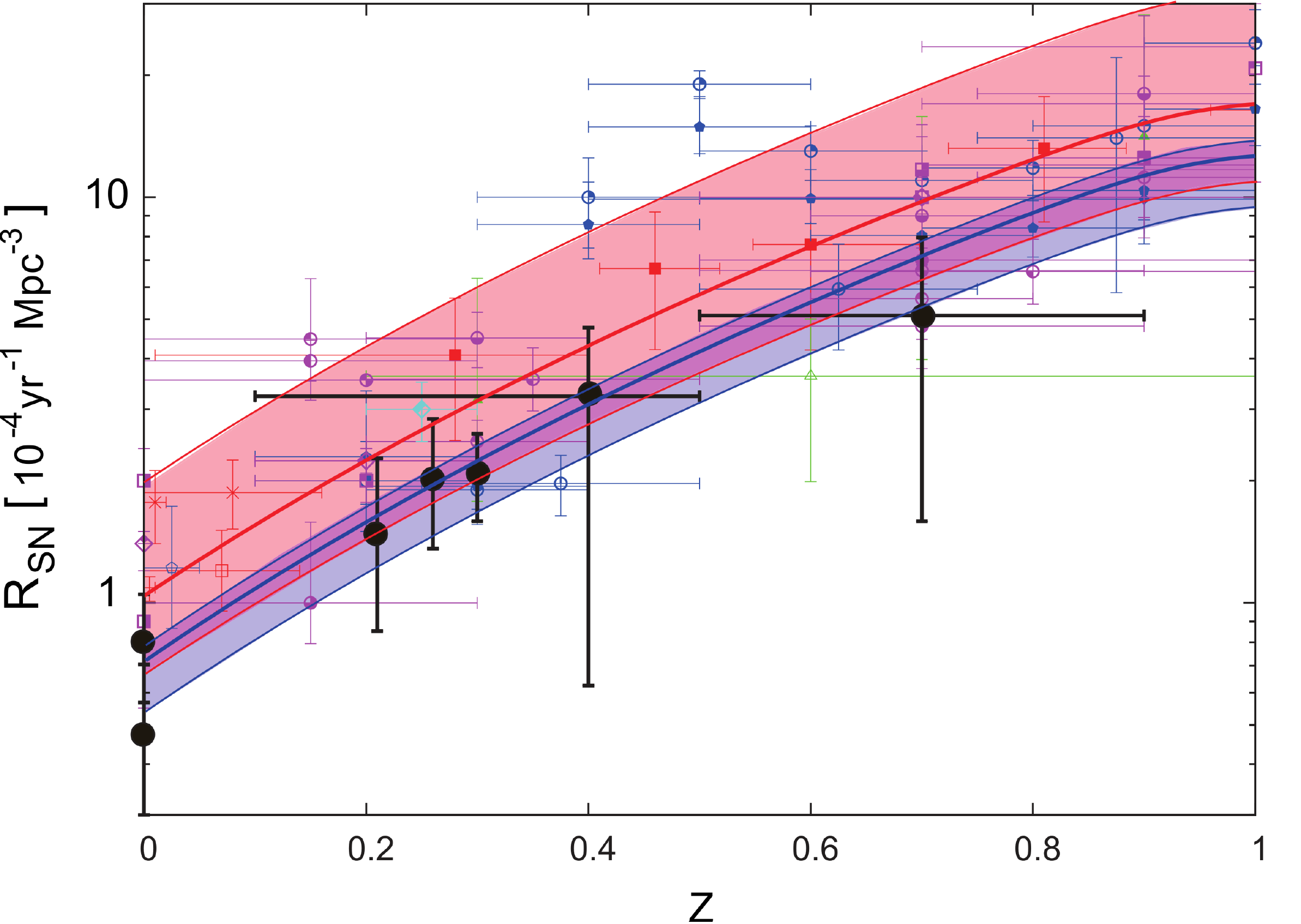}
\caption{[Color online] Supernova rate function $R_{SN}$  for the lower redshift interval $0 \leq z \leq 1$ deduced from the SFR shown in
 Figure \ref{fig:1} via Eq. (\ref{4-5-1}). This is compared with the observed supernova rate. Red thick solid line
and its associated band are the 
best fit $R_{SN}$ function and the  $\pm$ 1$\sigma$ error. Closed black circles show the measured SN rates (\citet{li11b}, \citet{cap99}, \citet{bot08},
\citet{cap05}, \citet{baz09}, \citet{dah04}).
The blue thick line shows the best fit SN rate function from these direct supernova measurements.
The blue thin lines show the upper and lower limits for a fit to the direct
observations  are taken from the LOSS measurement [see \citet{lea11}, \citet{li11a} and
\citet{li11b}].
\label{fig:2}}
\end{center}
\end{figure}

\clearpage

\begin{figure}[h]
\begin{center}
\includegraphics[angle=0, width=4.0in]{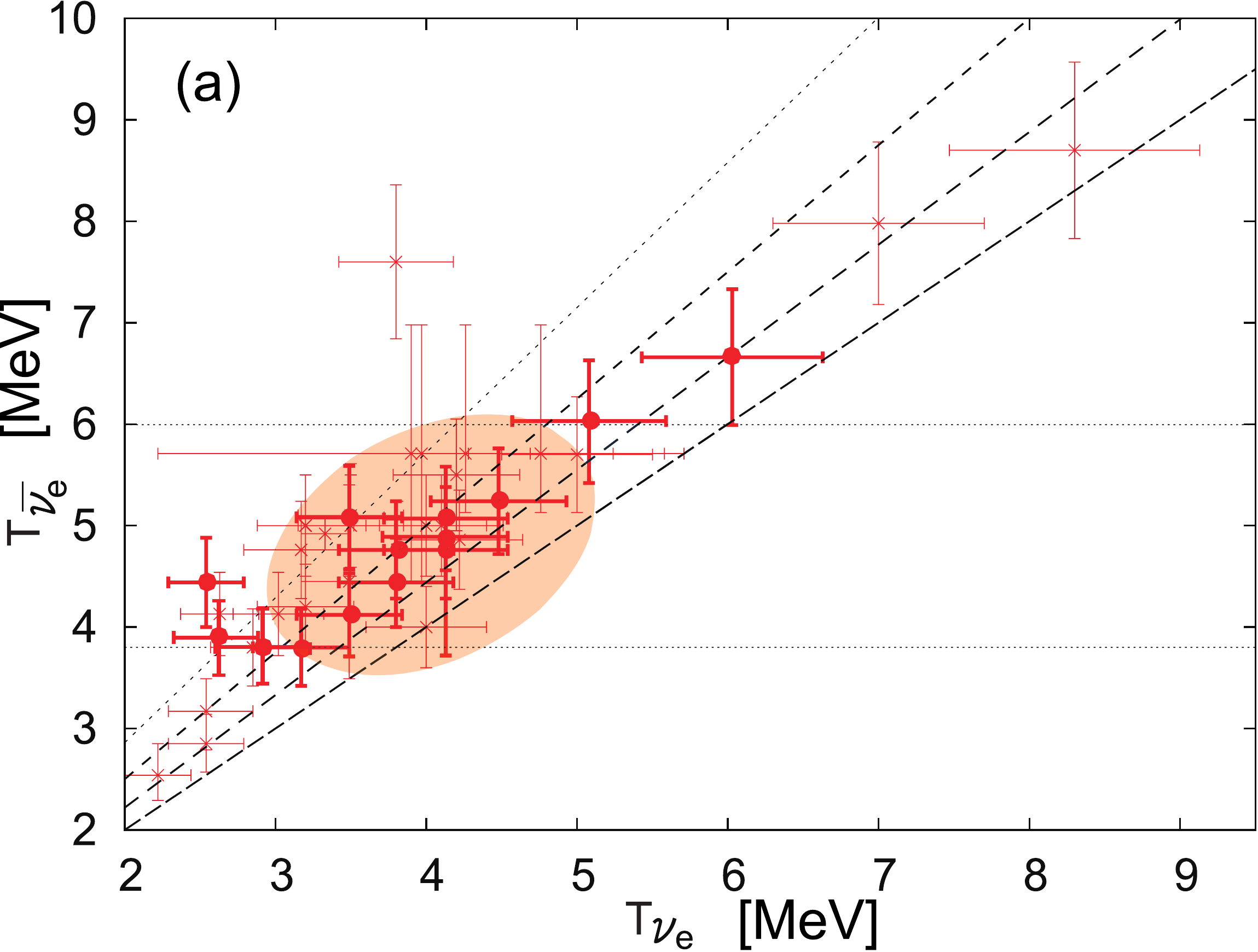}
\includegraphics[angle=0, width=4.1in]{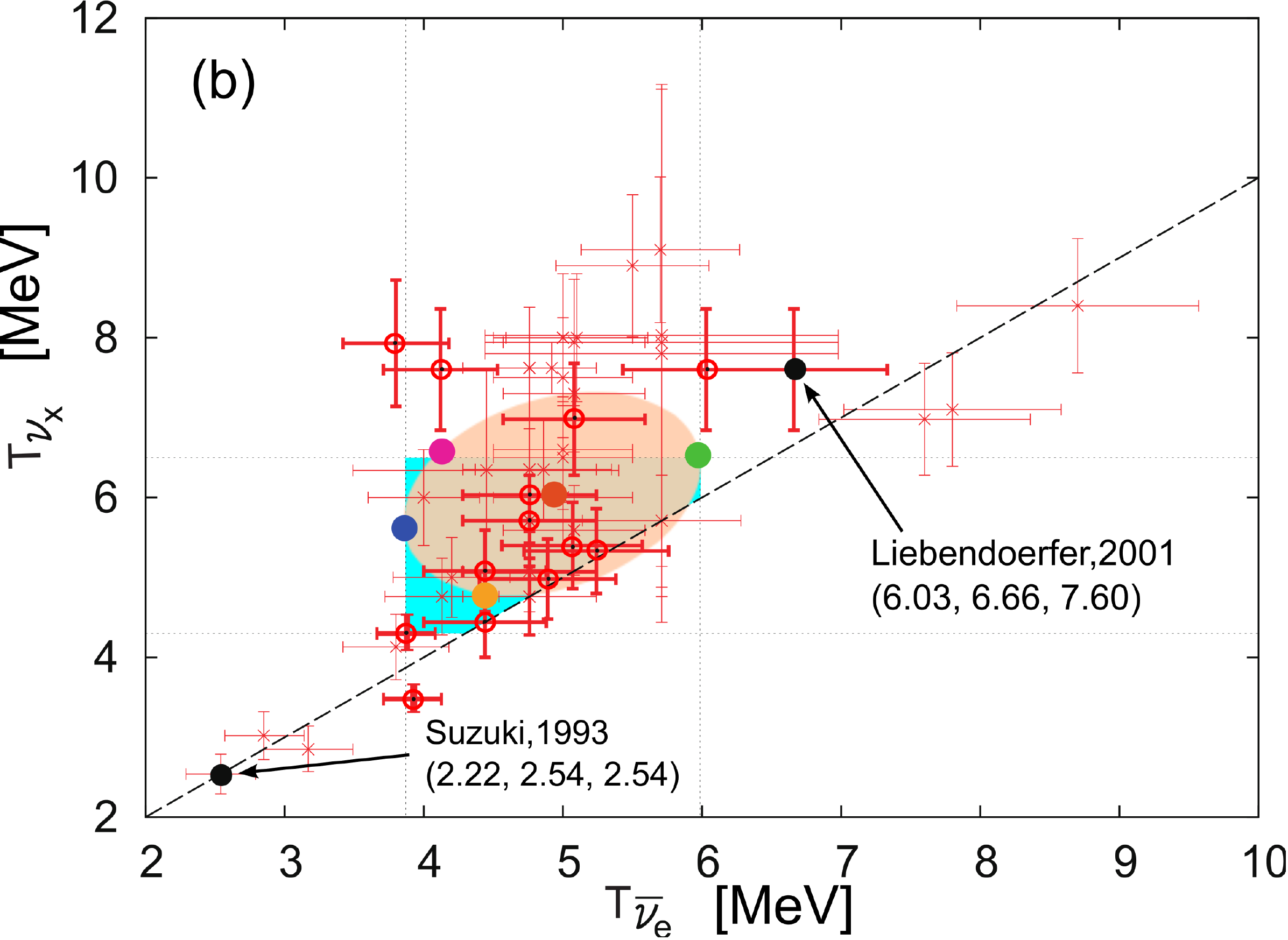}
\caption{[Color online] (a) The correlation between T$_{\nu_e}$ and T$_{\bar{\nu_e}}$. (b) The correlation between T$_{\bar{\nu_e}}$ and T$_{\nu_x}$.
 In (b) Green, red, magenta,
blue, yellow and two black circles show the 7 representative points with (T$_{\bar{\nu_e}}$,
T$_{\nu_x}$) = (6.0 MeV, 6.5MeV), (5.0 MeV, 6.0 MeV),
(4.1MeV, 6.5MeV), (3.9MeV, 5.6MeV), (4.5MeV, 4.7MeV), (6.7MeV, 7.6MeV) and
(2.5MeV, 2.5MeV), respectively. 
The Black circle at (6.7MeV, 7.6MeV) is  from \citet{lie01},
and the point at (2.5MeV, 2.5MeV) is  from \citet{suz93}. These 7 pairs of T$_\nu$ are
used in deriving the uncertainty in the SRN energy spectra.
In both figures the orange colored ellipse 
 indicates the 1$\sigma$ contour around the central value.  The  horizontal dotted lines in (a) [Vertical lines in (b)] show upper and
 lower limits of T$_{\bar{\nu_e}}$.  
Straight lines show T$_{\nu_x}$/T$_{\bar{\nu_e}} = 1 / 1.0, 1 / 0.9, 1 / 0.8,
1 / 0.7$. The light blue area in (b) shows the  allowed region of
T$_{\bar{\nu_e}}$-T$_{\nu_x}$ pairs as discussed
in the text.  Red closed circles with error bars show
T$_{\bar{\nu_e}}$-T$_{\nu_x}$ pairs derived from original 
numerical simulations.  The error bars denote the estimated $\pm$10\% change in neutrino temperatures with time as discussed in the text.
\label{fig:3}}
\end{center}
\end{figure}
%\end{landscape}
%\newpage

\clearpage

%\newpage
\begin{figure}[h]
\begin{center}
\includegraphics[angle=0,width=3.5in]{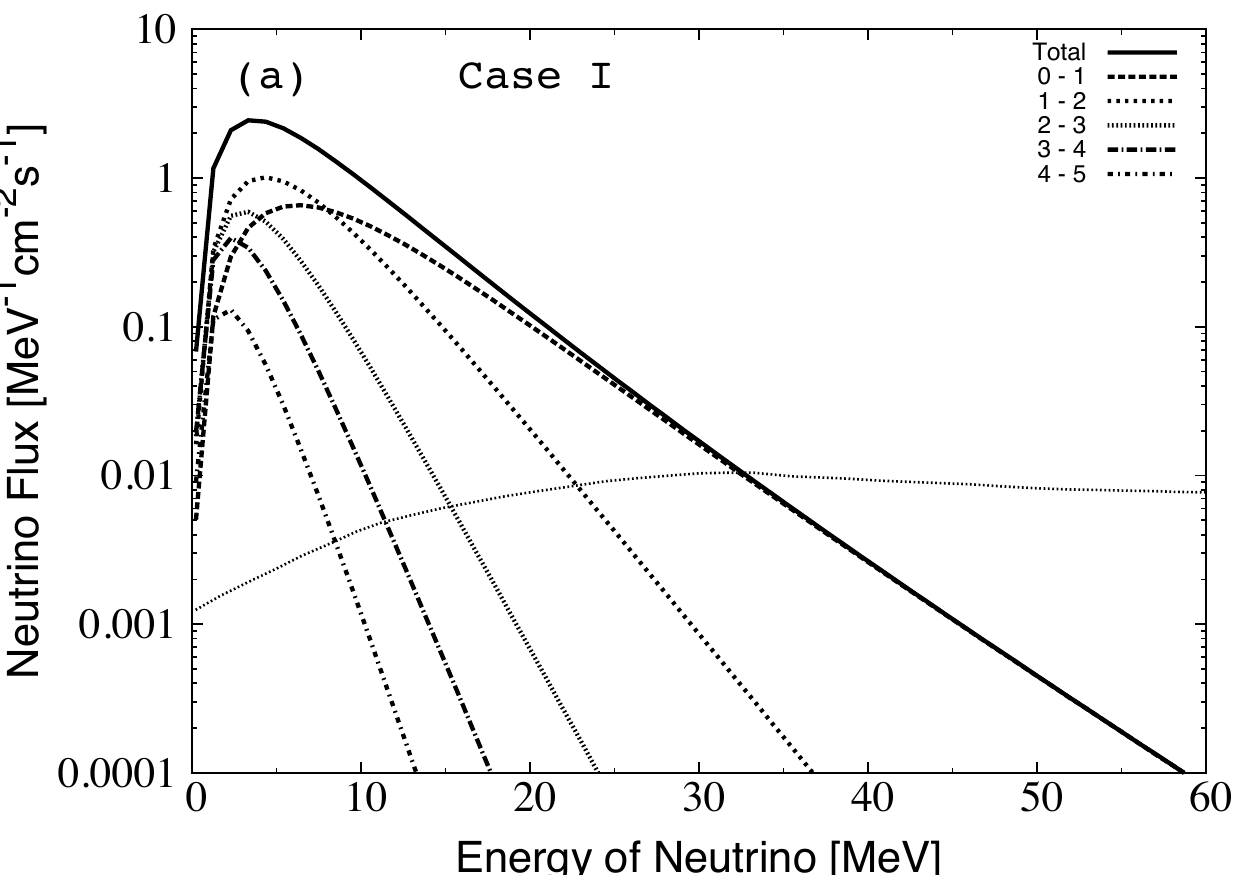}\\
\includegraphics[angle=0,width=3.5in]{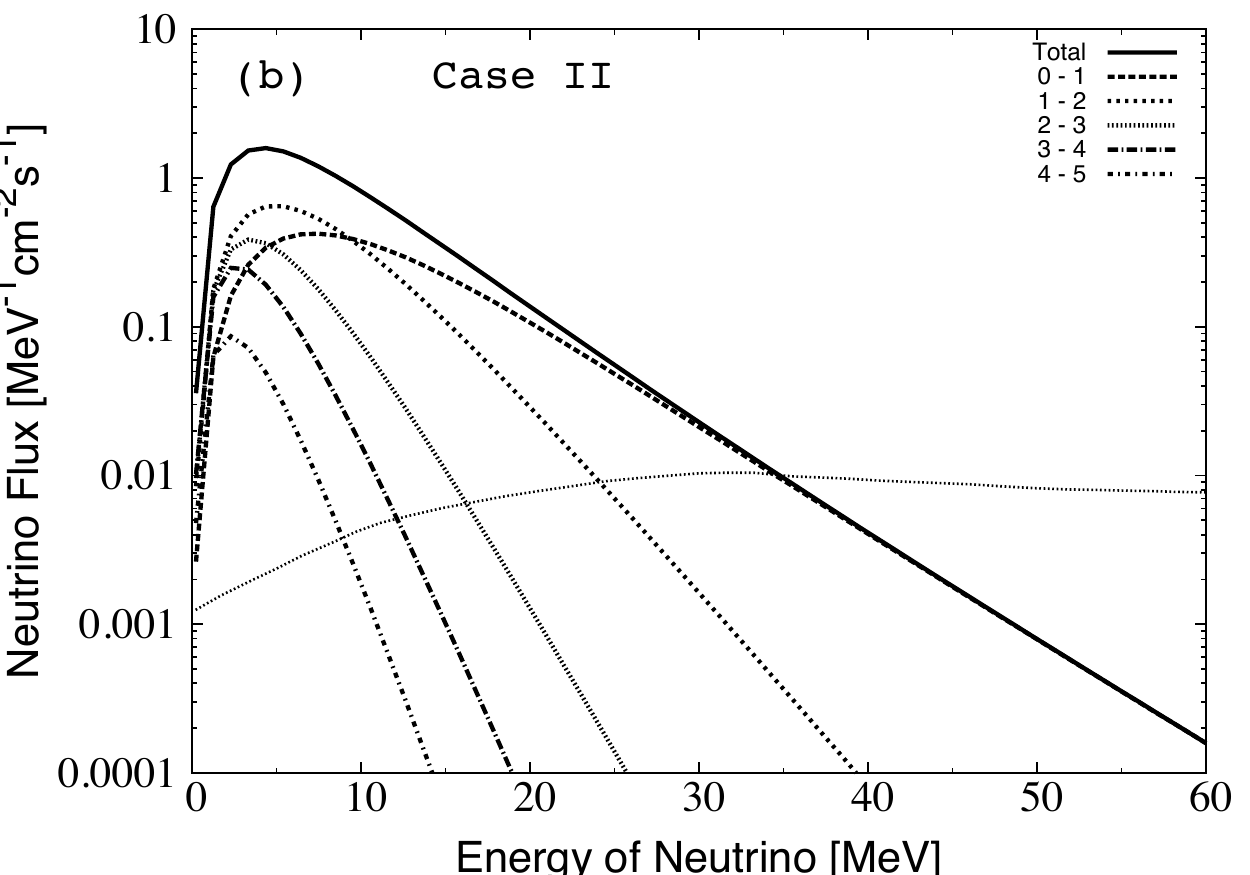}\\
\includegraphics[angle=0,width=3.5in]{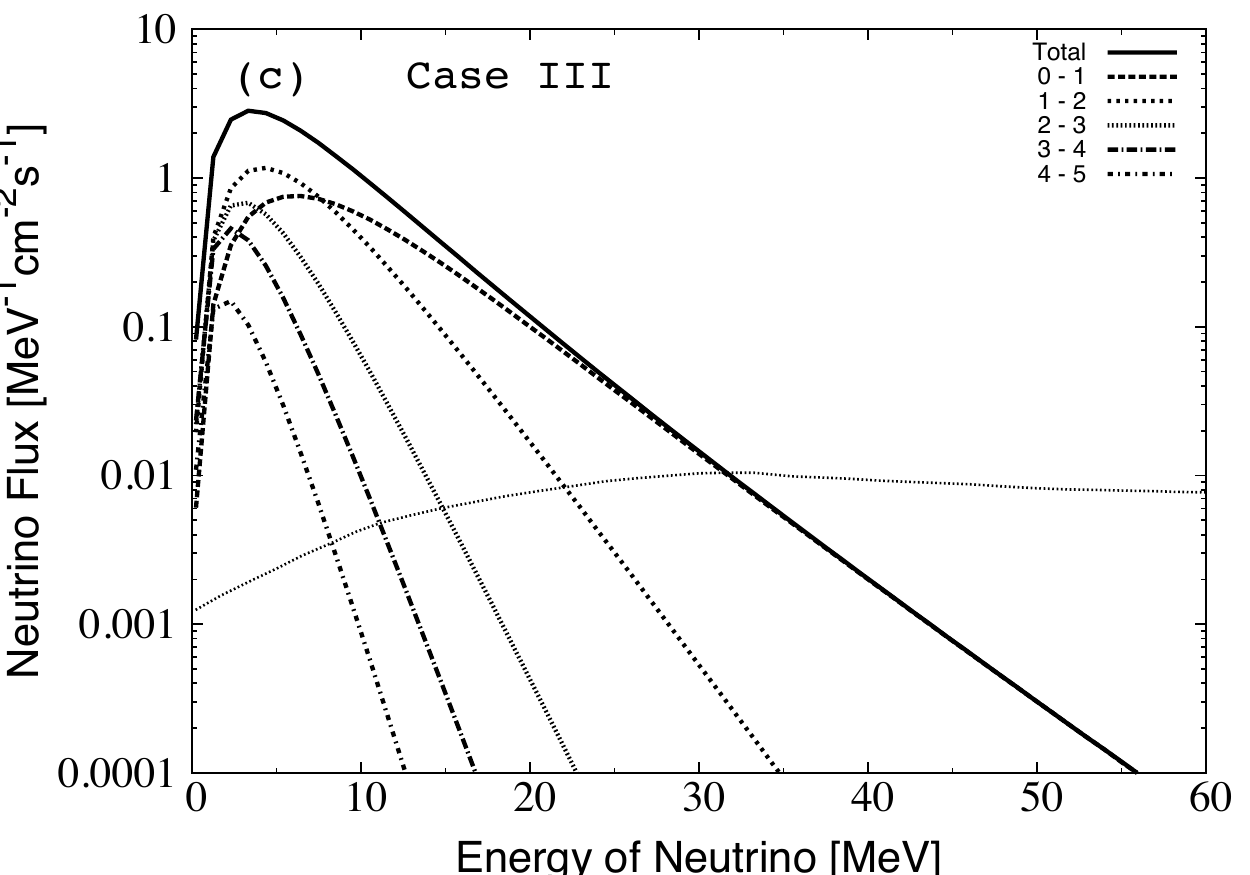}
\caption{Total number flux of SRN (thick  line) and flux arising from various redshift bins (as labeled) as a function of
  $\bar{\nu_e}$ energy, assuming (T$_{\bar{\nu_e}}$, T$_{\nu_x}$) = (5.0 MeV,
 6.0 MeV). Each panel presents a different  oscillation case.  Panel (a) is for oscillation Case {\it I} ($\bar{\nu_e}$ =
 0.7 $\times$ $\bar{\nu_e}^0$ + 0.3 $\times$ $\nu_x^0$).  Panel (b) is for oscillation Case {\it II} ($\bar{\nu_e}$ =
 ${\nu_x}^0$). Panel (c) is for no oscillations  (Case {\it III}).
The thin black line indicates  the  background noise from atmospheric neutrinos.  The intersections of the SRN flux with the background sets the upper detection limit at $\sim 33-37$ MeV for the total  neutrino spectrum.
\label{fig:4}}
\end{center}
\end{figure}

\clearpage

%\newpage
\begin{figure}[h]
\begin{center}
\includegraphics[angle=0,width=3.5in]{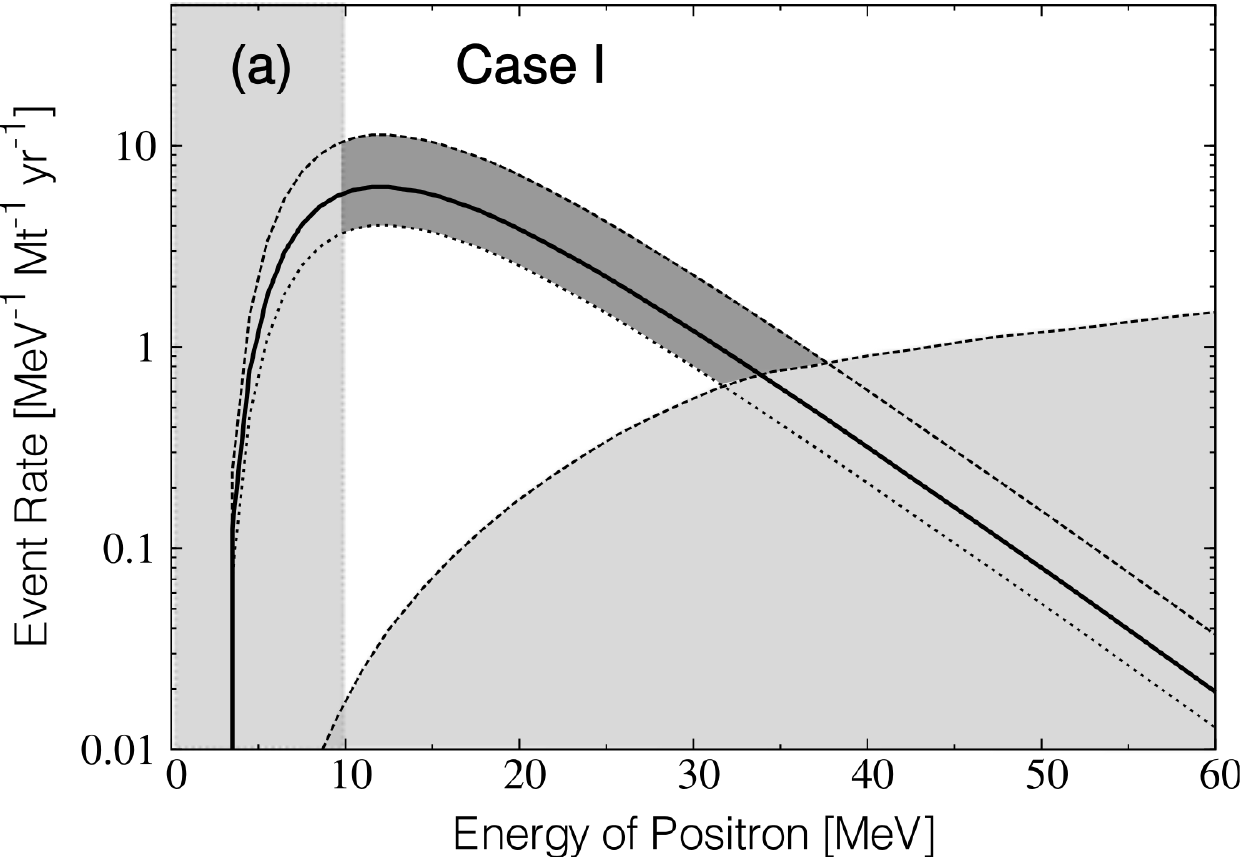}\\
\includegraphics[angle=0,width=3.5in]{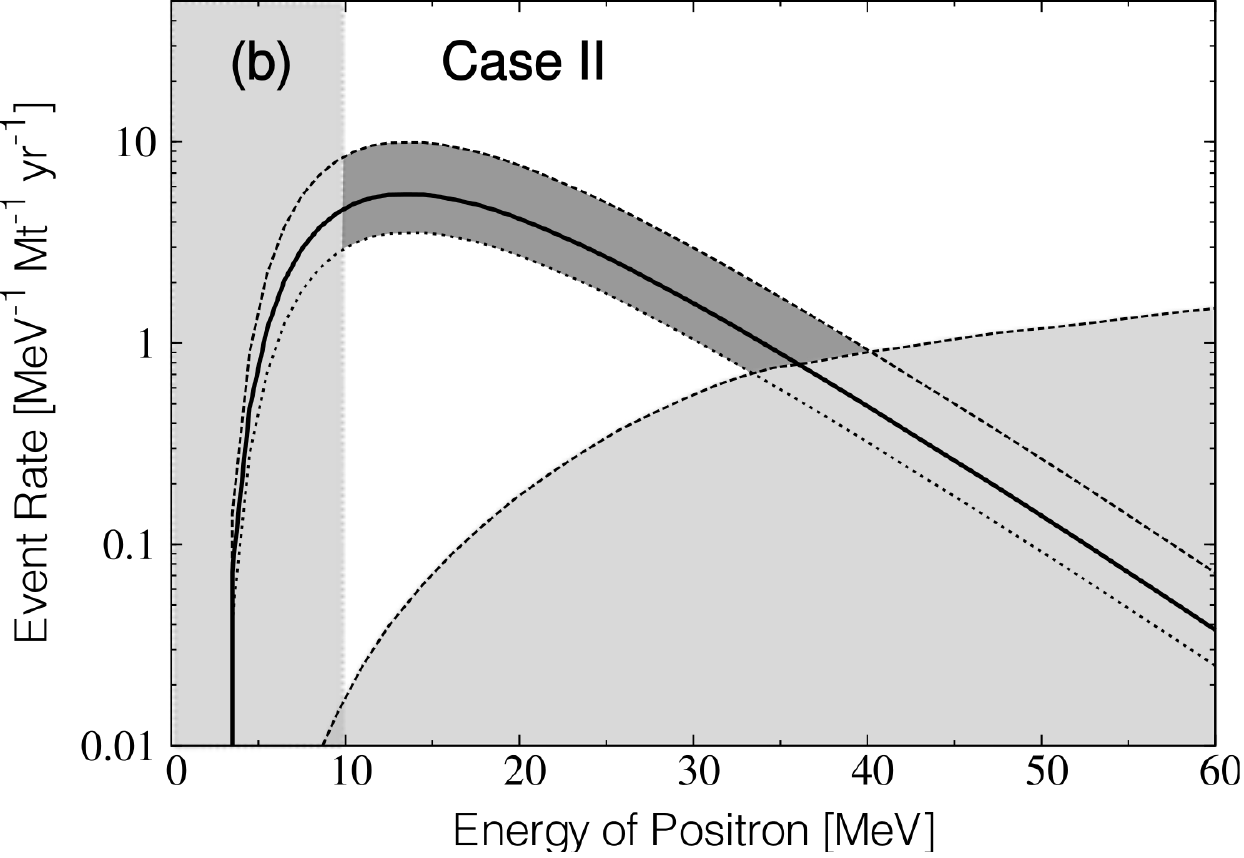}\\
\includegraphics[angle=0,width=3.5in]{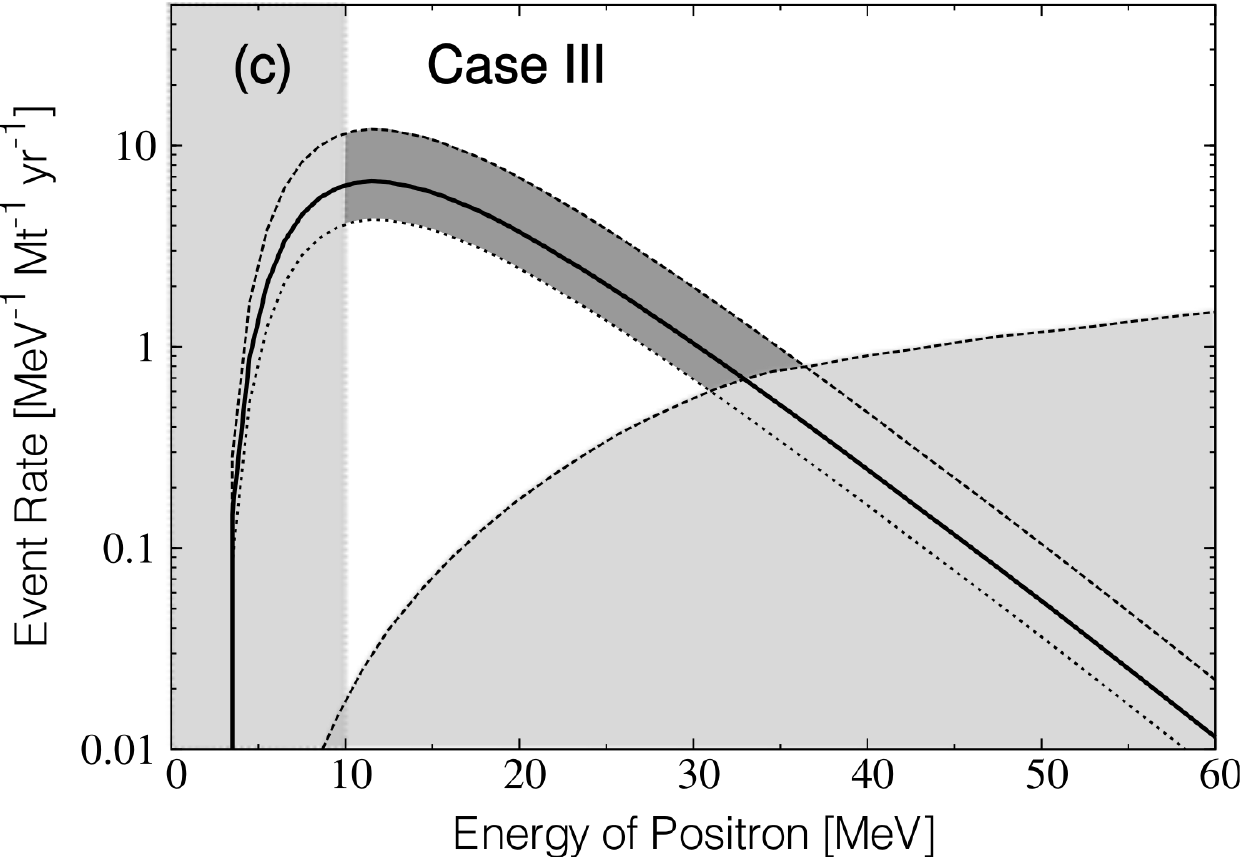}
\caption{Predicted $\bar{\nu_e}$ detection rate
as a function of e$^+$ energy for a  10$^6$ ton class water \v{C}erenkov detector
with 1 year of run time assuming the piecewise linear SFR from the present work assuming neutrino temperatures of 
(T$_{\bar{\nu_e}}$, T$_{\nu_x}$) = (5.0 MeV, 6.0 MeV).
The thick  line shows the predicted SRN detection rate for the piecewise linear  SFR. Thin  lines show the  $\pm 1 \sigma$ uncertainty due to
the uncertainty in the SFR fit. Panel (a) is for   oscillation Case {\it I} ($\bar{\nu_e}$ =
 0.7 $\times$ $\bar{\nu_e}^0$ + 0.3 $\times$ $\nu_x^0$).  Panel (b) is for  oscillation Case {\it II} ($\bar{\nu_e}$ =
 ${\nu_x}^0$), and panel (c) is for no oscillations (Case {\it III}).
The shaded region  in the energy range of 0 - 10 MeV indicates the region where  background
noise due to terrestrial nuclear reactor $\bar{\nu_e}$ may dominate. The shaded region that intersects the spectrum at  $\sim 33
$ to $ 37$ MeV indicates  the  background from atmospheric neutrinos.
\label{fig:5}}
\end{center}
\end{figure}

\clearpage

\begin{figure}[h]
\begin{center}
\includegraphics[angle=0,width=3.5in]{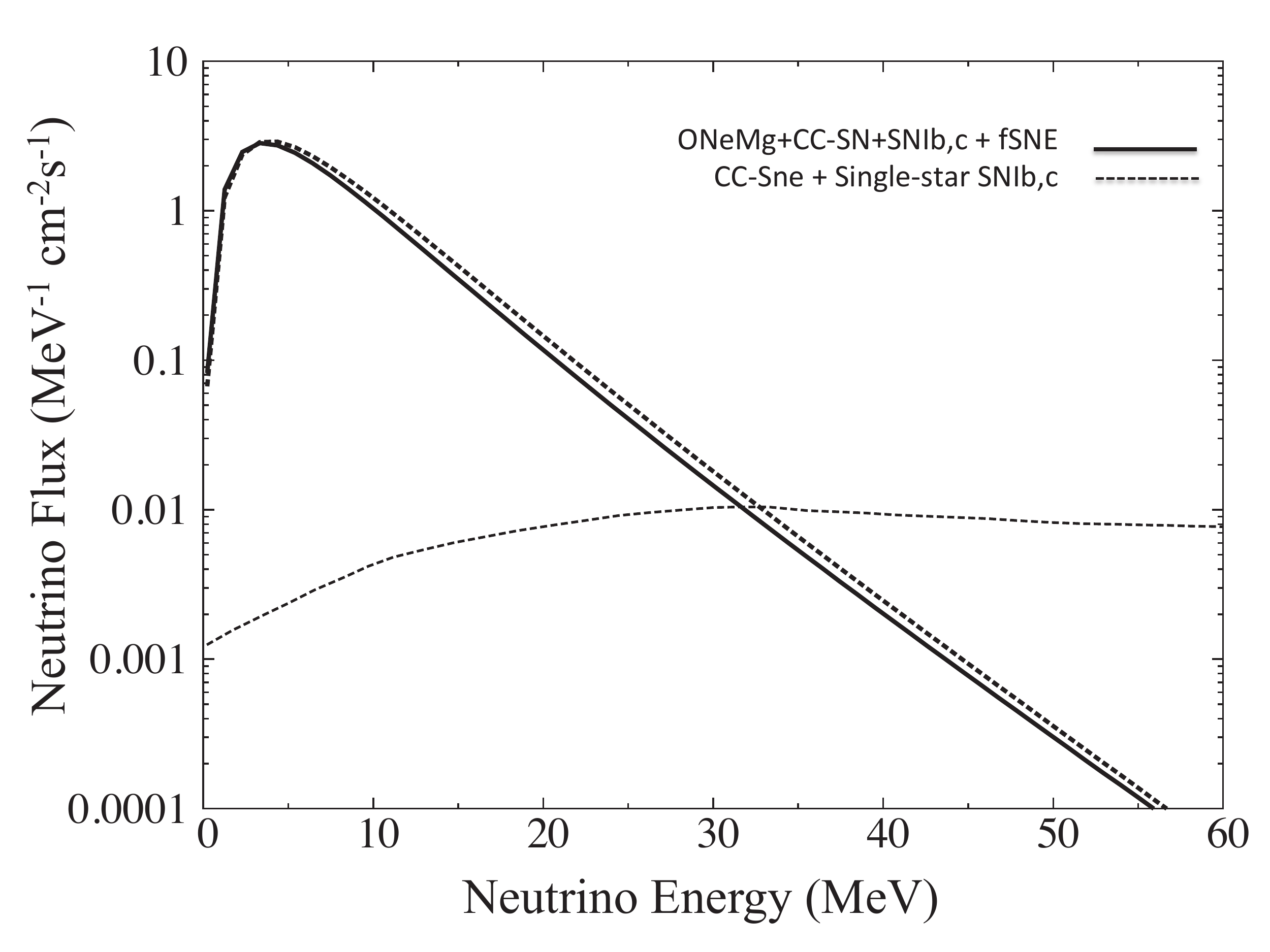}\\
\includegraphics[angle=0,width=3.5in]{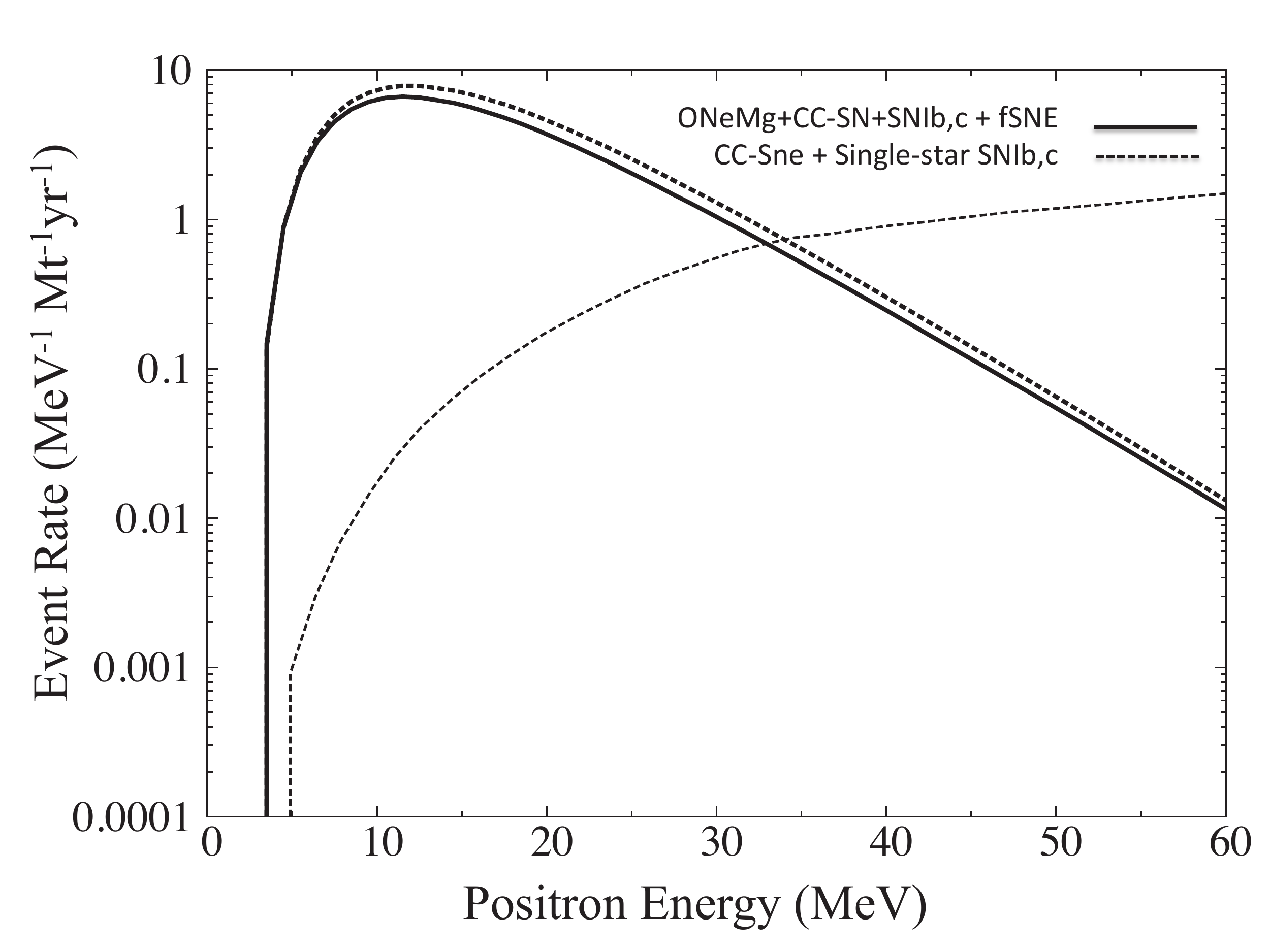}\\
\caption{ Dependence of the  predicted total SRN flux (upper panel) and the 
 detection rate (lower ) on the adopted models for SNIb,c, fSNe, and ONeMg SNe in the case of a fiducial neutrino temperature  of
 (T$_{\bar{\nu_e}}$, T$_{\nu_x}$) = (5.0 MeV, 6.0 MeV) and no oscillations.
 The solid line is for the model model with CC-SNe  plus ONeMg SNe and binary SNIb,c as described in the text.  The dashed line is for the single-star SNIb,c, as adopted in \cite{hor11}. The dotted line in the upper panel and the dark shaded regions  in  the lower panel show the
neutrino backgrounds as described in Fig.~\ref{fig:5}. 
\label{fig:6}}
\end{center}
\end{figure}  

 \clearpage

\begin{figure}[h]
\begin{center}
\includegraphics[angle=0,width=4.5in]{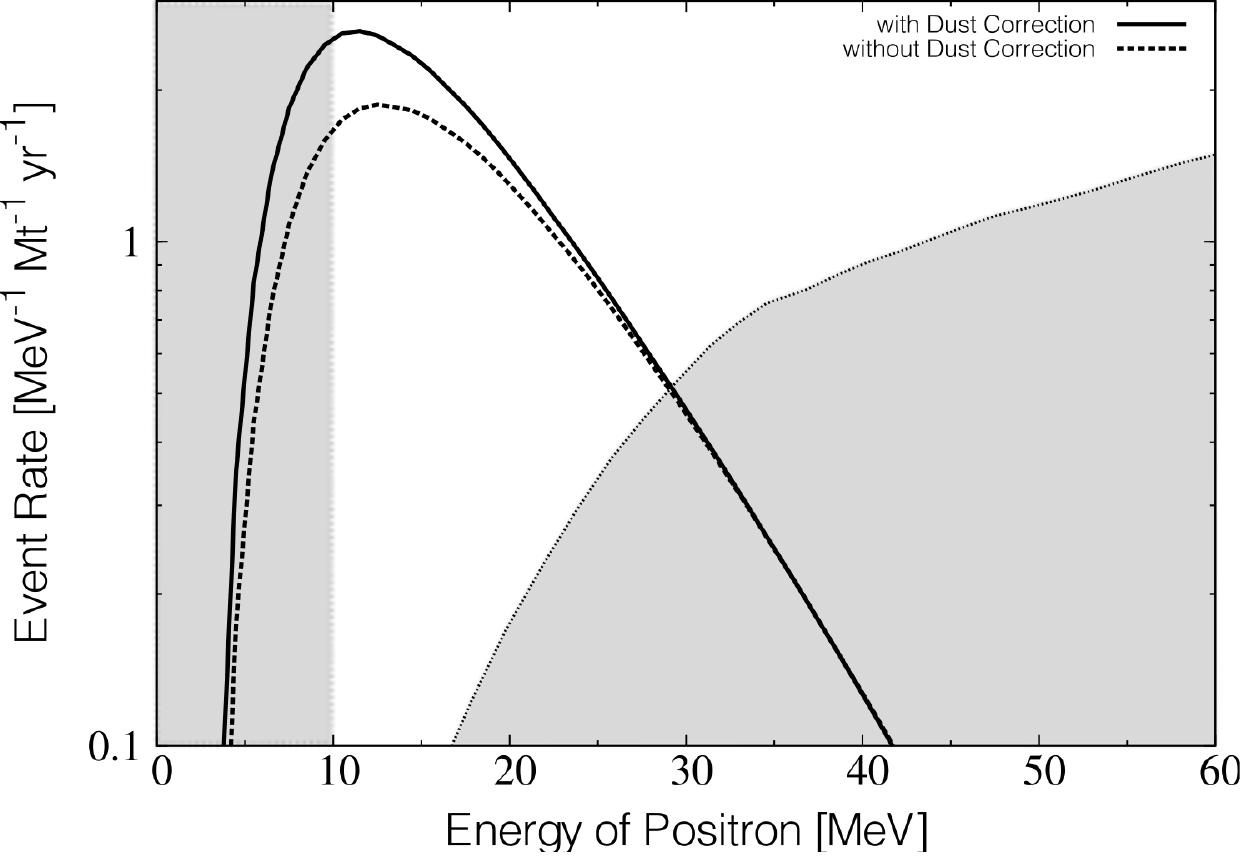}
\caption{Energy dependence of the SRN detection rate for two models of the SFR from
 \citet{kob00}. The solid  line shows the detection rate  based upon a model fit to the   SFR data
corrected for  dust  extinction.  The  dashed line is based upon a fit to the  SFR data without the dust correction.
Shaded regions indicate the neutrino backgrounds.
\label{fig:7}}
\end{center}
\end{figure}

\clearpage

\begin{figure}[h]
\begin{center}
\includegraphics[angle=0,width=3.5in]{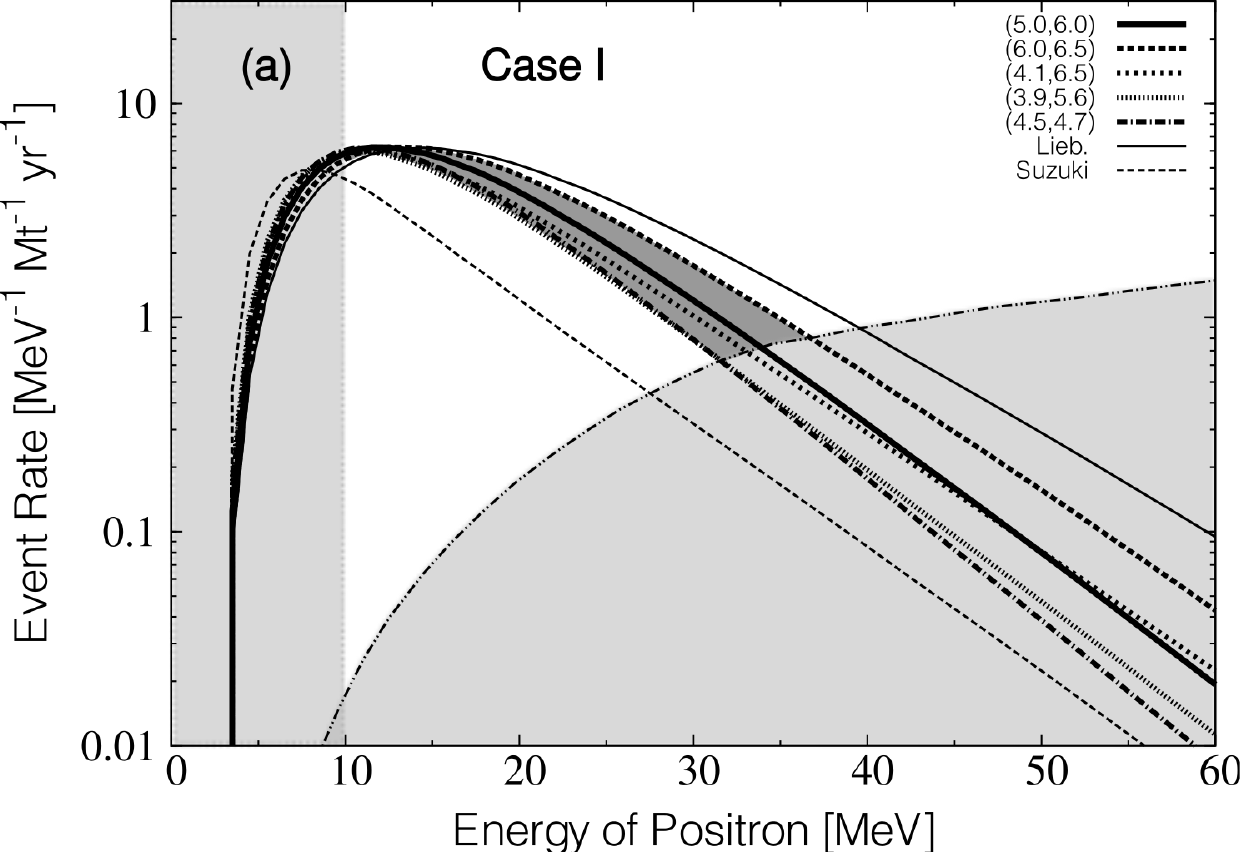}\\
\includegraphics[angle=0,width=3.5in]{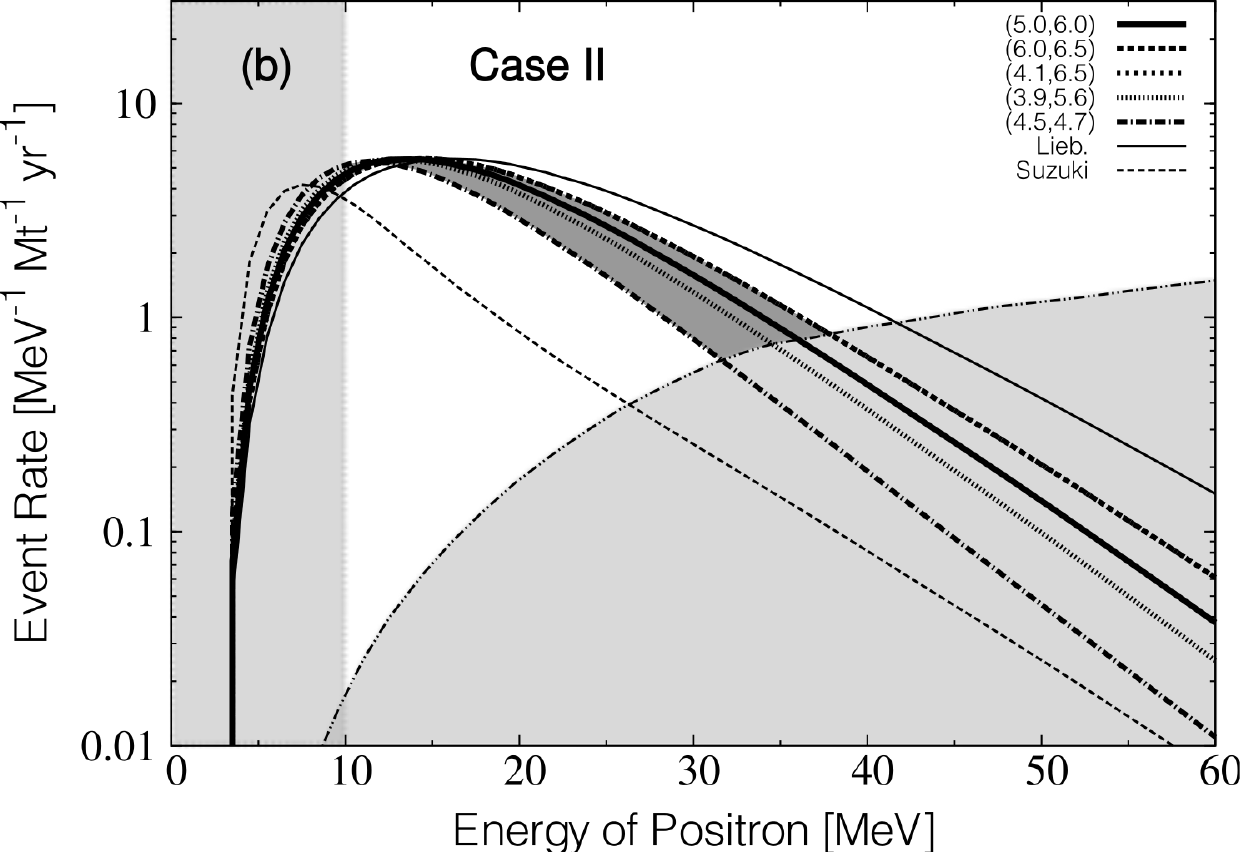}\\
\includegraphics[angle=0,width=3.5in]{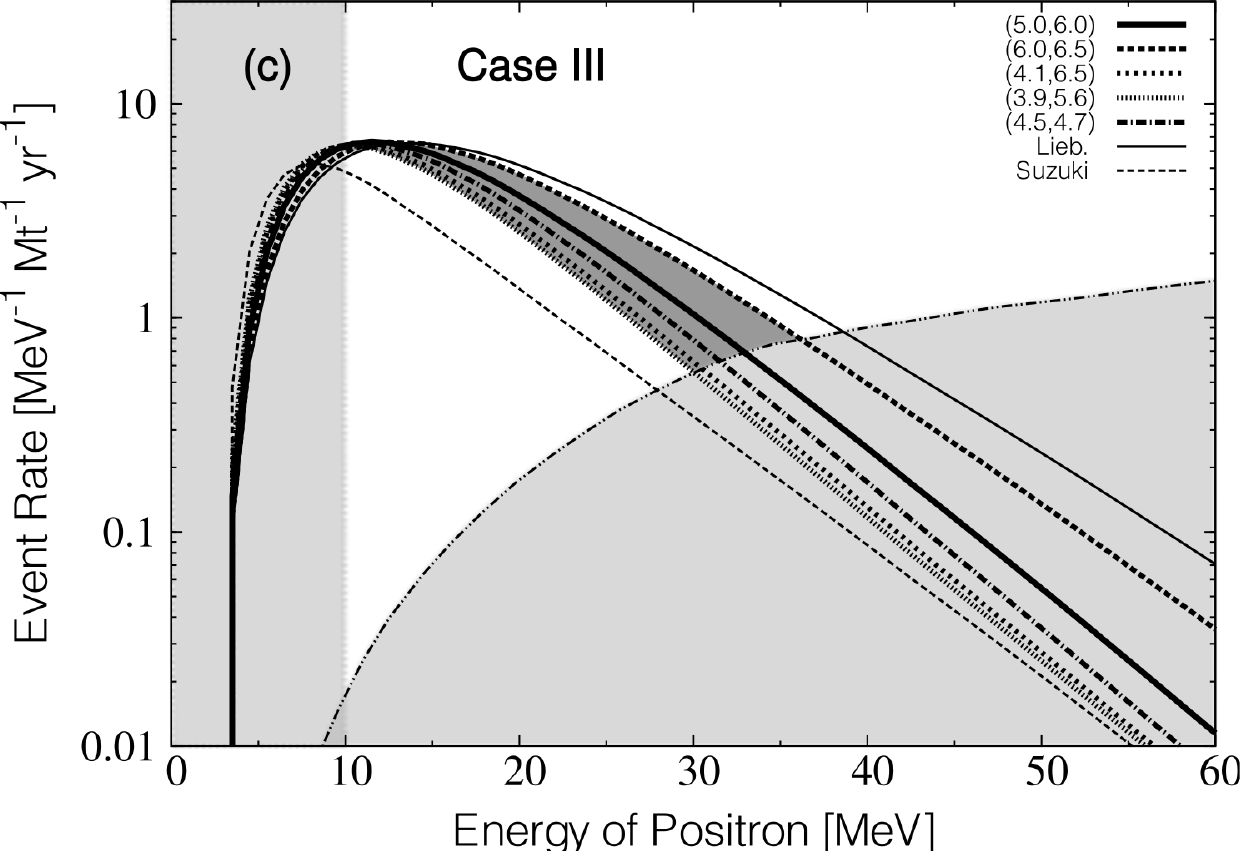}
\caption{ $T_\nu$ dependence of predicted SRN
 detection rate from normal core collapse supernovae as a function of e$^+$ energy for a  10$^6$ ton class water \v{C}erenkov detector
with 1 year of run time.  The three panels are for the three neutrino oscillation cases as labeled.  The shaded  energy range below10  MeV indicates the region where  the background
noise due  reactor $\bar{\nu_e}$ may dominate. The shaded energy range that intersects the spectrum at   $\sim 33
$ to $ 37$ MeV indicates the region where the  background may be dominated by noise from  atmospheric neutrinos.
The solid, short-dashed, long-dashed, long dash-dotted, short dash-dotted, thin line, and dotted lines as labeled  represent the energy
spectra for SRN detection rates assuming that (T$_{\bar{\nu_e}}$,
T$_{\nu_x}$) = (6.7MeV, 7.6MeV), (6.0 MeV, 6.5MeV), (5.0 MeV, 6.0 MeV),
(4.1MeV, 6.5MeV), (3.9MeV, 5.6MeV), (4.5MeV, 4.7MeV), (2.5MeV, 2.5MeV), respectively. 
Lines labeled as Lieb.~(6.7MeV, 7.6MeV) and Suzuki (2.5MeV, 2.5MeV) are  from \citet{lie01}
and  \citet{suz93}, respectively.
 The dark shaded region  in each figure shows the
uncertainty in the  detected SRN $e^+$ energy spectra due to the neutrino temperature uncertainty adopted  in this study. 
\label{fig:8}}
\end{center}
\end{figure}

\clearpage
\begin{figure}[h]
\begin{center}
\includegraphics[width=4.5in]{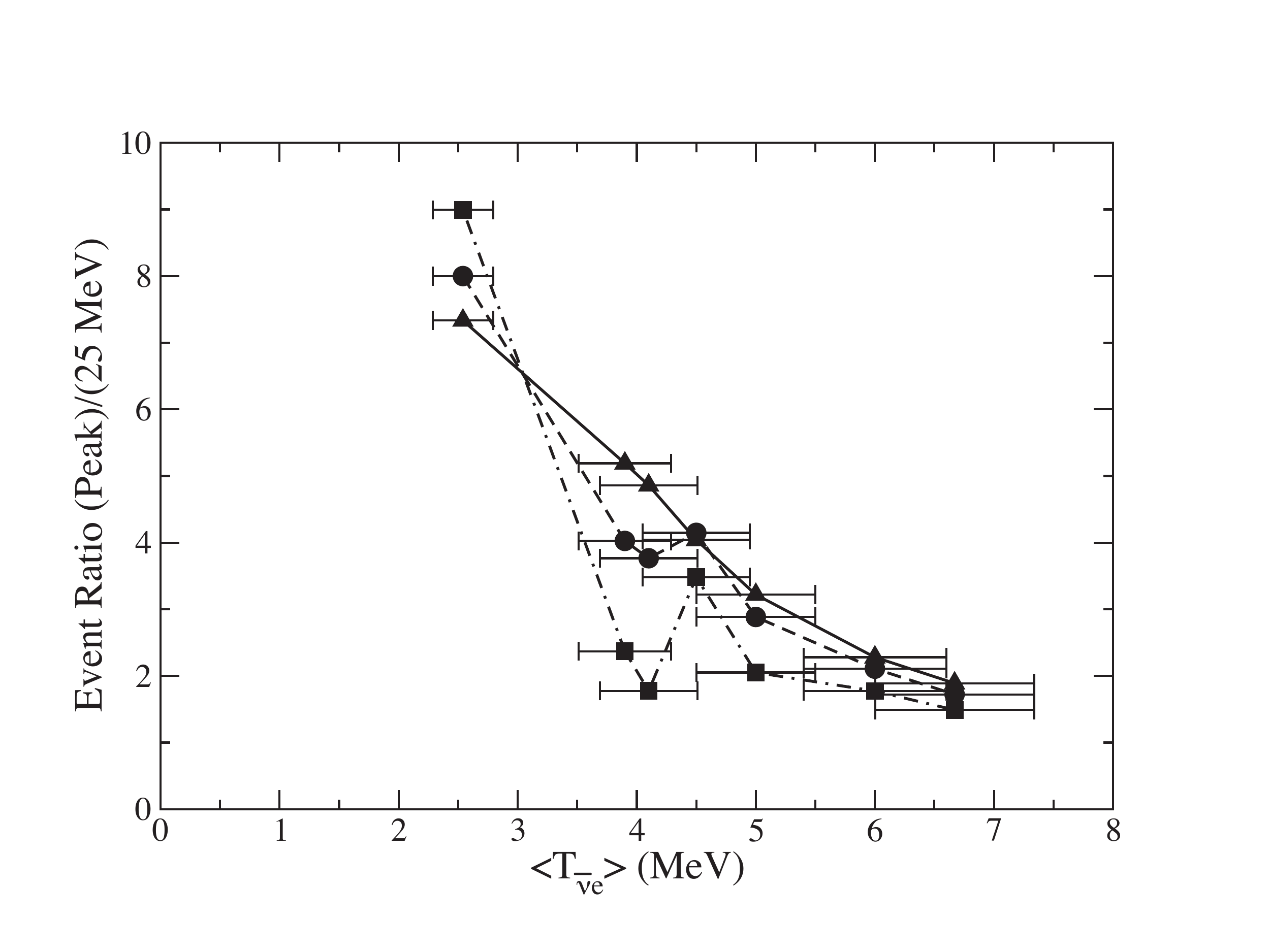}
\caption{Sensitivity or the ratio of events at the observed positron peak peak to events with a positron energy of 25 MeV, corresponding to the 7 fiducial SN collapse models considered here.  Circles and dashed line are from Fig.~\ref{fig:8}a (oscillation {\it Case I}).  Squares and dot-dashed line are from Fig.~\ref{fig:8}a (oscillation {\it Case II}).  Triangles and solid line are from Fig.~\ref{fig:8} (no oscillations {\it Case III}). This illustrates how
the observed positron spectrum might be used to infer the supernova neutrino temperature.  Error bars drawn indicate the adopted 10\% uncertainty in model neutrino temperatures
due to the variation of neutrino temperature with time during the explosion.  The vertical scatter in the points indicates the uncertainty due to oscillation parameters.
\label{fig:9}}
\end{center}
\end{figure}

\clearpage
\begin{figure}[h]
\begin{center}
\includegraphics[width=4.5in]{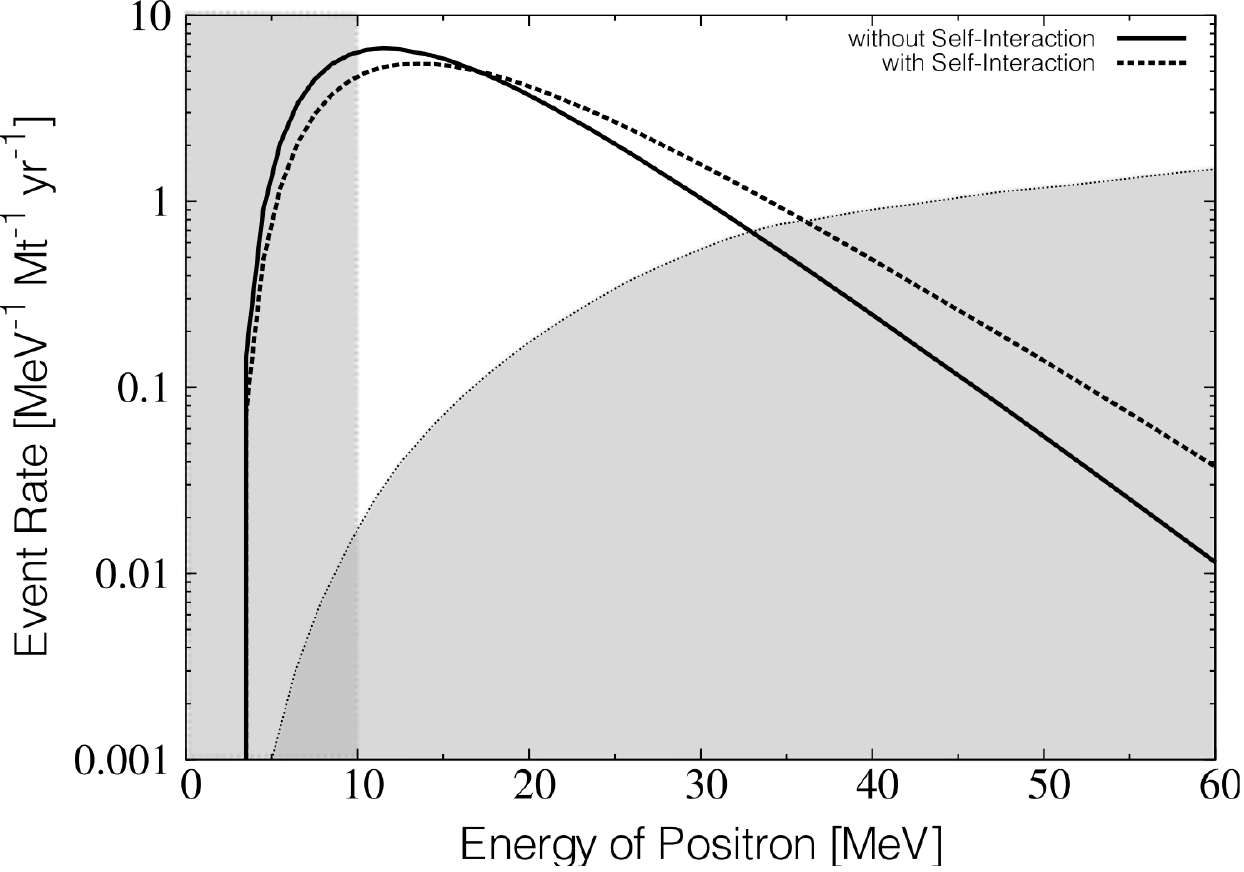}
\caption{Detected SRN energy spectrum  for our fiducial model in  cases with (dashed line) or without (solid line)  of a single-angle neutrino self interaction   and no neutrino oscillations. Shaded regions are the backgrounds as defined in Fig.~\ref{fig:5}.
\label{fig:10}}
\end{center}
\end{figure}

\clearpage
\begin{figure}[h]
\begin{center}
\includegraphics[angle=0,width=3.5in]{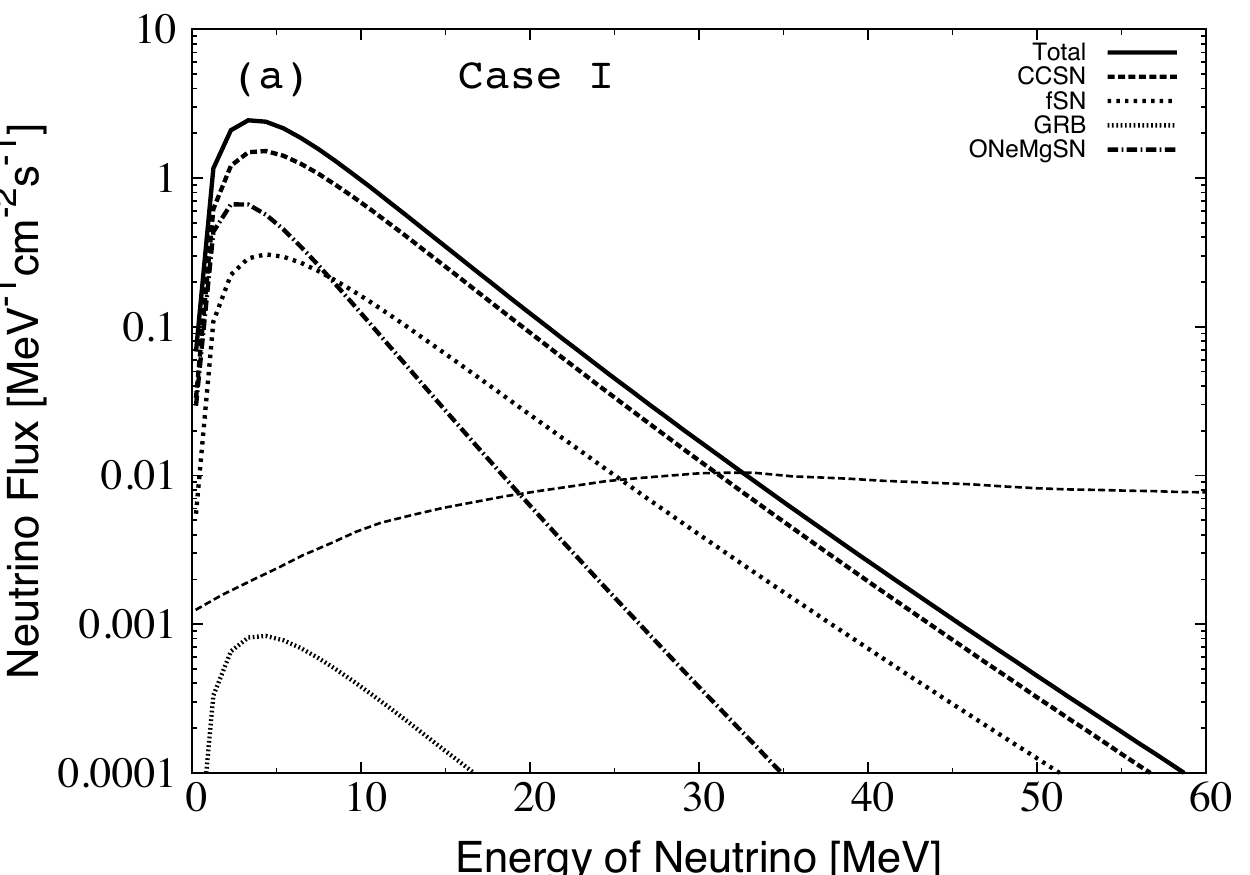}\\
\includegraphics[angle=0,width=3.5in]{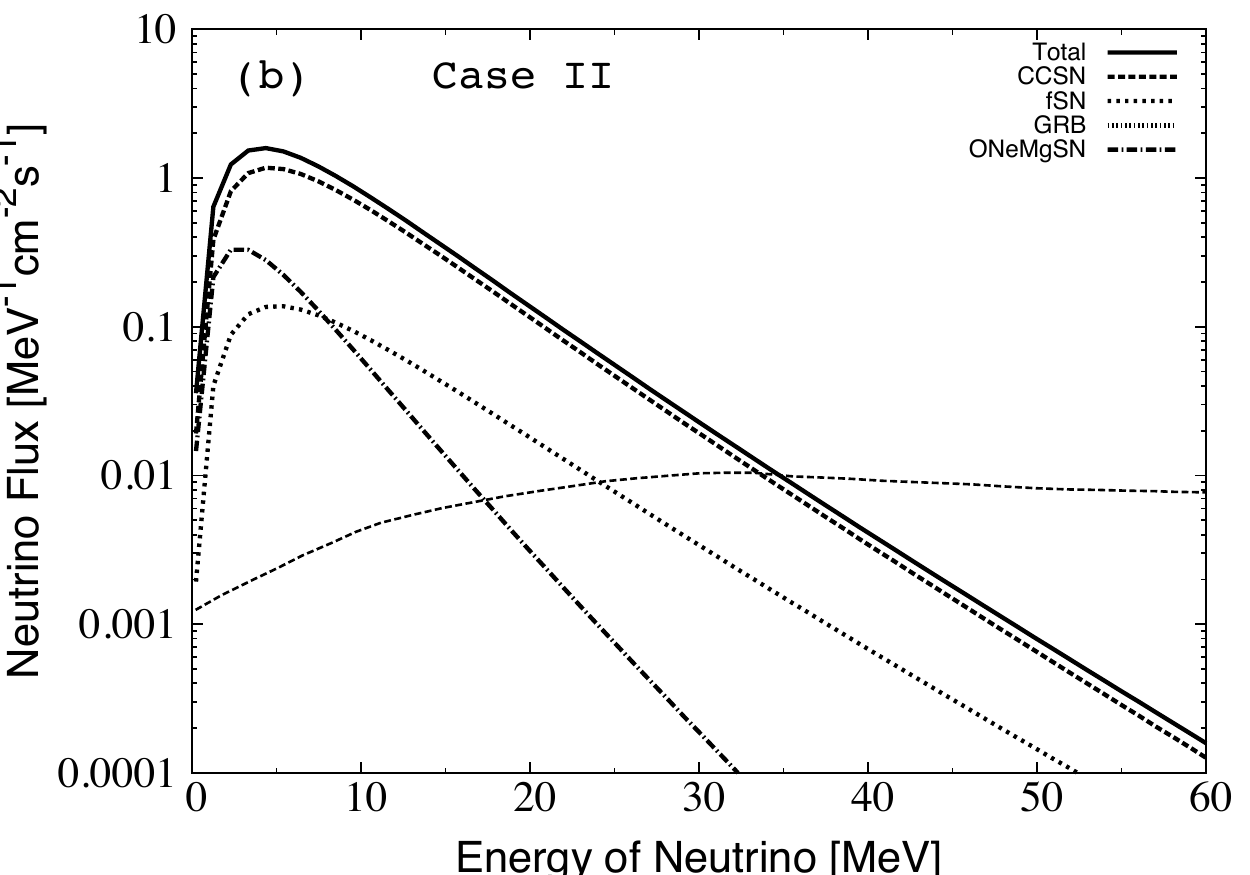}\\
\includegraphics[angle=0,width=3.5in]{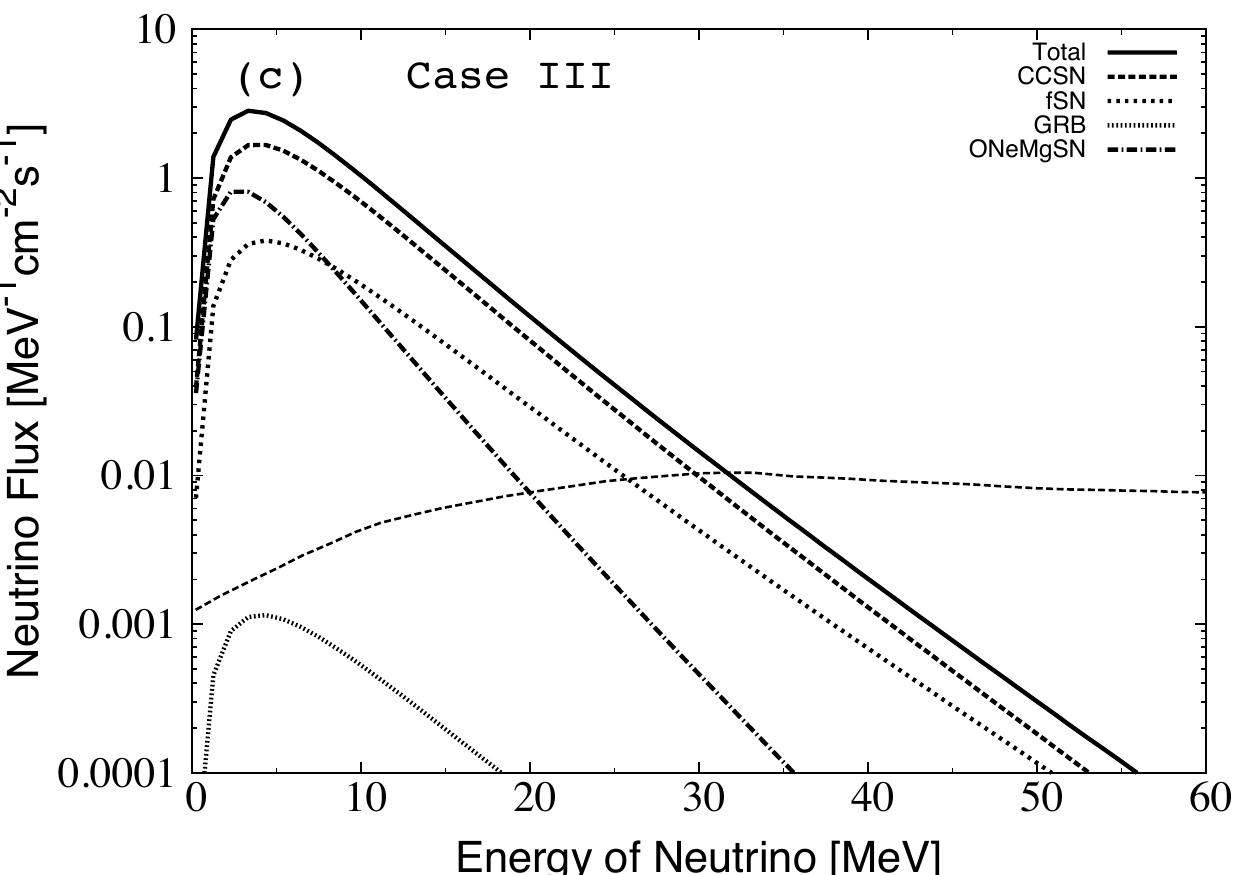}
\caption{Relative contributions to the total to the total
 predicted  SRN flux spectrum  (Solid line) from CC-SNe (short dashed line), ONeMg SNe (dotted line), fSNe (long dashed line)  and GRBs (dot-dashed line) for our fiducial model and three oscillation cases.  The background due to atmospheric neutrinos is also indicated as a short dashed line.
\label{fig:11}}
\end{center}
\end{figure}

\clearpage
\begin{figure}[h]
\begin{center}
\includegraphics[angle=0,width=3.5in]{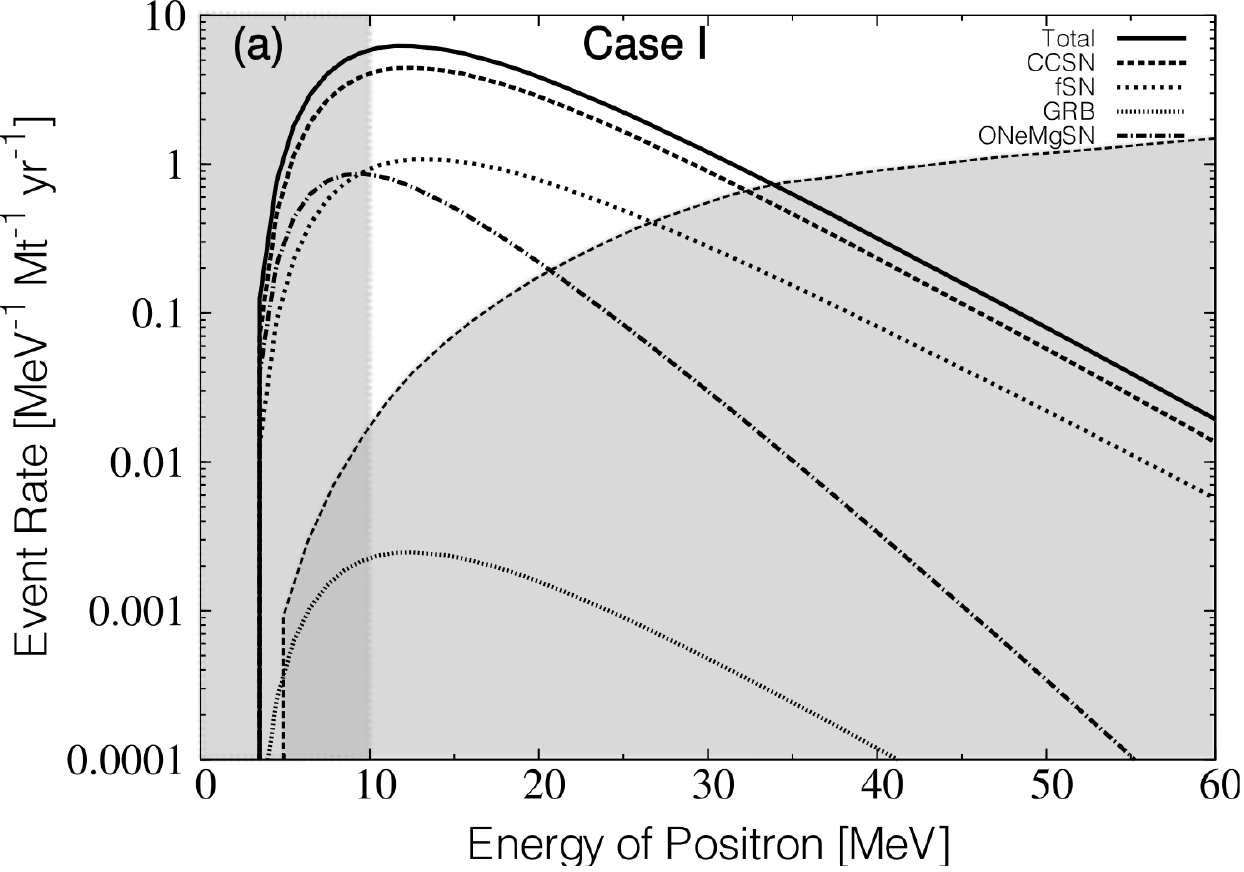}\\
\includegraphics[angle=0,width=3.5in]{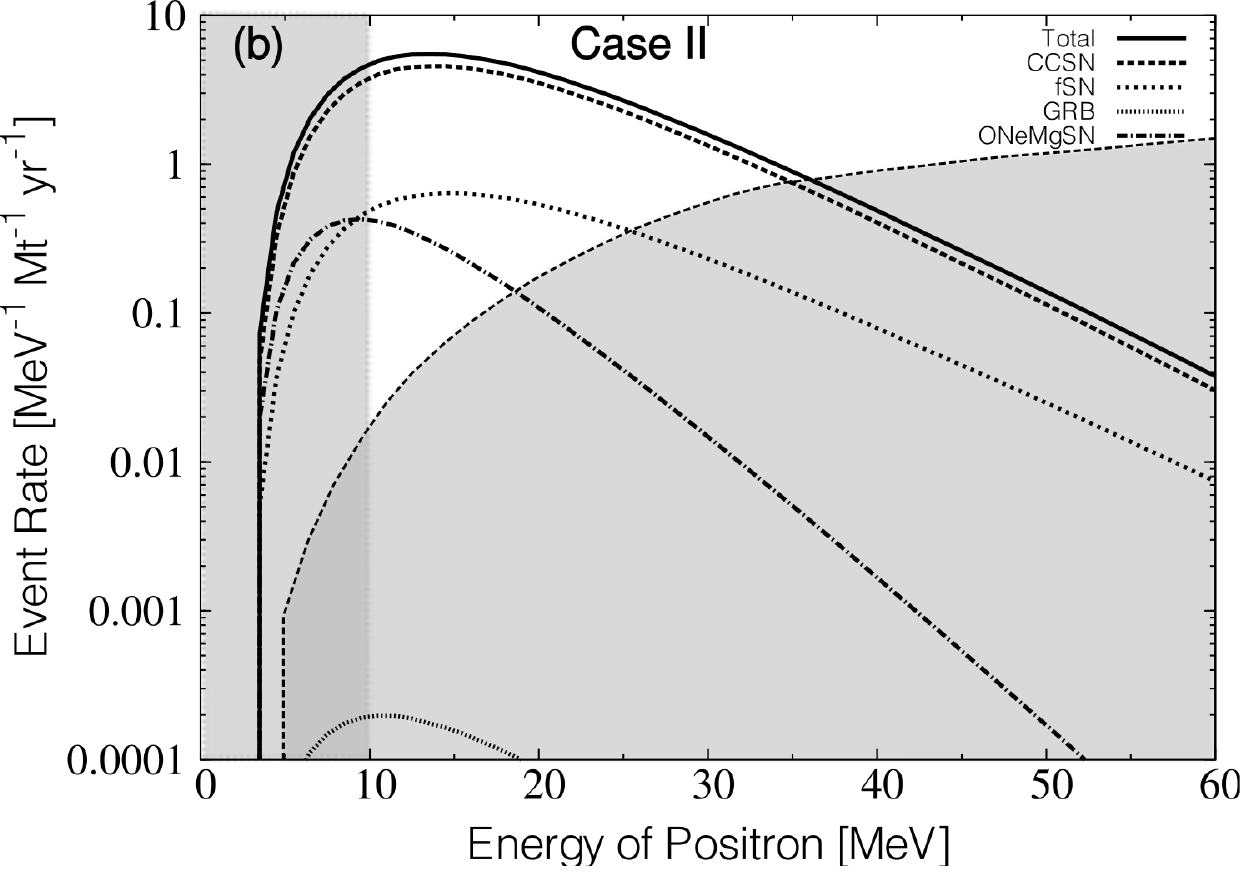}\\
\includegraphics[angle=0,width=3.5in]{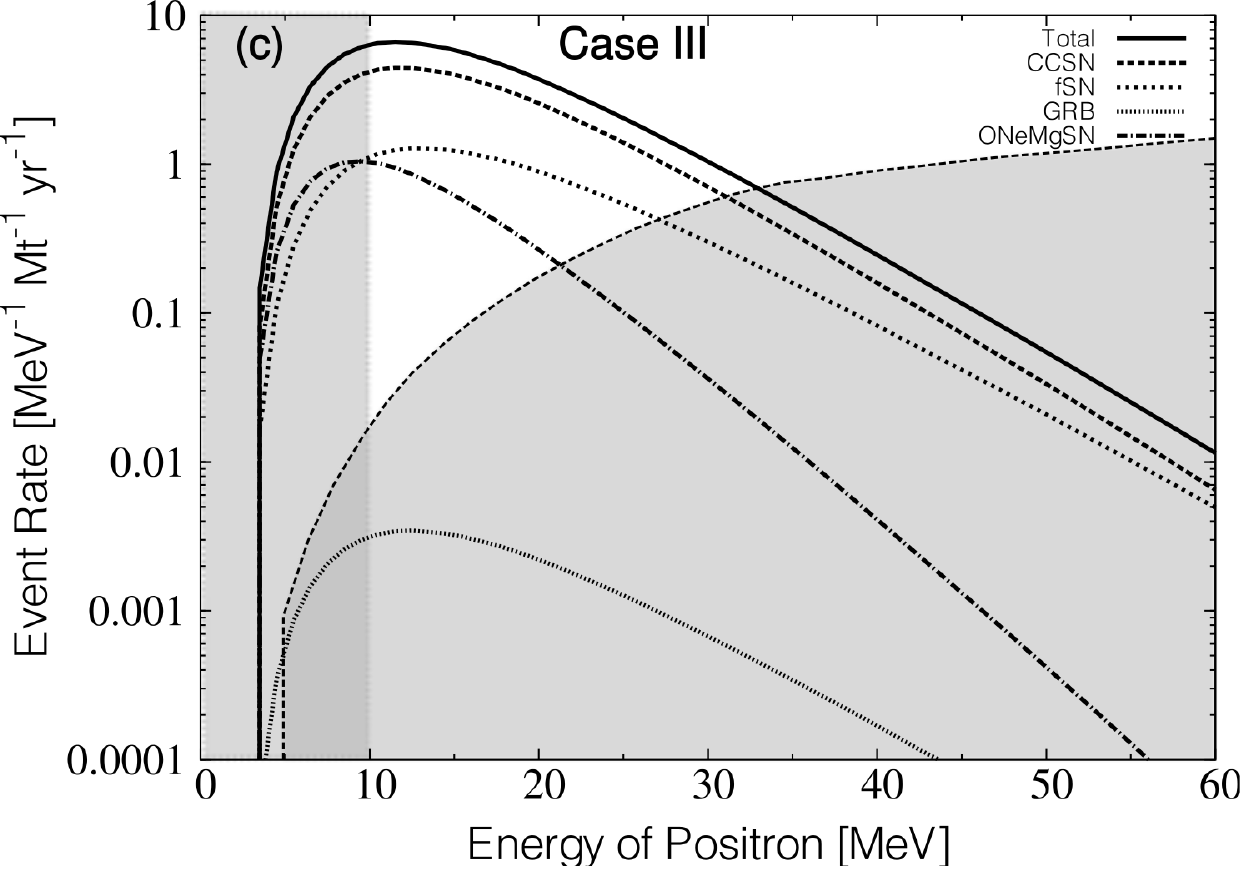}
\caption{Relative contributions to the total
 SRN detection rate (solid line) from rom CC-SNe (short dashed line), ONeMg SNe (dotted line), fSNe (long dashed line)  and GRBs (dot-dashed line) for our fiducial model and three oscillation cases.   Shaded regions are backgrounds as defined in Figure \ref{fig:5}.
\label{fig:12}}
\end{center}
\end{figure}

\clearpage
\begin{figure}[h]
\begin{center}
\includegraphics[angle=0,width=3.5in]{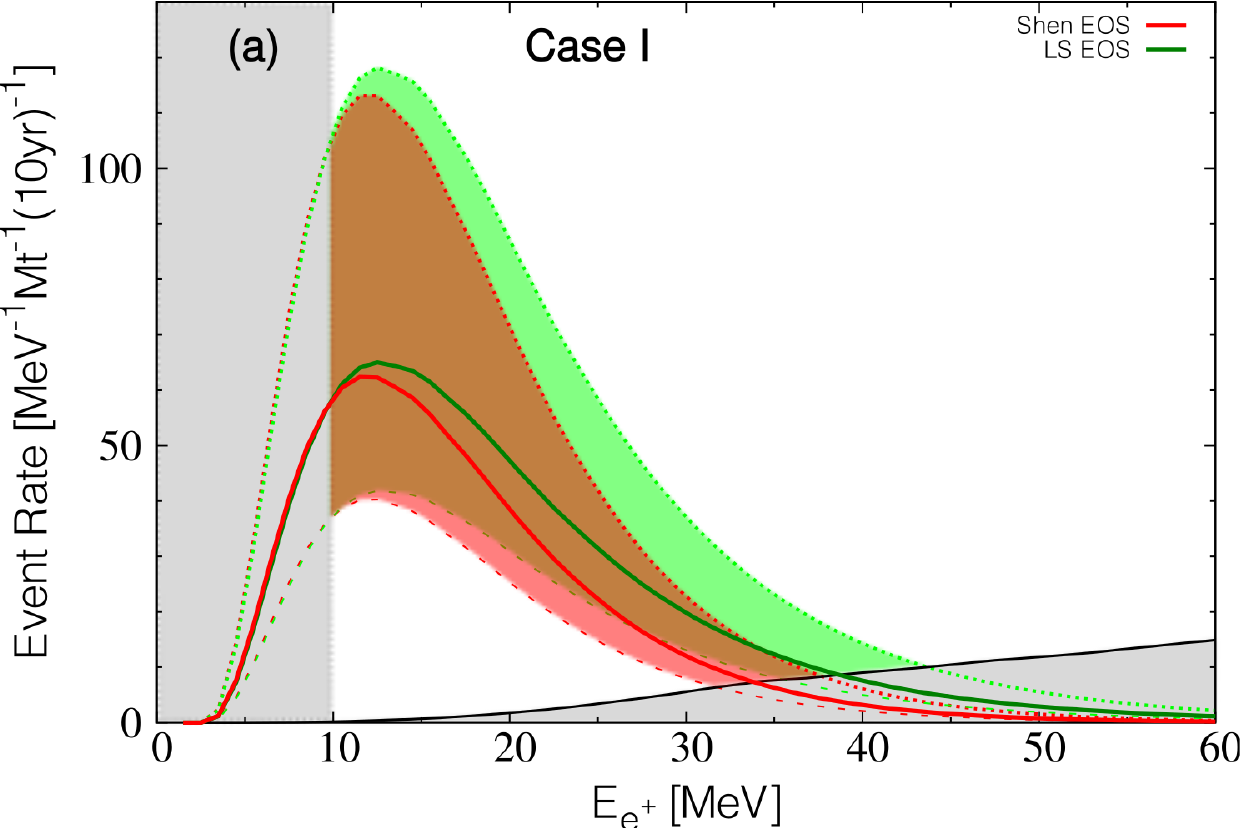}\\
\includegraphics[angle=0,width=3.5in]{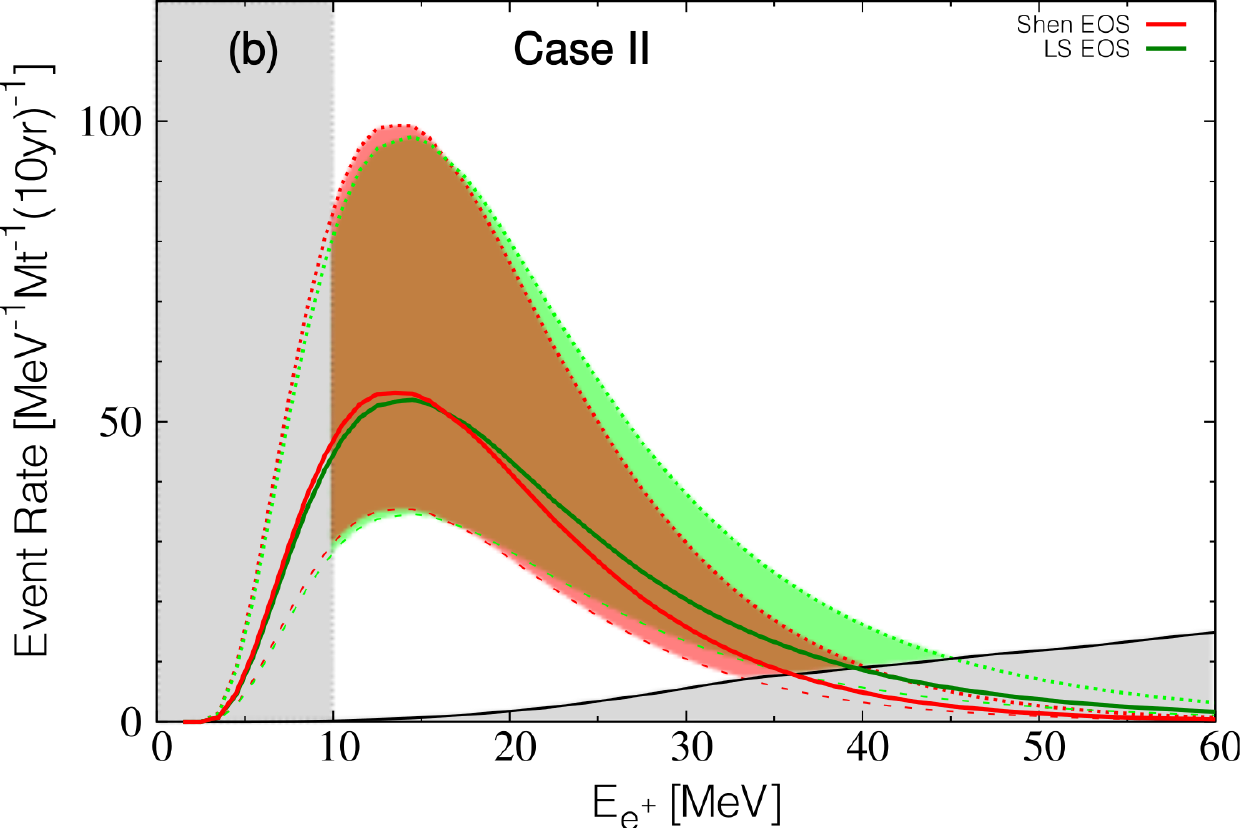}\\
\includegraphics[angle=0,width=3.5in]{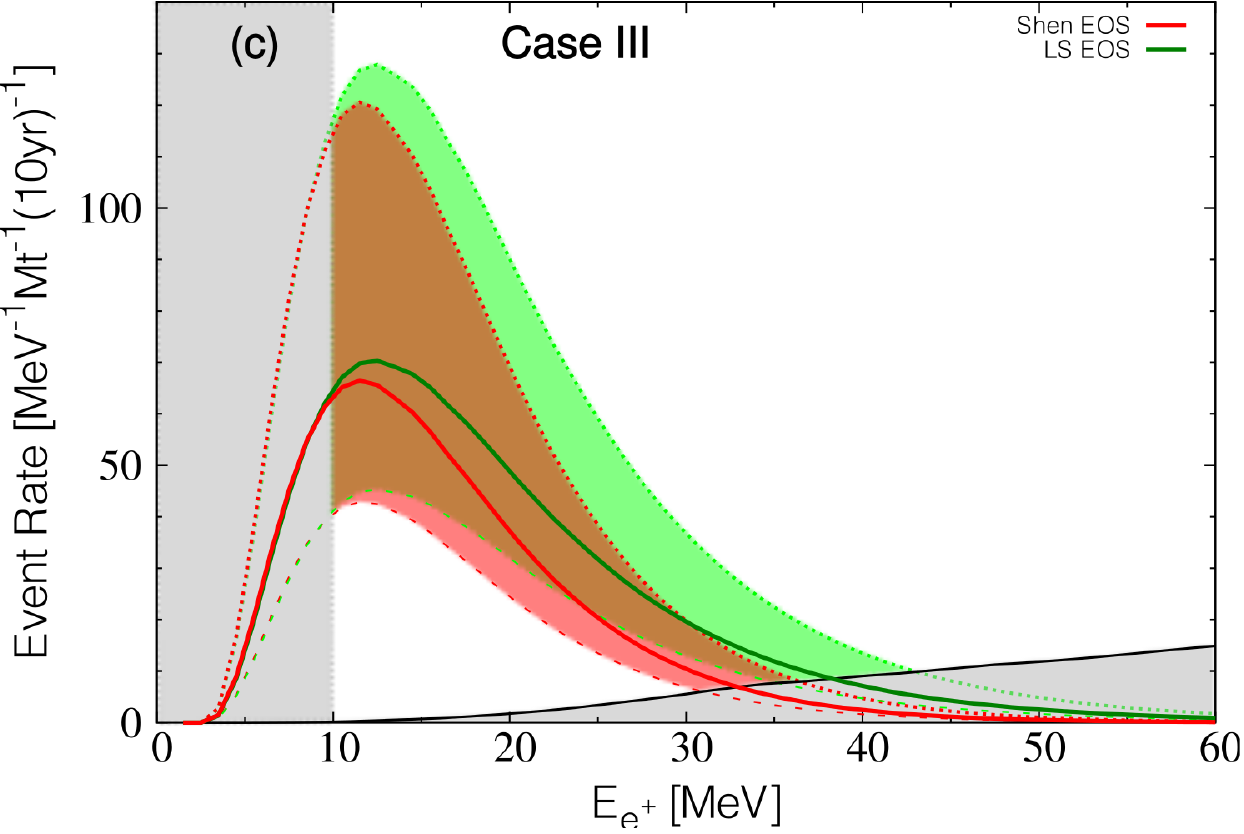}
\caption{Predicted $e^+$ energy spectra and uncertainty of the total SRN detections for the fiducial  model and massive SFR deduced here.  Uncertainty is
based upon the SFR and the detector statistics. 
Each panel corresponds to the case of: (a) non-adiabatic
oscillations (Case {\it I}); (b)  adiabatic oscillations (Case {\it II}); and (c)  no oscillations (Case {\it III}).
Red,  line and shaded region is for fSNe models based upon the stiff RMF EoS of \cite{she98}.  The green line and shaded region is for fSNe based upon the soft
EoS of \cite{lat91}.  Grey regions denote the backgrounds as defined in Figure \ref{fig:5}. 
\label{fig:13}}
\end{center}
\end{figure}

\clearpage
\begin{figure}[h]
\begin{center}
\includegraphics[angle=0,width=3.5in]{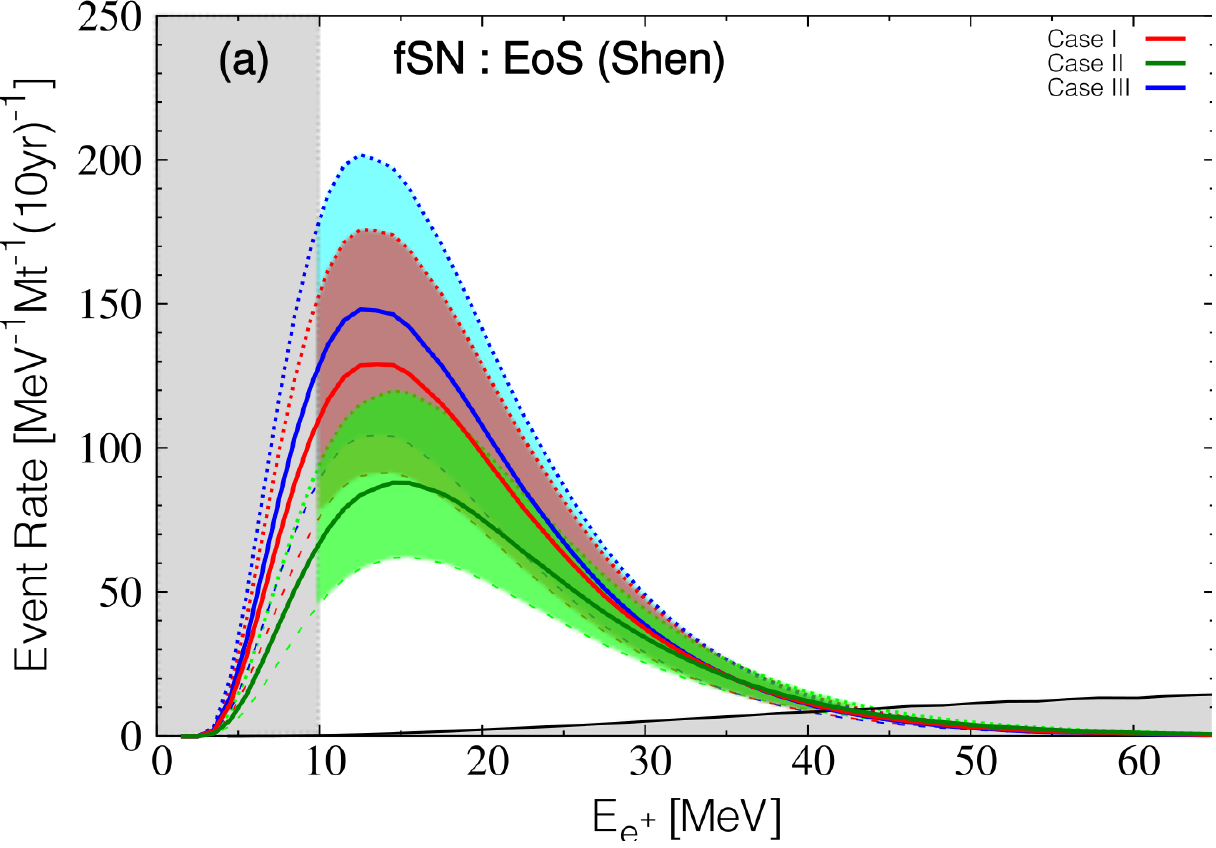}\\
\includegraphics[angle=0,width=3.5in]{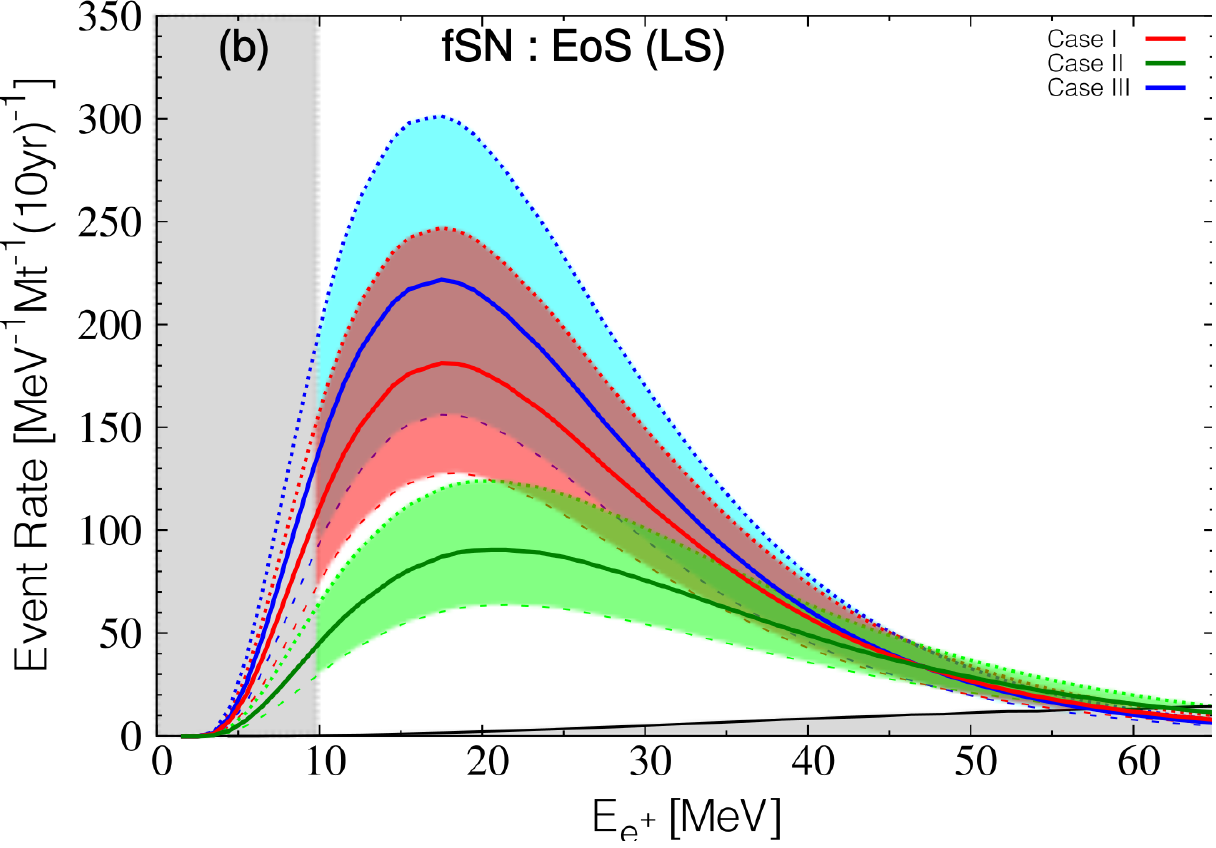}\\
\caption{Predicted $e^+$ energy spectra and uncertainty of the SRN detection rate as a function of $e^+$ energy for a $10^6$ ton water \v{C}erenkov detector with 10 years of run time.  Shown are cases in which failed  fSNe  account for a factor of two difference between supernova rate deduced from the massive SFR and the observed supernova rate.
Red, green, and blue lines shows the SRN detection rate in the case of non-adiabatic (Case {\it I}), adiabatic (Case {\it II}), or no neutrino  oscillations (Case {\it III}), respectively. 
Light red, light green, and light blue areas show the
uncertainty in  the SRN detection rate in each oscillation
case due to the SFR of \cite{hor11} and the detector statistics. The vertical light grey areas denote the  backgrounds. 
Each figure presents: (a) the case that all the missing SNe are fSN modeled with the stiff EoS of
\citet{she98}; (b) the case that all the missing SNe are fSN modeled with the soft EoS of
\citet{lat91}.
\label{fig:14}}
\end{center}
\end{figure}

\clearpage
\begin{figure}[h]
\begin{center}
\includegraphics[angle=0,width=3.5in]{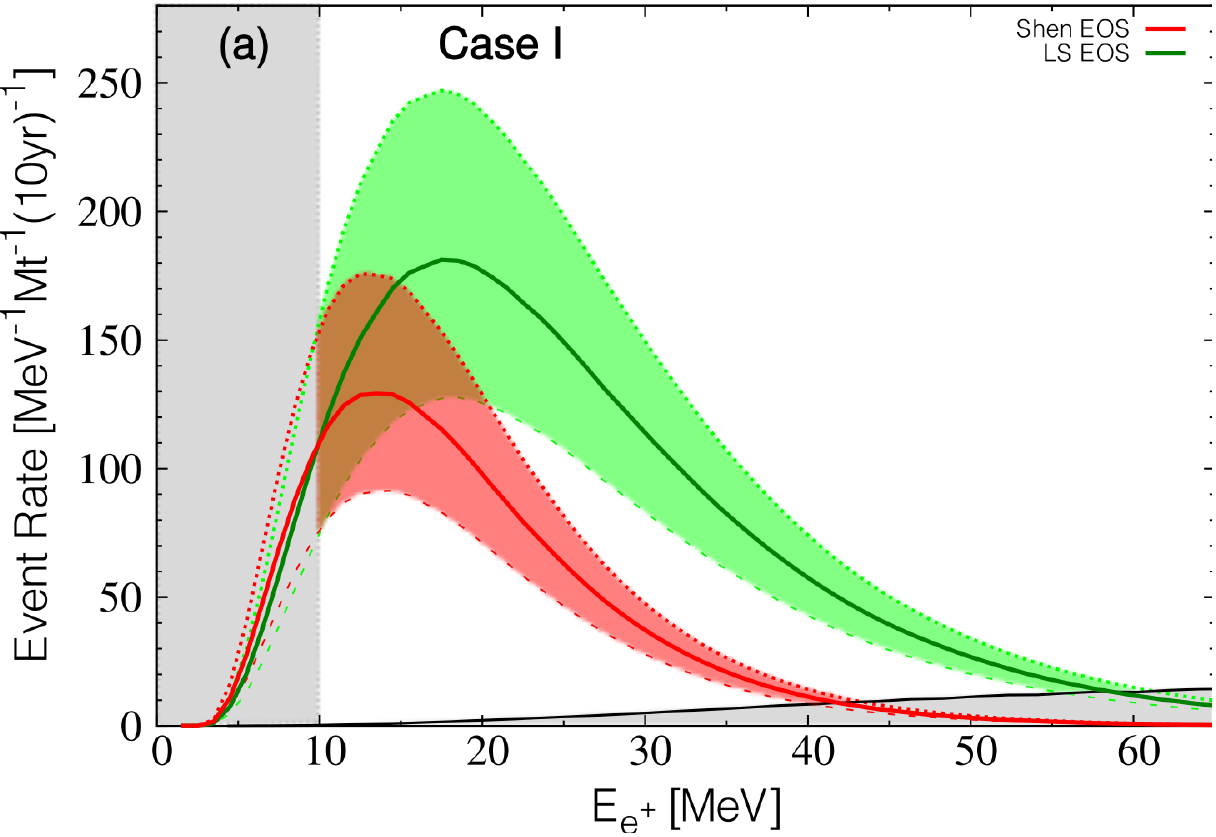}\\
\includegraphics[angle=0,width=3.5in]{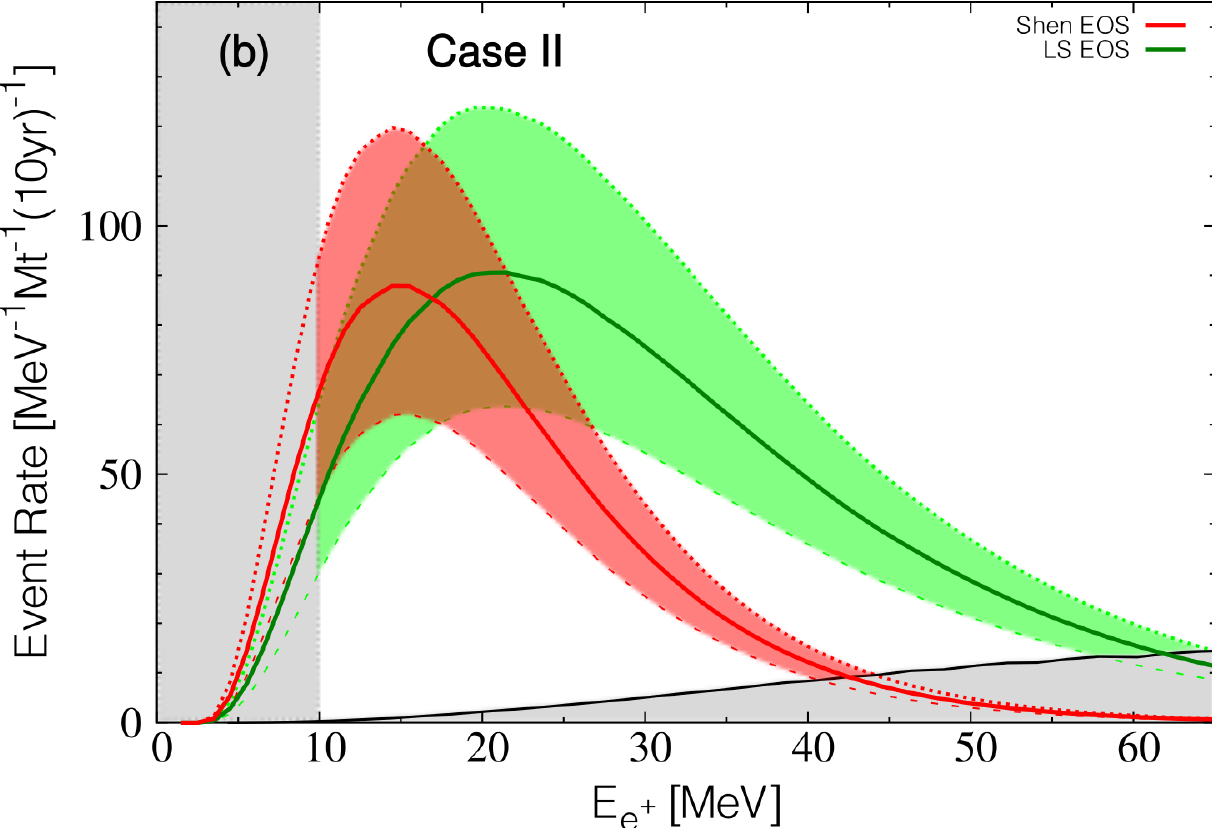}\\
\includegraphics[angle=0,width=3.5in]{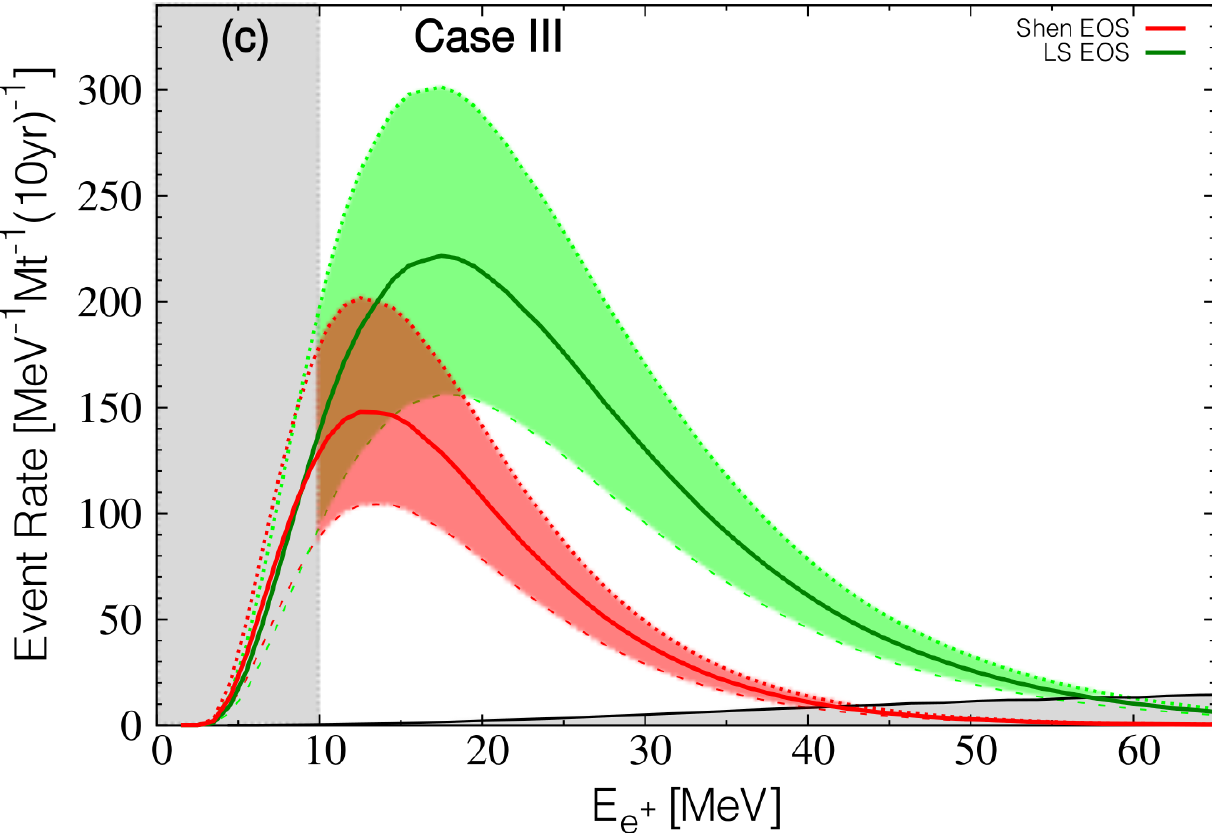}
\caption{Same as Fig.\ref{fig:12} but in this case  the three different cases of neutrino oscillations shown in separate panels.
Red, green, and blue lines show the SRN detection rate in the case that  all of the
missing SNe are fSN with the stiff RMF EoS of \citet{she98},  and the case that all the
missing SNe are fSN modeled with the soft EoS of \citet{lat91}.
 Light red, light green, and light blue areas show the
uncertainty in the  SRN detection rate based upon the SFR of \cite{hor11} and the detector statistics.
Each figure represents the case of: (a) non-adiabatic
oscillations (Case {\it I}); (b)  adiabatic oscillations (Case {\it II}); and (c)  no oscillations (Case {\it III}).
\label{fig:15}}
\end{center}
\end{figure}

\clearpage

\begin{table}[h]
\begin{center}
\caption{Coefficients for the piecewise linear  star formation rate\label{tbl.1}}
\begin{tabular}{ccccccc}
\tableline\tableline
 & $\dot{\rho}_0$ &  $\alpha$ & $\beta$ & $\gamma$ &  $B$ & $C$\\
 & (M$_\odot $Mpc$^{-3} y^{-1}$) &\\
% $z_1$ & $z_2$ & $\eta$ & B & C\\
%\multicolumn{1}{c}{$P$\tablenotemark{a}} & $P R_{maj}$ & $P R_{min}$ &
%\multicolumn{1}{c}{$\Theta$\tablenotemark{b}} \\
\tableline
Best Fit & 0.0104 &  4.22 &  -0.20 & -11.30 & 2.70$\times10^6$ & 6.37 \\
Upper limit & 0.0200 &  4.22 &  0.14 & -9.0 &  1.25$\times10^{-8}$ & 6.87 \\
Lower limit & 0.0069 &  4.22 &  -0.42 & -13.60 &  1.6$\times10^3$ & 5.90 \\
\tableline
\end{tabular}
%% Any table notes must follow the \end{tabular} command.
\tablenotetext{1}{Value of $\alpha = 4.22$ fixed at best-fit value.  Smoothing parameter  $\eta = -10$ adopted from \cite{hop06,yuk08}.  Breaks in 
the spectrum occur at $z_1 = B^{1/(1 - \alpha/\beta)} \approx 1$ and $z_2 =  [C^{1/(1-\beta/\gamma)} (1 + z_1)^{(\alpha-\beta)/(\gamma-\beta)}] \approx 4$
\citep{yuk08}.}
\end{center}
\end{table}

\clearpage

\begin{table}[h]
\begin{center}
\caption{Results of numerical simulations in previous work.\label{tbl.2-1}}
\begin{tabular}{lcccccccc}
%\rotate
%\tablewidth{0pt}
\tableline\tableline
{model }    &
 {time(s)}        &  {T$_{\nu_e}$}  &
 {T$_{\bar{\nu_e}}$} &  {T$_{\nu_x}$}  &
 {$\eta_{\nu_e}$}  &  {$\eta_{\bar{\nu_e}}$}  &
 {$\eta_{\nu_x}$} &  {$<e>$ or $\surd<e^2>$ } \\
 { }    &  { } &  {(MeV)}  &
  {(MeV)}  &   {(MeV)}  & 
  { }  &  {}  & {}   &  {} \\
\tableline
%\startdata
\citet{may87} & - & 3.80 & 7.60 & 6.98 & 0.33 & 0.3 & 0.33 & - \\
\citet{bru87}\tablenotemark{a} & - & 3.17 & 3.80 & 7.93 & - & - & - & - \\
\citet{jan89}\tablenotemark{a} & - & 2.54 & 4.44 & 5.08 & $\sim$0 & $\sim$0 &
$\sim$0 & $<e>$ \\
\citet{myr90}\tablenotemark{a} & time table & 3.49 & 4.12 & 7.60 & - & - & - &
$<e>$ \\
\citet{suz90}& 1.0 & 3.02 & 4.13 & 4.76 & - & - & - & - \\
\citet{suz90} &20.0 & 2.54 & 3.17 & 2.85 & - & - & - & - \\
\citet{suz91} & 1.0 & 2.63 & 4.13 & 4.76 & - & - & - & - \\
\citet{suz91} &15.0 & 2.54 & 2.85 & 3.02 & - & - & - & - \\
\citet{suz93} & 1.0 & 2.85 & 3.80 & 4.13 & - & - & - & - \\
\citet{suz93} &15.0 & 2.22 & 2.54 & 2.54 & - & - & - & - \\
\citet{tot98} & 0.3 & 3.80 & 4.76 & 6.03 & - & - & - & - \\
\citet{tot98}\tablenotemark{a} & 10.0 & 3.40 & 6.35 & 7.93 & - & - & - & - \\
\citet{lie01}\tablenotemark{a} & - & 6.03 & 6.66 & 7.60 & - & - & - &
$\surd<e^2>$ \\
\citet{mez01}\tablenotemark{a} & - & 5.08 & 6.03 & 7.60 & - & - & - & - \\
\citet{and03} & mean & 3.49 & 5.08 & 6.98 & - & - & - & $<e>$ \\
\citet{and03}\tablenotemark{a} & fulltime & 3.49 & 4.44 & 5.71 & - & - & - &
$<e>$ \\
\citet{bur03}\tablenotemark{b} & 0 & 4.48 & 5.24 & 5.33 & - & - & - &
$\surd<e^2>$ \\
\citet{bur03}\tablenotemark{a} & 0.15 & 4.13 & 5.08 & 7.30 & - & - & - &
  $\surd<e^2>$ \\
\citet{bea04} & - & 3.20 & 5.00 & 6.00 & - & - & - & - \\
\citet{gil03} 1& - & 3.49 & 5.08 & 7.94 & - & - & - & - \\
\citet{gil03} 2& - & 4.13 & 5.08 & 7.30 & - & - & - & - \\
\citet{gil03} 3& - & 4.13 & 5.08 & 5.59 & - & - & - & - \\ 
\citet{gil03} 4& - & 4.13 & 5.08 & 5.59 & - & - & - & - \\ 
\citet{gil03} 5& - & 4.13 & 5.08 & 5.59 & - & - & - & - \\
\tableline
\end{tabular}
\tablenotetext{1}{Utilized to  derive the $ \pm 1\sigma$ error ellipse.}
\tablenotetext{2}{based upon \citet{kei03b}}
\end{center}
\end{table}
\clearpage

\begin{table}[h]
\begin{center}
\caption{Results of numerical simulations in previous works(2).\label{tbl.2-2}}
\begin{tabular}{lcccccccc}
%\rotate
%\tablewidth{0pt}
\tableline\tableline
 {model}    &
 {time(s)}        &  {T$_{\nu_e}$}  &
 {T$_{\bar{\nu_e}}$} &  {T$_{\nu_x}$}  &
 {$\eta_{\nu_e}$}  &  {$\eta_{\bar{\nu_e}}$}  &
 {$\eta_{\nu_x}$} &  {$<e>$ or $\surd<e^2>$} \\
 { }    &  { } &  {(MeV)}  &
  {(MeV)}  &   {(MeV)}  & 
  { }  &  {}  & {}   &  {} \\
\tableline
Accretion Phase model 1\tablenotemark{a} & - & 4.13 & 4.76 & 5.71 & 2.8 & 3.4
& 1.1 & $<e>$ \\
-\citet{kei03b}\tablenotemark{a} & - & 3.80 & 4.44 & 4.44 & 1.4 & 2.7 &
0.6-1.6 & - \\
Accretion Phase model 2\tablenotemark{a} & - & 4.13 & 5.07 & 5.40 & 1.7 & 3.0
& 0.8 & - \\
-\citet{kei03b}\tablenotemark{a} & - & 4.13 & 4.76 & 5.07 & 2.1 & 3.2 & 0.8 &
- \\
\citet{lun03} & - & 2.22-5.58 & 4.44-6.98 & 4.88-11.17 &
0.0-3.0 & 0.0-3.0 & 0.0-2.0 & - \\
\citet{ros03} C& - & 4.20 & 5.50 & 8.90 & - & - & - & - \\
\citet{ros03} D& - & 4.10 & 5.00 & 6.60 & - & - & - & - \\
\citet{ros03} E& - & 5.00 & 5.70 & 9.10 & - & - & - & - \\
\citet{sel03} & - & 4.00 & 5.00 & 7.50 & - & - & - & - \\ 
\citet{vog03}& - & 3.50 & 5.10 & 8.00 & - & - & - & - \\
\citet{dig04} G& - & 3.81 & 4.79 & 5.71 & - & - & - & - \\
\citet{dig04} L& - & 3.81 & 4.79 & 7.62 & - & - & - & - \\
\citet{gil04} & - & 2.22-5.71 & 4.44-6.98 &
4.76-11.11 & 0.0 & 0.0 & 0.0 & - \\ 
\citet{tom04} L & - & 3.81 & 4.76 & 7.62 & - & - & - & - \\
\citet{tom04} G1 & - & 3.81 & 4.76 & 5.71 & - & - & - & - \\
\citet{tom04} G2 & - & 3.81 & 4.76 & 4.76 & - & - & - & - \\
\citet{yos04} & - & 3.20 & 5.00 & 6.00 & - & - & - & - \\
\citet{bal05} & - & 4.13 & 4.76 & 5.08 & 0.0 & 0.0 & 0.0 & - \\
\citet{bar05} & - & 2.22-6.29 & 4.44-6.98 & 4.44-11.17 & -
& - & - & - \\
\citet{ath05} & - & 3.17-3.49 & 4.76-5.08 & 7.30-7.94 & -
& - & - & - \\
\citet{sum05} SH& 1.0 & 7.00 & 7.80 & 7.10 & - & - & - & $\surd<e^2>$ \\
\citet{sum05} LS& 1.0 & 8.30 & 8.70 & 8.40 & - & - & - & $\surd<e^2>$ \\
\citet{bea06} & - & 4.00 & 5.00 & 8.00 & 0.0 & 0.0 & 0.0 & - \\
\citet{bek06} & - & 3.50 & 5.00 & 8.00 & 0.0 & 0.0 & 0.0 & - \\
\tableline
\end{tabular}
\tablenotetext{1}{Utilized  to derive the $ \pm 1\sigma$ error ellipse.}
\tablenotetext{2}{based upon \citet{kei03b}}
\end{center}
\end{table}

\clearpage
%\twocolumn

\begin{table}[h]
\begin{center}
\caption{Results of numerical simulations in previous works(3).\label{tbl.2-3}}
\begin{tabular}{lcccccccc}
%\rotate
%\tablewidth{0pt}
\tableline\tableline
 {model}    &
 {time(s)}        &  {T$_{\nu_e}$}  &
 {T$_{\bar{\nu_e}}$} &  {T$_{\nu_x}$}  &
 {$\eta_{\nu_e}$}  &  {$\eta_{\bar{\nu_e}}$}  &
 {$\eta_{\nu_x}$} &  {$<e>$ or $\surd<e^2>$} \\
 { }    &  { } &  {(MeV)}  &
  {(MeV)}  &   {(MeV)}  & 
  { }  &  {}  & {}   &  {} \\
\tableline
\citet{cho06} & - & 3.50 & 5.00 & 8.00 & 0.0-3.0 & 0.0-3.0 &
0.0-2.0 & - \\
\citet{oli06} & - & 4.22 & 4.86 & 6.35 & 0.0 & 0.0 & 0.0 & - \\
\citet{suz06} & - & 3.17 & 4.76 & 4.76-7.94 & 0.0 & 0.0 & 0.0 & - \\
\citet{yos06}a & - & 3.20  & 5.00 & 6.00 & - & - & - & - \\
\citet{yos06}b & - & 3.20  & 5.00 & 6.66 & - & - & - & - \\
\citet{aga07} & - & 3.17-3.81 & 3.49-5.40 & 4.76-7.62 &
0.0 & 0.0 & 0.0 & - \\
\citet{bak07}\tablenotemark{a} & - & 4.13 & 4.89 & 4.98 & - & - & - & - \\
\citet{cha08} LL& - & 3.81 & 4.76 & 7.62 & - & - & - & - \\
\citet{cha08} G1& - & 3.81 & 4.76 & 5.71 & - & - & - & - \\
\citet{cha08} G2& - & 3.81 & 4.76 & 4.76 & - & - & - & - \\
\citet{das08} & - & 3.17 & 4.76 & 6.35 & - & - & - & - \\
\citet{dig08} Liv& - & 3.81 & 4.76 & 7.62 & - & - & - & - \\
\citet{dig08} Gar& - & 3.81 & 4.76 & 5.81 & - & - & - & - \\
\citet{kne08} & - & 3.81 & 4.76 & 5.71 & - & - & - & - \\
\citet{yos08} 1& - & 3.20 & 5.00 & 6.00 & - & - & - & - \\
\citet{yos08} 2& - & 4.00 & 4.00 & 6.00 & - & - & - & - \\
\citet{yos08} LT& - & 3.20 & 5.00 & 6.50 & - & - & - & - \\
\citet{yos08} ST& - & 3.20 & 4.20 & 5.00 & - & - & - & - \\
\citet{das09} & - & 4.76 & 5.71 & 5.71 & - & - & - & - \\
\citet{Fischer12} & 1.0 & 2.9 & 3.8 & 4.3 & - & - & - & $<e>$ \\
\citet{Roberts12} & - & 2.6 & 3.9 & 3.5 & - & - & - & $<e>$ \\
\tableline
\end{tabular}
\tablenotetext{1}{Utilized  to derive the $ \pm 1\sigma$ error ellipse.}
\tablenotetext{2}{based upon \citet{kei03b}}
\end{center}
\end{table}
%\end{landscape}
\clearpage

%\newpage
\begin{table}
\begin{center}
\caption{Percentage contribution to the total neutrino flux and detected events for various redshift bins.}
\label{tbl.5} 
%\rotate
\begin{tabular}{ccccccc}
\tableline
 $z_{min}$ & $z_{max}$  &     \% events  &  \% flux \\ 
\tableline
0. &     5. &     100. & 100. \\ 
\\
0.0 &     0.1 &     5.9 &  3.7 \\ 
0.1 &     0.2 &     6.6 & 4.6 \\
0.2 &     0.3 &     7.2 &  5.3 \\
0.3 &     0.4 &     7.6 & 6.0 \\ 
0.4 &     0.5 &     7.8 &  6.7 \\
0.5 &     0.6 &     8.0 & 7.2 \\
0.6 &     0.7 &     7.9 & 7.6 \\
0.7 &     0.8 &     7.8 & 7.9 \\
0.8 &     0.9 &     7.6 & 8.0 \\
0.9 &     1.0 &     6.9 & 7.7 \\
1.0 &     1.1 &     5.7 & 6.6 \\ 
1.1 &     1.2 &     4.5 & 5.4 \\ 
1.2 &     1.3 &     3.5 & 4.4 \\ 
1.3 &     1.4 &     2.7 & 3.5 \\ 
1.4 &     1.5 &     2.1 & 2.9 \\
1.5 &     1.6 &     1.7 &  2.3 \\
1.6 &     1.7 &     1.3 & 1.9 \\ 
1.7 &     1.8 &     1.0 & 1.5 \\
1.8 &     1.9 &     0.8 & 1.2 \\
1.9 &     2.0 &       0.7 & 1.0 \\
2.0 &     2.1 &     0.5 & 0.8 \\ 
2.1 &     2.2 &     0.4 & 0.7 \\
2.2 &     2.3 &     0.3 & 0.6 \\ 
2.3 &     2.4 &     0.3 &  0.5 \\
2.4 &     2.5 &     0.2 & 0.4 \\
\\
2.5 &     5.0 &     1.0  &  1.6 \\
\tableline
\end{tabular}
%% Any table notes must follow the \end{tabular} command.
\tablenotetext{1}{This table is for  Case {\it III} (no oscillation).}
\tablenotetext{2}{The contributions to the  flux and event rate between z = 0 - 5.0  sum to  100$\%$.}
\end{center}
\end{table}

%\newpage
\begin{table}
\begin{center}
\caption{Percentage contributions to the total neutrino flux and detected events for various redshift bins using  the SFR models 
  of \citet{kob00} with  dust extinction corrected data  (DC) or data  without dust correction (no DC).}
  \label{tbl.6} 
\begin{tabular}{ccccccc}
\tableline
 $z_{min}$ & $z_{max}$  &     DC flux &  no DC flux & DC  Events  & no DC Event \\ 
\tableline
0.0 &  0.1 &  6.7 & 8.9 & 9.6 & 12.0 \\ 
0.1 &  0.2 &  6.4 & 8.4 & 6.7 & 10.7 \\ 
0.2 &  0.3 &  5.9 & 7.8 & 7.6 &  9.4 \\ 
0.3 &  0.4 &  5.5 & 7.2 & 6.7 & 8.3 \\ 
0.4 &  0.5 &  5.3 &  7.0 &  6.1 &  7.5 \\ 
0.5 &  0.6 &  5.2 &  6.8 &  5.7 &  7.0 \\ 
0.6 &  0.7 &  5.1 & 6.7 & 5.4 & 6.6 \\ 
0.7 &  0.8 &  5.1 &  6.7 & 5.1 &  6.3 \\ 
0.8 &  0.9 &  5.5 & 6.9 &  5.3 &  6.3 \\ 
0.9 &  1.0 &  6.6 &  6.3 & 6.1 & 5.4 \\ 
1.0 &  1.1 &  7.1 & 5.5 &  6.3 & 4.6 \\ 
1.1 &  1.2 &  6.5 &  4.4 &  5.5 &  3.6 \\ 
1.2 &  1.3 &  5.5 &  3.6 &  4.6 &  2.8 \\ 
1.3 &  1.4 &  4.6 &  2.9 &  3.7 & 2.2 \\
1.4 &  1.5 &  3.8 &  2.3 & 3.0 &  1.7 \\ 
1.5 &  1.6 &  3.1 &  1.8 & 2.3 &  1.3 \\
1.6 &  1.7 &  2.5 &  1.4 & 1.8 &  1.0 \\ 
1.7 &  1.8 &  2.0 &  1.1 &  1.4 & 0.8 \\ 
1.8 &  1.9 &  1.9 & 0.9 &  1.1 &  0.6 \\
1.9 &  2.0 & 1.3 & 0.7  & 0.9 &  0.5 \\
2.0 &  2.1 &  1.0 &  0.6 &  0.7 &  0.4 \\ 
2.1 &  2.2 &  0.8 &  0.4 &  0.5 &  0.3 \\ 
2.2 &  2.3 &  0.6 & 0.3 &  0.4 & 0.2 \\ 
2.3 &  2.4 &  0.5 &  0.3 &  0.3 &  0.2 \\ 
2.4 &  2.5 &  0.4 &  0.2 &  0.3 &  0.1 \\
\\
2.5 &  5.0 &  1.5 &  0.7 &  0.9 &  0.4 \\ 
\tableline
\end{tabular}
%% Any table notes must follow the \end{tabular} command.
\tablenotetext{1}{This table is for  Case {\it III} (no oscillation).}
\tablenotetext{2}{The contributions to the  flux and event rate between z = 0 - 5.0  sum to  100$\%$.}
%\tablenotetext{3}{above: without dust correction, under: with dust correction}
\end{center}
\end{table}
%\newpage

\begin{table}
\begin{center}
\caption{SRN detected events for 10 yr running in a  10$^6$ ton water \v{C}erenkov detector
for various redshift bins and  oscillation cases {\it I, II, III}. \label{tbl.7}}
\begin{tabular}{ccccccccccc}
\tableline\tableline
$(T_{\bar \nu_e}, T_{\nu_x}$) & total & z=0-1 & 1-2 &  2-3 & 3-4 & 4-5\\
%\multicolumn{1}{c}{$P$\tablenotemark{a}} & $P R_{maj}$ & $P R_{min}$ &
%\multicolumn{1}{c}{$\Theta$\tablenotemark{b}} \\
\tableline
\\
Oscillation Case {\it I}\\
\tableline
(6.7,7.6) & 1092 & 744 & 299 & 41 & 8.3 & 0.8 \\
(5.0,6.0) & 814 & 587 & 202 & 22 & 3.0 & 0.3 \\
(6.0,6.5) & 963 & 674 & 252 & 31 & 4.9& 0.5 \\
(4.1,6.5) & 716 & 522 & 173 & 18 & 2.6 & 0.3 \\
(3.9,5.6) & 649 & 483 & 150 & 14 & 1.9 & 0.2 \\
(4.5,4.7) & 682 & 508 & 156 & 15 & 1.8 & 0.1 \\
(2.5,2.5) & 314 & 245 & 63 & 5.8 & 0.8 & 0.1 \\
\tableline
\\
Oscillation Case {\it II}\\
\tableline
(6.7,7.6) &  1102 & 728 & 316 & 47 & 9.2 & 1.1 \\
(5.0,6.0) & 851 & 592 & 226 & 28 & 4.4 & 0.4 \\
(6.0,6.5) & 934 & 638 & 255 & 34 & 5.8 & 0.6 \\
(4.1,6.5) & 934 & 638 & 255 & 34 & 5.8 & 0.6 \\
(3.9,5.6) & 782 & 552 & 203 & 24 & 3.5 & 0.3 \\
(4.5,4.7) & 622 & 457 & 148 & 14.6 & 1.9 & 0.2 \\
(2.5,2.5) & 2281 & 178 & 44 & 4.6 & 0.71 & 0.1\\
\tableline
\\
Oscillation Case {\it III}\\
\tableline
(6.7,7.6) & 10891 & 752 & 292 & 38 & 6.4 & 0.7 \\
(5.0,6.0) & 800 & 587 & 191 & 19 & 2.4 & 0.2 \\
(6.0,6.5) & 975 & 689 & 251 & 30 & 4.5 & 0.4 \\
(4.1,6.5) & 633 & 481 & 138 & 12 & 1.3 & 0.1 \\
(3.9,5.6) & 595 & 456 & 127 & 11 & 1.2 & 0.1 \\
(4.5,4.7) & 708 & 523 & 162 & 15 & 1.7 & 0.1 \\
(2.5,2.5) & 350 & 273 & 70 & 6.4 & 0.8 & 0.1 \\
\tableline
\end{tabular}
%% Any table notes must follow the \end{tabular} command.
\tablenotetext{1}{Each case represents the same pairs of neutrino temperature
  shown in Fig.\ref{fig:8}.}
\end{center}
\end{table}
%\newpage

\begin{table}
\begin{center}
\caption{Neutrino temperature dependence of the total SRN detection events for a  10$^6$ton water \v{C}erenkov detector 
with a 10 yr run time for  oscillation cases {\it I, II, III}. \label{tbl.8}.}
\begin{tabular}{lccccc}
\tableline\tableline
{($T_{\bar \nu_e}, T_{\nu_x}$)}&  {$N(I)$}&  {$N(II)$}&  {$N(III)$}\\    
\tableline
( 6.7 , 7.6 ) & 836 & 826 &  842 \\
( 6.0 ,	6.5 ) & 745 & 709 &  761 \\
( 5.0 ,	6.0 ) & 638 & 650 &  636 \\
( 4.1 ,	6.5 ) &  567 & 709 & 515 \\
( 3.9 ,	5.6 ) &  519 & 601 & 487 \\
( 4.5 ,	4.7 ) & 545 & 487 &  570 \\
( 2.5 ,	2.5 ) & 268 & 191 &  300 \\
\tableline
\end{tabular}
%% Any table notes must follow the \end{tabular} command.
\tablenotetext{1}{Each case represents the same pairs of neutrino temperature
  shown in Fig.\ref{fig:8}.}
\end{center}
\end{table}
%\newpage

 \clearpage
\begin{table}
\begin{center}
\caption{Parameters for the neutrino sources considered in this work.
\label{tbl.9}}
\begin{tabular}{lcccccc}
\tableline\tableline
{detail} &  {ONeMg SN} &
 {SNII} &  {SNIb,c} &  {fSN(SH EoS)}  &
 {fSN(LS EoS)} &  {GRB}\\
%\multicolumn{1}{c}{$P$\tablenotemark{a}} & $P R_{maj}$ & $P R_{min}$ &
%\multicolumn{1}{c}{$\Theta$\tablenotemark{b}} \\
\tableline
mass(M$_{\sun}$) & (8 - 10) & 8 - 25(10-25)$^1$ & 8-40(10-40)$^1$ &
 25 - 125 (99.96$\%$) & 25 - 125 (99.96$\%$) &  25 - 125 (0.04$\%$) \\ 
Remnant & Neutron Star & Neutron Star &  Neutron Star & Black Hole & Black Hole & Black Hole \\
Phenomenon & Dim Supernova & Supernova & Supernova & Failed Supernova & Failed Supernova & Gamma-Ray Burst \\
T$_{\nu_e}$(MeV) & 3.0 & 3.2 & 3.2 & 5.5 & 7.9 & 3.2 \\
T$_{\bar{\nu_e}}$(MeV) & 3.6 & 5.0 & 5.0 &5.6 & 8.0 & 5.3 \\
T$_{\nu_x}$(MeV) & 3.6 & 6.0 & 6.0 &6.5 & 11.3 & 4.4 \\
E$_{\nu_e}^{total}$(erg) & 3.3$\times 10^{52}$ & 5.0$\times 10^{52}$ & 5.0$\times 10^{52}$ & 5.5$\times 10^{52}$ &
8.4$\times 10^{52}$ & 1.7$\times 10^{53}$ \\
E$_{\bar{\nu_e}}^{total}$(erg) & 2.7$\times 10^{52}$ & 5.0$\times 10^{52}$ & 5.0$\times 10^{52}$ & 4.7$\times10^{52}$ &
7.5$\times 10^{52}$ & 3.2$\times 10^{53}$ \\
E$_{\nu_x}^{total}$(erg) & 1.1$\times 10^{52}$ & 5.0$\times 10^{52}$ & 5.0$\times 10^{52}$ & 2.3$\times 10^{52}$ &
2.7$\times 10^{52}$ & 1.9$\times 10^{52}$ \\
$\triangle$ t & few $s$ & few $s$ & few $s$ & - 0.5$s$ & - 0.5$s$ & -
10$s$
 \\
%Rate against total explosion number & (33$\%$) & 80$\%$ (47$\%$) & 20$\%$ &
%20$\%$ &  
\tableline
\end{tabular}
%% Any table notes must follow the \end{tabular} command.
\tablenotetext{1}{In the case that  ONeMg SNe are not
considered, the lower mass range for CC-SNe is taken to be 8 M$_\sun$. Otherwise,  the lower mass range is 10 M$_\sun$.}
%  with different flavors in fig.2 and fig.3}
\end{center}
\end{table}
\clearpage

\begin{table}
\begin{center}
\caption{SRN detection rate contributions  from core collapse SNe, ONeMg  SNe, failed SNe,  and GRBs  in a 10$^6$t  water \v{C}erenkov detector  with 10 yr run time. }
\label{tbl.10} 
\begin{tabular}{crrrrrrrrrrr}
\tableline\tableline
& total & z = 0-1 & 1-2 &  2-3 & 3-4 & 4-5\\
\tableline
Case {\it I} \\
\tableline
 Total & 814 & 586 & 201 & 21.8 & 3.01& 0.275 \\
SNe & 593 & 425 & 149 & 16.2& 2.21 & 0.200 \\
ONeMg  SNe & 62.7 & 51.1 & 10.9 & 0.612 & 0.0406 & 0.0019 \\
Failed SNe & 158 & 110 & 42 & 5.00 & 0.749 & 0.073 \\
GRB & 0.327 & 0.235 & 0.0815 & 0.0087 & 0.0012 & 0.00010 \\
\tableline
Case {\it II} \\
\tableline
 Total & 851 & 592 & 226 & 28.0 & 4.40 & 0.45 \\
 SNe & 713 & 494 & 191 & 23.6 & 3.69 & 0.37 \\
ONeMg SNe & 31.2 & 25.5 & 5.40 & 0.30 & 0.02& 0.00035 \\
Failed SNe & 107 & 72 & 30.2 & 4.07 & 0.69 & 0.08 \\
GRB & 0.0190 & 0.0152 & 0.0035 & 0.00028 & 0.000027 & 0.000002 \\
\tableline
Case {\it III} (no osc.)\\
 Total & 637 & 487 & 134 & 13.1 & 1.61 & 0.182 \\
 SNe & 412 & 315  & 87.3 & 8.41 & 0.992 & 0.107 \\
ONeMg SNe & 62.7 & 52.9 & 9.26 & 0.505 & 0.0324 & 0.0019 \\
failed SNe & 161 & 119 & 37.80 & 4.21 & 0.584 & 0.0738 \\
GRB & 0.020 & 0.016 & 0.004 & 0.0004 & 0.0003& 0.000002 \\
\tableline
%\multicolumn{1}{c}{$P$\tablenotemark{a}} & $P R_{maj}$ & $P R_{min}$ &
%\multicolumn{1}{c}{$\Theta$\tablenotemark{b}} \\
\tableline
\end{tabular}
%% Any table notes must follow the \end{tabular} command.
\end{center}
\end{table}
%\clearpage

\end{document}